\documentclass[a4paper,fleqn]{cas-dc}

\usepackage[authoryear,longnamesfirst]{natbib}
\usepackage{placeins}
\usepackage{caption}
\usepackage{etoolbox}
\usepackage[table,svgnames, x11names, dvipsnames]{xcolor} 
\usepackage{balance,color}
\usepackage{multirow}
\usepackage{pifont}
\usepackage{newunicodechar}
\newunicodechar{✓}{\ding{51}}
\usepackage{booktabs}
\usepackage{amsmath}
\usepackage{subcaption}
\usepackage{rotating}
\usepackage{float}
\usepackage{pdflscape}

\usepackage{xurl}     
\usepackage{hyperref}

\usepackage{wrapfig}

\usepackage{booktabs,tabularx}
\usepackage{pifont} 
\usepackage{amssymb} 
\newcommand{\simplecircledmark}{\textcircled{\ding{51}}}
\newcommand{\ignore}[1]{}

\def\tsc#1{\csdef{#1}{\textsc{\lowercase{#1}}\xspace}}
\tsc{WGM}
\tsc{QE}
\tsc{EP}
\tsc{PMS}
\tsc{BEC}
\tsc{DE}

\begin{document}
\let\WriteBookmarks\relax
\def\floatpagepagefraction{1}
\def\textpagefraction{.001}

\shorttitle{SMS Phishing Attacks and Defenses: A Systematic Review}

\shortauthors{Pritom et~al.}

\title [mode = title]{Short Message Service (SMS) Phishing Attacks and Defenses: A Systematic Review}                      



%
\author[1]{Mir Mehedi A. Pritom}[type=editor,orcid=0000-0000-0000-0000]

\cormark[2]


\ead{mpritom@tntech.edu}

\ead[url]{https://www.mirmehedipritom.com}


\affiliation[1]{organization={Department of Computer Science, Tennessee Tech University},
    addressline={1 William L Jones Dr}, 
    city={Cookeville},
    postcode={TN 38505}, 
    country={USA}}

\author[1]{Seyed Mohammad Sanjari}
\cormark[1]
\author[1]{Maraz Mia}[%
]


\author[1]{Ashfak Md Shibli}[%
]

\author[1]{S M Mostaq Hossain}[%
]

\author[1,2]{Muhammad Ismail}[%
]

\affiliation[2]{organization={Cybersecurity Education, Research, and Outreach Center, Tennessee Tech University},
    addressline={1 William L Jones Dr}, 
    city={Cookeville},
    postcode={TN 38505},
    country={USA}}

\author[3]{Shouhuai Xu}
\ead{sxu@uccs.edu}
\ead[URL]{https://xu-lab.org/}

\affiliation[3]{organization={Department of Computer Science, University of Colorado at Colorado Springs},
    addressline={1420 Austin Bluffs Parkway}, 
    city={Colorado Springs},
    postcode={CO 80918}, 
    country={USA}}

\cortext[cor1]{Corresponding author}
\cortext[cor2]{Principal corresponding author}



\begin{abstract}
SMS Phishing (also known as `smishing') is a growing deceptive social engineering (SE) attack that leverages mobile SMS 
to conduct cybercrimes such as stealing sensitive information or spreading malware by tricking users into interacting
with attackers' messages (e.g., responding to or clicking URLs). This threat has increased rapidly in recent years, causing \$470M in financial losses for United States users in 2024 alone. This threat is also evolving rapidly, meaning that attackers continually adapt their tactics, reshaping the landscape. There is a significant body of literature on investigating smishing attacks and defenses. However, there is no systematic review that reflects the current attack and defense landscape along with available resources (i.e., relevant datasets). 
This motivates us to systematize the current smishing research efforts, 
including the following four research pillars: (a) \textit{user perception and susceptibility}, (b) \textit{attack characterization}, (c) \textit{defense landscape}, and
(d) \textit{smishing datasets}. 
This leads us to propose novel future research directions towards effectively mitigating smishing attacks. 

\end{abstract}




\maketitle

\section{Introduction}
\label{sec:introduction}
Mobile devices are ubiquitous 
as we rely on them for daily communications, banking, shopping, and other online activities. However, this reliance has made mobile users a prime target for cybercriminals attempting to steal \textit{personally identifiable information} (PII), financial data and other sensitive information through fraudulent activities. This study focuses on one particular form of these cyber social engineering attacks, known as Short Message Service (SMS) phishing or `smishing' for short \citep{goel2018smishing}.

These attacks often use urgency or fear tactics to deceive recipients into clicking malicious links or revealing personal information, by making incoming fraudulent messages believable to end-users. 
The threat of smishing attacks is real, 
as evidenced by the Federal Trade Commission (FTC) report of \$470M in financial losses in 2024 due to these attacks
(5.5 times increase from \$86M in 2020) \citep{FTC_latest_report_2025}. 
Moreover, a recent Anti Phishing Working Group (APWG) report \citep{apwg_phishing_trend2024} shows that phone-based phishing attacks, including smishing and vishing (voice phishing) have grown by 
$30$\% 
in the first quarter of 2024 when compared with the previous quarter. 
This makes it imperative to keep up with the rapidly evolving smishing threats, which motivates the present study.

The rapid evolution of smishing attacks can be justified as follows.
Firstly, the available commercial anti-smishing tools are not robust enough 
\citep{commerial_anti_smishing_timko_2023}. Next, defenses such as blacklisting (based on message contents, including  URLs within texts) are ineffective as attackers continuously adapt their tactics in employing stealthy techniques 
\citep{jain2022content,aleroud2017phishing}. Moreover, the introduction of 
Artificial Intelligence (AI)-based chatbots, such as 
ChatGPT, 
has ushered in a new  threat vector at the disposal of smishing attackers \citep{shibli2024abusegpt,gupta_threatgpt,roy2024chatbots}. 
Recent studies on new perspectives of smishing attacks, such as 
users' perception and susceptibility to smishing attacks \citep{lutfor_23_codaspy_users_smished,ribeiro2024factors,tabassum2024drives_uncc,sarno2024gets_users,clasen2021friend,edwards2023smishing,alabdan2020phishing,blancaflor2023case}, further highlights the need for a systematic understanding of the smishing threat and defense landscape. 


\noindent{\bf Our Contributions}. Despite the impact of these evolving threats and the urgent need for updated defense strategies, the literature lacks systematic reviews of smishing attack tactics, existing defense strategies, and future research directions. To address this gap, we focus on four major areas in this paper: (a) \textit{user perception and susceptibility}, (b) \textit{attack characterization}, (c) \textit{defense landscape}, and (d) \textit{smishing datasets}. First, we aim to understand current users' perception of, and susceptibility to, smishing based on past user studies. Next, we characterize smishing threats and outline a bird's-eye view of the smishing threat landscape. Then, we study existing defense and detection mechanisms and describe their strengths and weaknesses, which helps identify open challenges. Lastly, we collect and analyze all the major smishing and spam SMS datasets available and draw insights through a data-driven investigation, which can help researchers in selecting the right dataset based on their research needs. 
Our analysis reveals several cross-cutting trends across these four pillars. First, studies on user perception and susceptibility largely focus on demographic and behavioral factors, with limited longitudinal and real-world validation. Second, attack characterization research has primarily emphasized URL-based and theme-based campaign analysis, while emerging LLM-driven and adaptive tactics remain underexplored. Third, the defense literature is dominated by machine learning and deep learning-based detection approaches, with comparatively less emphasis on explainable, user-centric, and infrastructure-level solutions. Finally, existing smishing datasets are often limited in scale, diversity, and temporal coverage. This makes a constraint for reproducibility and long-term evaluation of proposed defenses.
To summarize, this study makes the following key contributions:

\ignore{
{\color{olive}This research also discusses subtleties such as real-world mass data collection \citep{timko2024smishing}, model degradation over time with certain features \citep{sidhpura2023_fedspam:}, and the necessity of culturally sensitive diverse models for a global approach to this problem \citep{ghourabi2020hybrid}. Faklaris \citep{faklaris2024mitigating} discusses the challenges and future work in mitigating smishing attacks, emphasizing the need for continued research and innovation in this domain. }Ribeiro et al. \citep{ribeiro2016should} propose a method for explaining the predictions of any classifier, which can be applied to smishing detection models to enhance user trust and understanding of the detection process.
}


\begin{itemize}
\item Deep-dive into the existing literature on user-study driven smishing research to understand current users' perception, behavior, and susceptibility when exposed to smishing messages.  

\item Characterize the SMS phishing threat and attack landscape via classification schemes that are applicable to existing smishing messages.

\item Critically analyze the existing detection methods and approaches to defend against this evolving threat. 

\item Analyze the currently available public SMS datasets, their key features/characteristics, and models that leverage those datasets, and characterize their performances. We provide a {\em Smishing Data Hub} on GitHub for helping future smishing researchers.  


\end{itemize}

\noindent{\bf Related Work}. 
A brief summary comparison of related survey coverage on smishing research are provided in table \ref{tab:related_work_comparison} to illustrate the need for this study.
As depicted, Al Saidat et al. \citep{al2024advancements} presented an overview of SMS spam detection;
Edwards et al. \citep{edwards2023smishing} and Clasen et al. \citep{clasen2021friend} studied the human element of smishing by analyzing end-user behaviors;
Siddiqi et al. \citep{siddiqi2022study} examined the psychological mechanisms underlying social engineering attacks and reviewed existing countermeasures, including smishing-related threats; Sugunaraj et al. \citep{sugunaraj2022cyber_laws} provided a short review of cyber fraud targeting U.S. seniors, discussing scam types, communication vectors (including SMS), and preventive resources; Faklaris et al. \citep{faklaris2023preliminary} analyzed the demographic susceptibility of United State's users to smishing; Longtchi et al. \citep{XuPIEEE2024} provided a systematic examination of internet-based social engineering (SE) attacks, including smishing, through a psychological lens where various psychological factors (PFs) are exploited by attackers using a set of psychological techniques (PTs).

Recently, Rahaman et al. \citep{rahaman2025defending} has presented a comprehensive survey of smishing attacks, reviewing machine learning, deep learning, hybrid, and IoT-based detection techniques, along with associated challenges and future research directions.

In addition, several surveys have focused on phishing website detection. For example, Vijayalakshmi et al. \citep{vijayalakshmi2020web} presented a taxonomy and performance analysis of URL- and webpage-based detection techniques, while Safi and Singh \citep{safi2023systematic} conducted a systematic review of machine learning and deep learning approaches for phishing websites. However, these studies primarily address web-based attacks rather than SMS-based social engineering threats.
These studies touch upon some components of the smishing landscape, 
but lacking systematic treatment of the smishing landscape. For instance, 
the studies mentioned above did not cover new developments in the application of Large Language Models (LLMs) and Explainable AI (XAI) to enhance smishing detection and transparency. Moreover, there is no systematic comparison of publicly available smishing datasets, which is crucial to investigating defenses against smishing attacks. 
By contrast, the present study aims to present the first systematic treatment of smishing attacks and defenses.

\noindent{\bf Paper Outline}. The rest of the paper is organized as follows. 
Section \ref{sec:research_questions} describes the methodology. 
Section \ref{sec:user_study} systematizes user perceptions and susceptibility to smishing attacks.
Section \ref{sec:threat_landscape} characterizes the smishing attack/threat landscape. 
Section \ref{sec:defense_landscape} systematizes the smishing defense landscape and existing methodologies.
Section \ref{sec:datasets} assesses existing smishing and SMS 
datasets available for research. 
Section \ref{sec:future_directions} discusses future research directions while section \ref{sec:conclusion} concludes the paper.


\begin{table*}[t]
\centering
\footnotesize
\caption{Comparison of Related Survey Works Across Key Dimensions}
\label{tab:related_work_comparison}
\rowcolors{2}{white}{gray!10}

\resizebox{\textwidth}{!}{%
\begingroup
\renewcommand{\arraystretch}{1.5}
\setlength{\tabcolsep}{10pt}
\begin{tabular}{|c|c|c|c|c|c|c|c|}
\toprule
\textbf{Ref.} &
\makecell{\textbf{User Perception} \\ \textbf{\& Susceptibility}} &
\textbf{Attack} &
\textbf{Defense} &
\textbf{Dataset Analysis} &
\makecell{\textbf{Regulatory} \\ \textbf{\& Policy}} &
\textbf{PF/PT} &
\textbf{Literature Review} \\
\midrule

\citep{al2024advancements} & x & x & ✓ & x & x & x & ✓ \\
\citep{XuPIEEE2024} & ✓ & ✓ & ✓ & x & x & ✓ & ✓ \\
\citep{siddiqi2022study} & ✓ & ✓ & ✓ & x & x & ✓ & ✓ \\
\citep{sugunaraj2022cyber_laws} & ✓ & ✓ & ✓ & x & ✓ & x & ✓ \\
\citep{rahaman2025defending} & x & ✓ & ✓ & x & x & x & ✓ \\
This Work & ✓ & ✓ & ✓ & ✓ & ✓ & ✓ & ✓ \\
\bottomrule
\end{tabular}
\endgroup
}
\end{table*}

\ignore{

\begin{table*}[!ht]
\centering
\footnotesize
\caption{Comparison of Related Survey Works Across Key Dimensions}
\rowcolors{2}{white}{gray!10}

\label{tab:related_work_comparison}
\resizebox{\linewidth}{!}{
\begingroup
\renewcommand{\arraystretch}{1.5} 
\setlength{\tabcolsep}{10pt} 
\begin{tabular}{|c|c|c|c|c|c|c|c|c|c|}
\hline
\toprule
\textbf{Ref.} & 
\textbf{User Perception \& Susceptibility} & 
\textbf{Attack} & 
\textbf{Defense} & 
\textbf{Dataset Analysis} & 
\textbf{Regulatory \& Policy} & 
\textbf{PF/PT} & 
\textbf{Literature Review} \\

\midrule
Al Saidat et al. (2024) \citep{al2024advancements}&    x & x & ✓ & x & x & x & ✓  \\
\hline

Longtchi et al. (2024) \citep{XuPIEEE2024} & ✓ & ✓ & ✓ & x & x & ✓ & ✓   \\
\hline

 Siddiqi et al. (2022) \citep{siddiqi2022study} & ✓ & ✓ & ✓ & x & x & ✓ &  ✓  \\
\hline

 Sugunaraj et al. (2022) \citep{sugunaraj2022cyber_laws}
 & ✓ & ✓ & ✓ & x & ✓ & x &  ✓  \\
\hline

Rahaman et al. (2025) \citep{rahaman2025defending} & x & ✓ & ✓ & x & x & x &  ✓   \\
\hline

 This Work  & ✓ & ✓ & ✓ & ✓ & ✓ & ✓ & ✓   \\
\hline
 
\end{tabular}
\endgroup
}
\end{table*}
}


\section{Scope and Methodology}
\label{sec:research_questions}
This paper focuses on smishing attacks, defenses, and datasets, meaning that other kinds of cyber social engineering attacks (e.g., Internet-based cyber social engineering attacks, vishing, web phishing)
are deemed out of scope. 

\begin{figure*}[t]
    \centering
\includegraphics[width=1\textwidth]{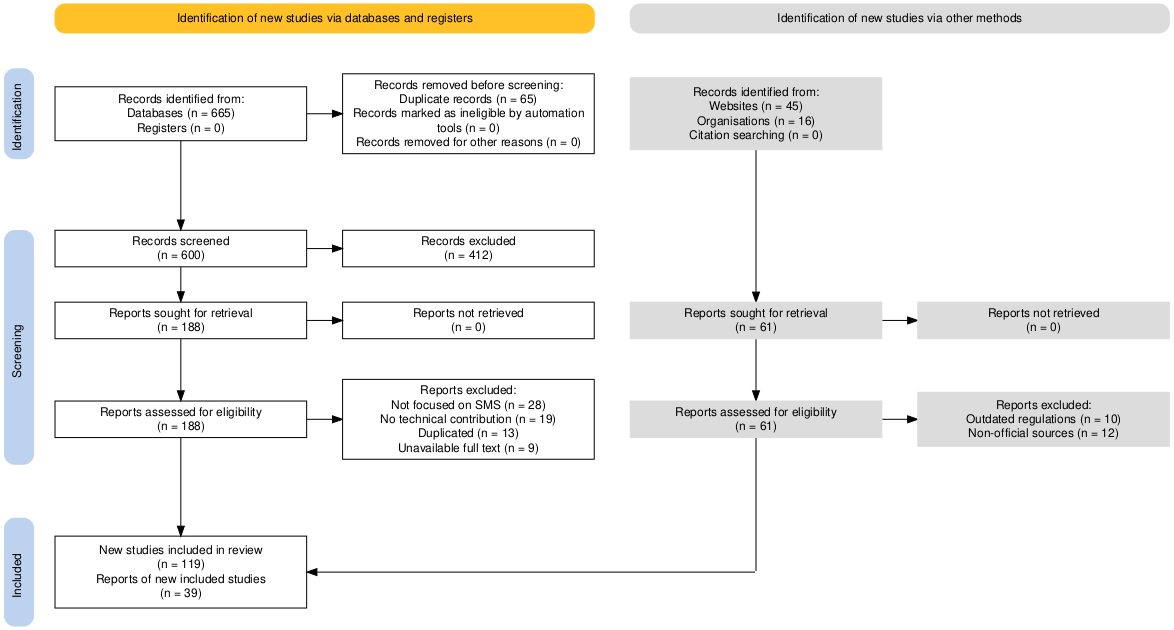}
    \caption{PRISMA flowchart for systematic literature review of SMS phishing research}
    \label{fig:prisma}
\end{figure*}

Figure \ref{fig:thematic-diagram} presents a pictorial summary of the scope of this paper where we systematically study the four pillar areas of smishing research such as the \textit{user perception and susceptibility}, \textit{attack characterization}, \textit{defense landscape}, and \textit{smishing datasets}, which are elaborated below. Accordingly, this work follows a systematic literature review (SLR) methodology and presents a taxonomy-based synthesis supported by a structured search strategy, defined inclusion and exclusion criteria, and a PRISMA-based screening process (Figure \ref{fig:prisma}).


\begin{center}
  \includegraphics[width=\columnwidth, height=1.75\columnwidth]{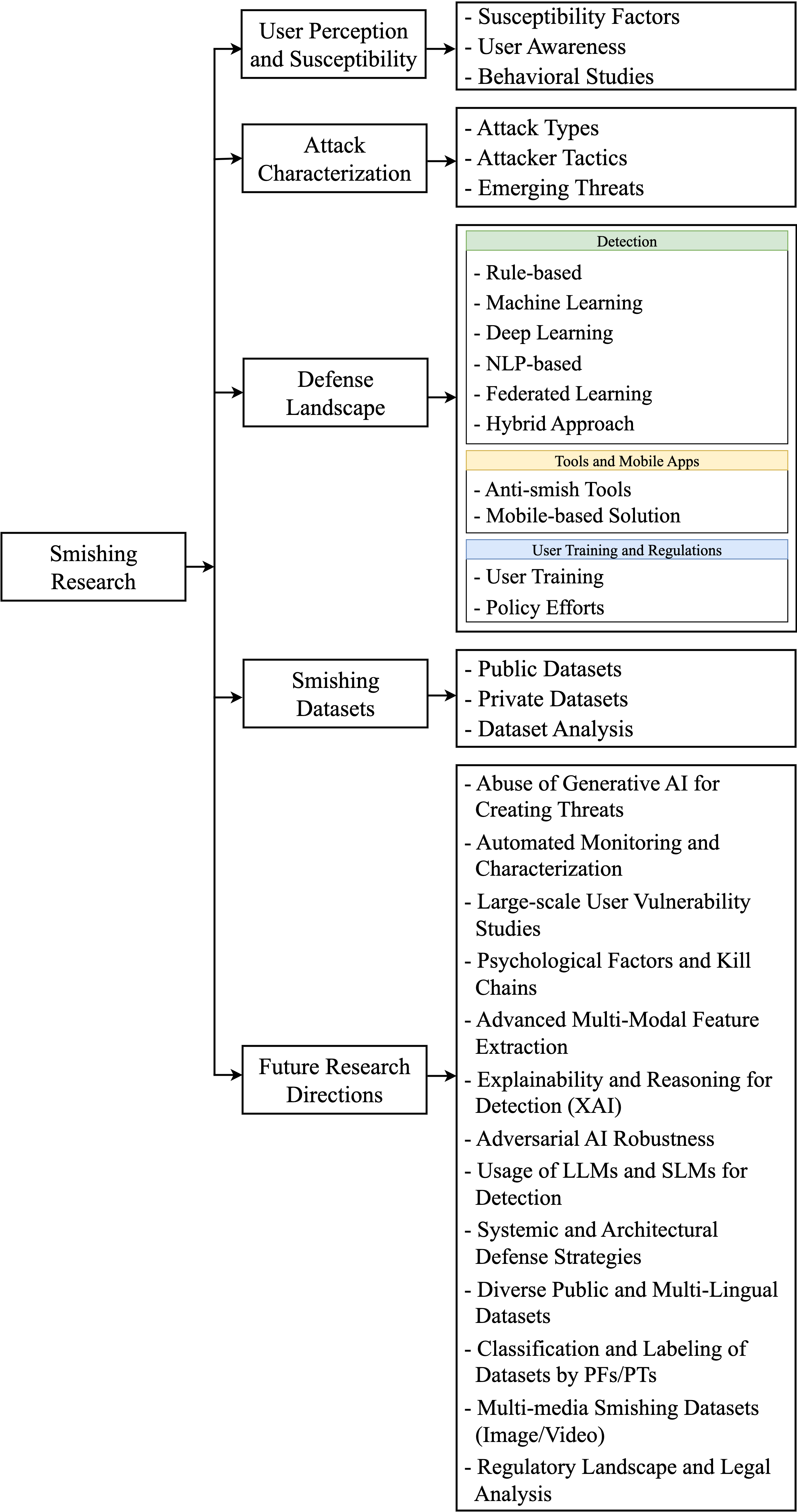}
  \captionof{figure}{Scope of the present study.}
  \label{fig:thematic-diagram}
\end{center}


\begin{itemize}


\item \textbf{User Perception and Susceptibility:} This 
pertains to user behaviors, perceptions, and especially user susceptibilities when exposed to smishing threats. 

\item \textbf{Attack Characterization:}
This includes 
attack types, attack tactics, emerging threats, and PTs (psychological techniques). 

\item \textbf{Defense Landscape:} This systematizes the strengths, weaknesses, and transparency of existing detection mechanisms reported in the literature, 
while considering practical defense countermeasures, anti-smish tools, mobile-based solutions, user training, awareness modules, and existing regulatory efforts. 

\item \textbf{Smishing Datasets:} This characterizes 
publicly available SMS datasets and better summarizes the contents therein. 
This allows us to 
create a unified Smishing Data Hub provided at \url{https://github.com/MarazMia/SMISH\_DT}.
\end{itemize}

\noindent{\bf Search Strategy and Selection Criteria.} Given the defined scope of this study, we conducted a systematic literature review to identify relevant research on smishing across the Google Scholar and DBLP databases. The search queries utilized the following Boolean combinations of keywords: (``smishing'' OR ``SMS phishing'') AND (``detection'' OR ``defense'' OR ``dataset'' OR ``characterization'' OR ``user susceptibility''). 

We applied the following Inclusion Criteria (IC) and Exclusion Criteria (EC) to filter the initial search results:
\begin{itemize}
    \item \textbf{IC1 (Relevance):} The study must primarily address SMS-based phishing (smishing). Studies focusing on general phishing or social engineering were included only if they provided essential theoretical frameworks (e.g., psychological factors) that are directly transferable to the smishing domain.
    \item \textbf{IC2 (Language):} Only articles published in English were included.
    \item \textbf{IC3 (Source Type):} Peer-reviewed conference proceedings, journal articles, and reputable technical reports were included.
    \item \textbf{EC1 (Duplicates):} Duplicate entries across databases were removed.
    \item \textbf{EC2 (Scope):} Articles that discussed "mobile security" broadly without a specific focus on social engineering via SMS were excluded.
\end{itemize}

This process has initially yielded a large set of publications. To manage the volume of results, we limited the screening to the first 10 pages of search results (sorted by relevance) for each keyword combination to ensure the most significant contributions were captured. To ensure the quality and relevance of the selected papers, we deployed a multi-step screening protocol where papers initially were screened based on their titles and abstracts to filter out irrelevant works. Subsequently, we reviewed the full text of these articles to ensure the alignment of these papers with the four research pillars. We conducted an expert review (by \textit{five} co-authors) to verify the selected papers and reduce bias. Any disagreements regarding inclusion or categorization were resolved by unanimous decision among the research group. Finally, a total of 119 papers have been retained after a careful manual screening for relevance based on the defined scope for this paper.

\begin{figure}
    \centering
    \includegraphics[width=1\linewidth]{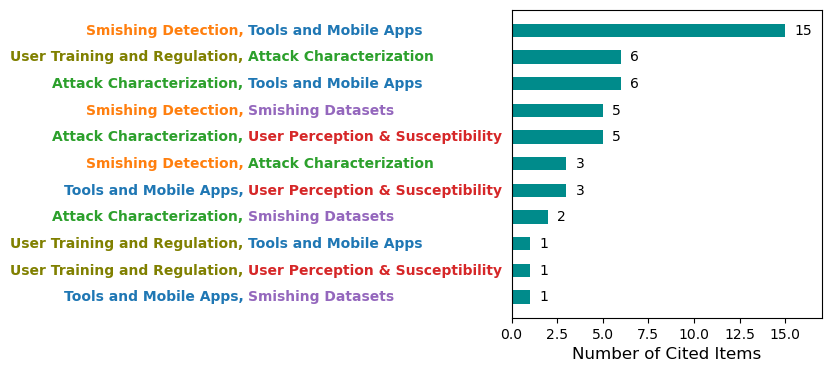}
    \caption{Count of overlapping papers and materials between different categories}
    \label{fig:overlapping_papers}
\end{figure}

In addition, to examine existing regulatory measures and law enforcement efforts addressing smishing, we conducted a targeted \textit{Grey Literature} search separated from our academic literature review. 
We queried official government databases (e.g., Federal Register, EUR-Lex) and legal repositories using specific policy-related keywords such as \textit{GDPR}, \textit{TCPA}, \textit{Telecommunications Act}, \textit{smishing penalties}, \textit{caller ID spoofing regulations}. This search targeted documents from major regulatory providers, including the U.S. Federal Communications Commission (FCC), the Federal Trade Commission (FTC), and the European Commission. This search resulted in the identification of 23 legal documents and 16 online reports, which were subsequently analyzed to extract insights into current regulatory frameworks and policy initiatives. Moreover, we have used 2 webpages or APIs (Application Programming Interfaces) and 3 online repositories containing SMS datasets along with 12 other datasets having their individual papers. 
We also identified 10 applications and anti-smishing tools within the scope of this review. 
Figure \ref{fig:paper_coverage} presents a bar graph illustrating the yearly distribution of papers and other available materials (e.g., reports, repositories, apps, legal documents) under our defined categories. Figure \ref{fig:paper_coverage} shows a significant increase in smishing research starting in 2020, likely due to more mobile communication during the COVID-19 pandemic. Publications on the \textit{Defense Landscape} spiked in 2023 and 2024, in such it covers over 40\% of the total paper. This reflects the community's response to new AI-driven threats and the push for more accurate detection tools. It is important to note that some papers and materials are considered under multiple categories (i.e., a paper can be both in \textit{User Perception and Susceptibility} category and \textit{Attack Characterization} category) based on their contents as shown in Figure \ref{fig:overlapping_papers}. This figure highlights the interdisciplinary nature of recent research. The large overlap between \textit{Attack Characterization} and \textit{Defense Landscape} suggests that modern defenses are increasingly built on insights into attack tactics rather than in isolation. In contrast, the minimal overlap between \textit{User Training and Regulation} and Technical Defenses (\textit{Defense Landscape (Detection)} and \textit{Defense Landscape (Tools \& Mobile Apps)}) reveals a gap between policy and technical implementation. It also shows that a single study often contributes to multiple areas due to the interdisciplinary nature of smishing research. For example, papers that categorize attack characterization often propose detection methods and place them in both \textit{Attack Characterization} and \textit{Defense Landscape}. To capture this, we used a multi-label tagging approach. We assigned papers to multiple categories based on their substantial contributions to pillars, instead of limiting them to a single primary label. It is important to note that, no single paper has contributed to more than 2 pillar categories.

\noindent\begin{minipage}{\columnwidth}
  \centering
  \includegraphics[width=0.88\columnwidth]{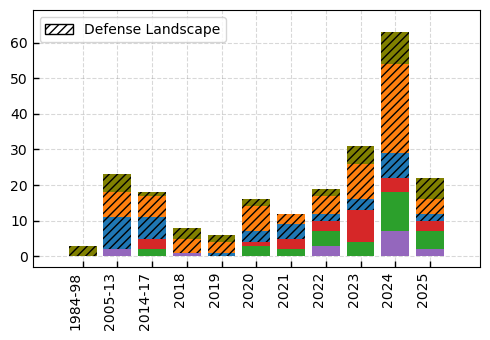}
  \captionof{figure}{Selected 119 papers, 23 legal documents, 16 online reports, 10 applications, 3 repositories, and 2 APIs with specific categories and publication years (until 2025), some items fall under multiple paper categories;}
  \label{fig:paper_coverage}
\end{minipage} 
\raisebox{0.4ex}{\colorbox[HTML]{d62728}{\color[HTML]{d62728}\rule{0.1cm}{0.1cm}}} \textbf{User Perception \& Susceptibility (25 papers, 1 report)} 
\raisebox{0.4ex}{\colorbox[HTML]{2ca02c}{\color[HTML]{2ca02c}\rule{0.1cm}{0.1cm}}} \textbf{Attack Characterization (22 papers, 9 reports)} 
\raisebox{0.4ex}{\colorbox[HTML]{ff7f0e}{\color[HTML]{ff7f0e}\rule{0.1cm}{0.1cm}}} \textbf{Defense Landscape (Detection) (Detection 7 Apps, 66 paper, 1 report)} 
\raisebox{0.4ex}{\colorbox[HTML]{1f77b4}{\color[HTML]{1f77b4}\rule{0.1cm}{0.1cm}}} \textbf{Defense Landscape (Tools \& Mobile Apps) (10 Apps, 24 papers, 2 Web/API)} 
\raisebox{0.4ex}{\colorbox[HTML]{808000}{\color[HTML]{808000}\rule{0.1cm}{0.1cm}}} \textbf{Defense Landscape (User Training and Regulation) (23 legal documents, 2 papers, 13 reports)} 
\raisebox{0.4ex}{\colorbox[HTML]{9467bd}{\color[HTML]{9467bd}\rule{0.1cm}{0.1cm}}}  \textbf{Datasets (12 papers, 3 repositories)} 


\subsection{Methodology}

Our methodology is driven by Research Questions (RQs) in association with the user perceptions and susceptibility, smishing attack characterization, drawing smishing defense landscape, and findings insights from smishing datasets. 
This systematization paves the way for proposing future research directions towards effectively mitigating smishing threats. 


\subsubsection{RQs Related to User Perception and Susceptibility} 

Understanding user perceptions and susceptibility are important to comprehend smishing threats. Hence, we propose the following three RQs to understand user's perception and susceptibility when exposed to smishing attacks based on existing user-studies:
\begin{itemize}
    \item \textbf{RQ-B1:} How much susceptible are users of different backgrounds when encounter smishing messages?
    \item  \textbf{RQ-B2:} Does the literature address a diverse user groups when conducting behavioral or quantitative survey studies? 
    \item \textbf{RQ-B3:} What are cues used by users to manually identify smishing messages? 

\end{itemize}


\subsubsection{RQs Related to Smishing Attack Characterization}
Attack characterization is important to completely understand smishing attack classifications, types, themes, and psychological aspects (i.e., techniques, factors). 
These key findings are essential for creating better  
user training materials, effective user notification system, and robust smishing detection and defense mechanisms.  
Thus, we propose the following two RQs:
\begin{itemize}
    \item \textbf{RQ-A1:} What are the current known threat classification and attack types for smishing threats? 
    \item \textbf{RQ-A2:} What are the various themes, topics, psychological factors (PFs), and psychological techniques (PTs) in the context of smishing threats? 
\end{itemize}


\subsubsection{RQs Related to Smishing Defense Landscape}


After understanding the smishing threat landscape, we systematize the existing smishing defense landscape, covering both advanced detection methodologies, practical app and tool based defense solutions, user trainings, and policy-based regulatory efforts. We propose the following eight RQs:
\begin{itemize}

    \item \textbf{RQ-C1:} What types of detection approaches and models are proposed?

    \item \textbf{RQ-C2:} What are the key indicators (or features) use in existing detection methods?
    
    \item \textbf{RQ-C3:} What are the strengths and weaknesses of existing detection mechanisms? 

    \item \textbf{RQ-C4:} Are explainability and transparency considered to enhance trust within existing detection models?


    \item \textbf{RQ-C5:} What is the extent of current studies focusing on the defense capabilities and anti-smish tools? 
    
    \item \textbf{RQ-C6:} How do mobile-based solutions, such as mobile applications and user alert/notification system contribute to preventing smishing threats? 

    \item  \textbf{RQ-C7:} How effective are user-training and use awareness modules to defend against smishing attacks? 
    
    \item  \textbf{RQ-C8: } Are there any existing laws, legal frameworks, and effective regulatory policies around the globe to mitigate smishing threats and what contributions do they have in certain geographical regions?
    
\end{itemize}

\subsubsection{RQs Related to Investigate Smishing Datasets}
The availability of SMS datasets is crucial specifically for understanding current attack tactics as well as machine learning (ML) and AI-driven detection research. We propose the following three RQs to assess existing smishing datasets:

\begin{itemize}

    \item \textbf{RQ-D1:} What are the publicly available datasets for SMS phishing that are leveraged in current research?
    \item \textbf{RQ-D2:} What are the characteristics (e.g., number of instances, message length, containing URLs or not, live websites, etc.) and coverages (e.g., ham, spam, smish) of each of these smishing datasets? 
    \item \textbf{RQ-D3:} How these datasets are leveraged in existing literature and what types of detection models are adopted using such datasets? 
    
\end{itemize}

\ignore{
\begin{itemize}

    \item How do different machine learning algorithms, such as Naive Bayes, Random Forest, and Support Vector Machines, compare in terms of accuracy and efficiency in smishing detection?
    \item Can a comparative analysis of various smishing detection tools identify best practices and guide the development of more effective solutions?
    
\end{itemize}
}

\subsubsection{RQs Related to Future Research}
As one of the primary goal of this paper is to guide future research efforts to collectively defend against smishing threats and create a more secure mobile messaging ecosystem for the end-users, we propose the following five RQs in terms of future directions:
\begin{itemize}
    \item  \textbf{RQ-E1:} What are the potential new promising avenues of research in attack characterization and landscape of smishing attacks?
    \item \textbf{RQ-E2:} What can we do further to continue assess diverse users' susceptibility and behaviors to understand the needs for tailored defense? 
    \item \textbf{RQ-E3:} What are the future of detection and defense landscapes that can safeguard users from smishing attacks?
    \item \textbf{RQ-E4:} What type of large-scale and diverse datasets can further help improve the defense posture in fighting this threat?
    \item \textbf{RQ-E5:} What can we do further to understand and improve current regulatory and policy efforts including law enforcements actions in fighting smishing threats?   
     
\end{itemize}

\ignore{
\textbf{Legal and Regulatory Aspects}

\begin{itemize} 
    \item What are the existing legal frameworks and legislations specific to phishing, and how do they contribute to the mitigation of smishing attacks?
    \item What gaps exist in current legal approaches, and how can legislation be strengthened to address emerging threats in the domain of SMS phishing?
\end{itemize}
}



\section{User Studies On Smishing Susceptibility and Psychological Awareness of Users}
\label{sec:user_study}
\subsection{Users' Susceptibility To Smishing Attacks}
User behavior and perception are important in dealing with smishing attacks. Since a lack of knowledge in mobile security makes users easy targets for attackers \citep{yan2021awareness}, the illusion of knowledge is also dangerous in this field. The term ``security expertise bias" refers to a situation in which users with more security awareness are overly suspicious, which decreases their ability to determine legitimate and phishing messages. This indicates that users with advanced security training are making mistakes when classifying benign messages and smishing. Users also showed vulnerabilities when they believed themselves to be account holders of the entity named in the message. In this case, they disregard fraudulent cues and fall to smishing. This suggests that having a sense of familiarity with security issues and the sender leads to the opposite result \citep{faklaris2023preliminary}. On the other hand, involvement with smishing is sometimes only for curiosity, even past experiences may not influence falling for it again, as the two-round setup of experiment results shown in \citep{lutfor_23_codaspy_users_smished}. 
Users’ routines also influence their susceptibility to smishing. Research shows that frequent use of social media or banking apps makes messages in those contexts more trustworthy. For example, Users spending more time on social media, are more vulnerable to social media-related smishing \citep{kumarasinghe2023user}. Finally, based on a habit, users often enter credentials or respond to prompts without careful consideration \citep{mishra2019content}.

\subsection{Visual Cues To Detect Smishing}

Visual cues are elements designed by attackers to intimidate or emotionally manipulate users. Users can also determine whether a message is phishing or legitimate using these cues. The visual cues help users to detect suspicious patterns which are commonly integrated into phishing attempts \citep{clasen2021friend}. We have reviewed these visual cues and divided them into four main categories. In what follows, we discuss some of those important visual cues:

\subsubsection{Meta-Data-based Cues}

\noindent \textbf{(A) Fake URLs:} indicates Fraud. These URLs often do not match other message attributes. Users who usually focus on these URLs are more likely to detect phishing messages \citep{timko2023quantitative}. Research particularly found that a mismatch between URL and claimed service is frequent \citep{clasen2021friend}.

\noindent \textbf{(B) Shortened URLs:} Using a shortened URL in a message is a common strategy in smishing. This cue is widely identified as suspicious and helps to detect phishing messages \citep{clasen2021friend}.

\noindent \textbf{(C) Suspicious URLs:} Attackers may use a shortened or deformed shape that looks similar to the original URL but they modified it partially to trick users into clicking on it \citep{dsmish_mishra_2023}. These URLs often redirect users to a phishing website that also looks similar to a legitimate website and motivate them to enter their personal data, for example login information or credit card numbers. Another form of suspicious URL which is known as Self-Answering Links (SAL), pretends to be a subscribe-unsubscribe link for a service but includes link to a malicious website \citep{mishra2019content}.

\noindent \textbf{(D) Spoofed area code:} In this process, attackers use local area codes that are familiar to users. They are more likely to open and trust the unsolicited message, believing it to be from a local contact or business 
corporation. These messages appear to have been sent from a close geographical location, so users are more likely to trust them and respond accordingly. For instance, an attacker who sends a smishing message using a spoofed area code identical to a target's region, such as Tennessee's (931), is more likely to capture the attention of a user within that same area \citep{lutfor_23_codaspy_users_smished}.

\subsubsection{Content-based Cues}

\noindent \textbf{(A) Messages containing forms in URLs:}
Similar to suspicious URLs, these messages contain URLs that redirect users to phishing websites that are designed to motivate users to enter their sensitive data such as login information or financial data through forms \citep{mishra2019content}.

\noindent \textbf{(B) Messages asking for personal information:}
These kinds of messages ask their recipients to send back their sensitive data such as passwords or credit card credentials \citep{jain2019feature}. Attackers sometimes ask for this information indirectly; they first provide a phone number or email address and ask recipients to contact them. When they are contacted, attackers may ask them to send their personal and sensitive information for a fake reward like a coupon or gift \citep{mishra2019content}.

\noindent \textbf{(C) Messages asking the user to download an APK file:}
These smishing messages contain a link for downloading an Android Package Kit (APK) file. Once downloaded, they are able to do different harmful activities on the destination device  \citep{mishra2019content}.

\noindent \textbf{(D) Call to action (CTA):} Benign SMSs usually have a Transparent and Explicit Call to Action \citep{timko2023quantitative}. However, using a call to action was discovered on some smishing campaigns. These messages encourage recipients to click on a suspicious link, contact the phone number mentioned in the message to have a conversation with attackers, or ask them to send back valuable information \citep{lutfor_23_codaspy_users_smished}.

\noindent \textbf{(E) Entity identification:}
Entities, also known as organization names or brands, are a potential factor to identify smishing. The study conducted by Timko et al. \citep{timko2023quantitative} showed that paying attention to the entity helped participants correctly identify real messages. Attackers use this feature to lure users. They pretend as a known brand or organization like `Walmart' or a `Bank' to get the user's trust. This sense of trust enables users to respond to these phishing messages \citep{lutfor_23_codaspy_users_smished}. 




\noindent \textbf{(F) Urgency related text contents:}
Any kind of words or using specific linguistic terms that trigger a sense of urgency will make users act quickly. As we discussed before, those reward and fear-based scenarios have impacts on users to give them a sense of urgency \citep{jain2019feature}.  

\noindent\textbf{(G) Suspicious phrase or keywords as red flags:} Several studies have employed different methods, such as ML and text processing, to identify commonly used phrases and keywords in smishing attacks. A set of phrases or terms that are recurrently found is as follows: \{`award', `congratulation', `winner', `alert', `claim', `activate', `verify', `attempts', `gift voucher', `blocked', `suspend', `unlock', `won', `prize', `subscribe', `activity', `update', `coupon', `refund', `call', `free', `text', `claim', `ur(your)', `like', `get', `cash', `you have been selected', `£'\} \citep{jain2019feature,jain2022content,smdetector_ghourabi}.

\noindent \textbf{(H) Poor grammar, misspelled words, and special characters as red flags:}
Legitimate messages usually use well-written structure and good grammar. In contrast, phishing messages have a poor linguistic structure with grammatical and spelling problems \citep{clasen2021friend}. These linguistic mistakes, combined with replacing letters with symbols, special characters, and numerical values are sometimes used by attackers intentionally to bypass automatic spam detection systems and create ambiguous words to lure recipients \citep{dsmish_mishra_2023}.

\noindent \textbf{(I) Incentive/reward lures:}
Reward-based lures have become a major part of smishing attacks. Data from AhnLab’s Q4-2023 mobile threat report \citep{ahnlab2023} shows that job scam texts (e.g., “earn \$200 a day from home”), making up around $61.2\%$ of all malicious SMS traffic, surpassed credit card and government impersonation scams. However, some kinds of attacks, such as tax rebate offers, increase whenever those themes are in the news \citep{ahnlab2023}. 
In a recent study \citep{lutfor_23_codaspy_users_smished}, 265 volunteers were sent eight (8) fake text messages, and $16.9\%$ of users responded to at least one. Among the incentive and reward-based scams, a `fake Walmart gift card win' theme proved highly successful with around $30.00\%$ response rate. Additionally, a `free iPhone prize' theme also yielded a significant response rate of $18.33\%$. While these reward-based messages were effective, the overall analysis of the study shows no significant difference in success rates when comparing fear-motivated and reward-motivated messages. 

Another interview study on smishing susceptibility \citep{tabassum2024drives_uncc} provided insight into how incentive-based lures can be effective. While the study does not explicitly label the psychological principles, their findings offer clear examples of attackers leveraging persuasion tactics. The principle of reward is evident in deceptive offers for gift cards and fraudulent business promotions. These lures can be especially impressive when they exploit a user's existing expectations. It shows that one participant fell victim to a gift card scam because they were already anticipating the arrival of a legitimate reward. Scarcity is also used to create urgency through time limit demands, with phrases like ``within 4 hours'' being used in fraudulent messages to pressure a response. 
The authors also recommend that users should be cautious of such tactics, advising them to verify messages related to a prize or lottery wins \citep{tabassum2024drives_uncc}. 

\noindent \textbf{(J) Threat/fear appeals: }
Threat and/or fear appeals play a crucial role in the effectiveness of smishing attacks by benefiting from users' concerns about their financial security, account access, and personal data \citep{faklaris2023preliminary}. Smishing messages often use these techniques to pressure victims into acting quickly and make them respond without checking if the message is a legitimate one \citep{lutfor_23_codaspy_users_smished}. A common tactic involves account suspension or lockout warnings where attackers pretend to be banks or other service providers, to falsely claim that the user's account has been compromised or will be suspended due to suspicious activity or a failure to update a critical information. These messages often urge users to click on a provided link or call a given number to verify their details or \textit{`reactivate'} their account, leading them to phishing websites or in direct contact with the cyber-criminals who can then steal their credentials. The fear of losing access to essential services or money can significantly increase the probability of a user falling for such a scam as identified in previous studies \citep{mishra2019mit, faklaris2023preliminary}. Another common strategy is to present the threat of a coming financial penalty or fine. Smishing messages might claim that the recipient owes money to a government agency, bank, toll booths or other organization and will face quick penalties, like a fine or legal action, if they do not make a payment \citep{faklaris2023preliminary}. These messages often create panic and put pressure on users into quickly clicking on malicious links or giving away sensitive financial information to avoid the supposed negative consequences \citep{mishra2019mit}. In practice, we also observe data breaches or identity theft alerts to be used as baits to create fear and push users to act quickly. Attackers may send messages claiming that the user's personal information has been exposed in a data breach and that they must take immediate steps to secure their accounts or stop identity theft \citep{faklaris2023preliminary}. These alerts often send users to fake websites where they are asked to enter their login details or other personal information, which the attackers collect and use for other frauds \citep{mishra2019content}. The fear of their personal information being stolen or abused makes users more likely to fall for these types of smishing scams \citep{mishra2019mit}. The success of these fear-based tactics grows even more when the message mentions a company or service that is known by the recipient or may believe they already have an account with, which makes them less cautious \citep{faklaris2023preliminary}.

\begin{figure*}[!t]
\centering
\begin{subfigure}{0.48\textwidth}
    \includegraphics[width=\textwidth]{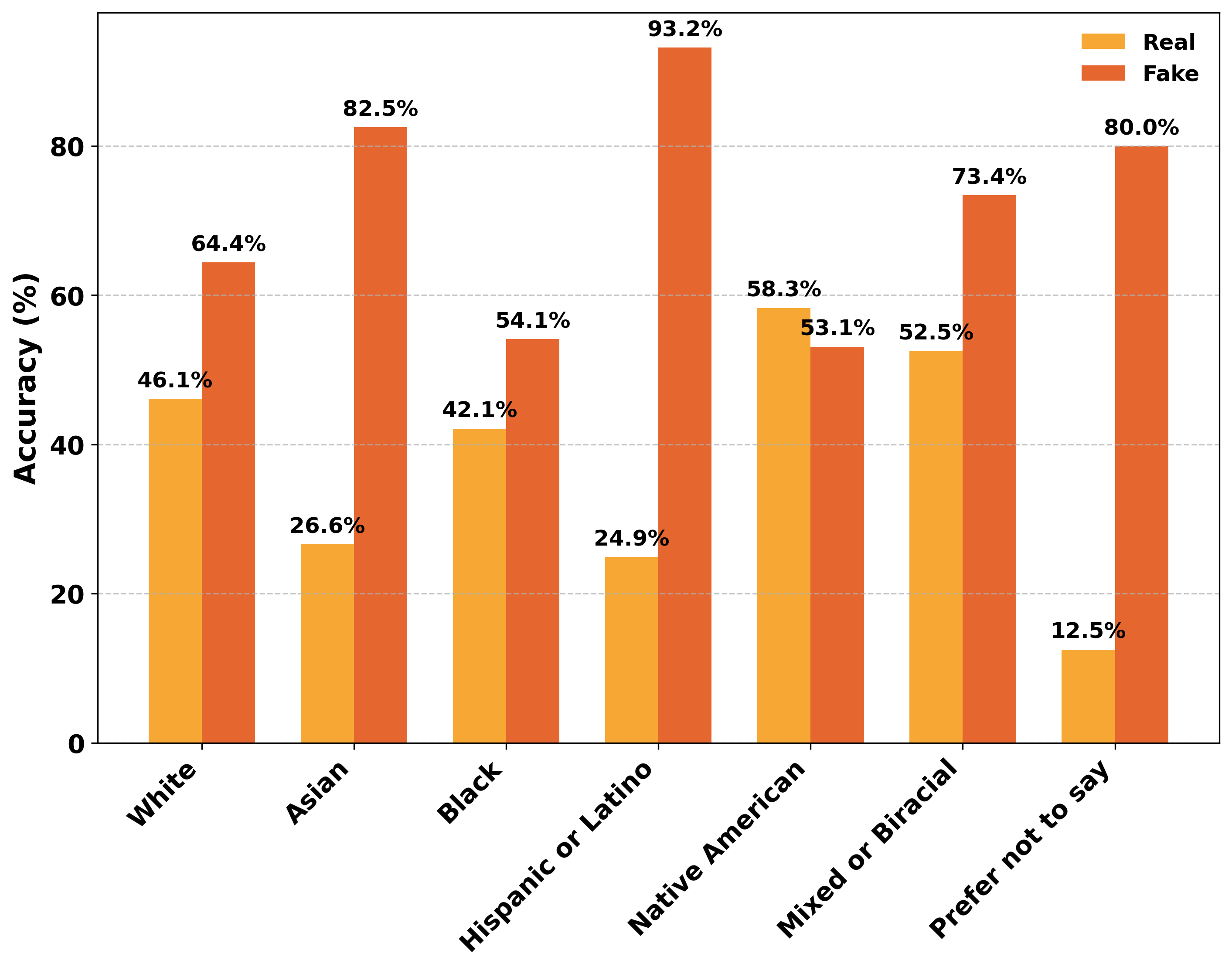}
    \caption{Detection accuracy (\%) by ethnicity}
    \label{fig:first}
\end{subfigure}
\hfill
\begin{subfigure}{0.48\textwidth}
    \includegraphics[width=\textwidth]{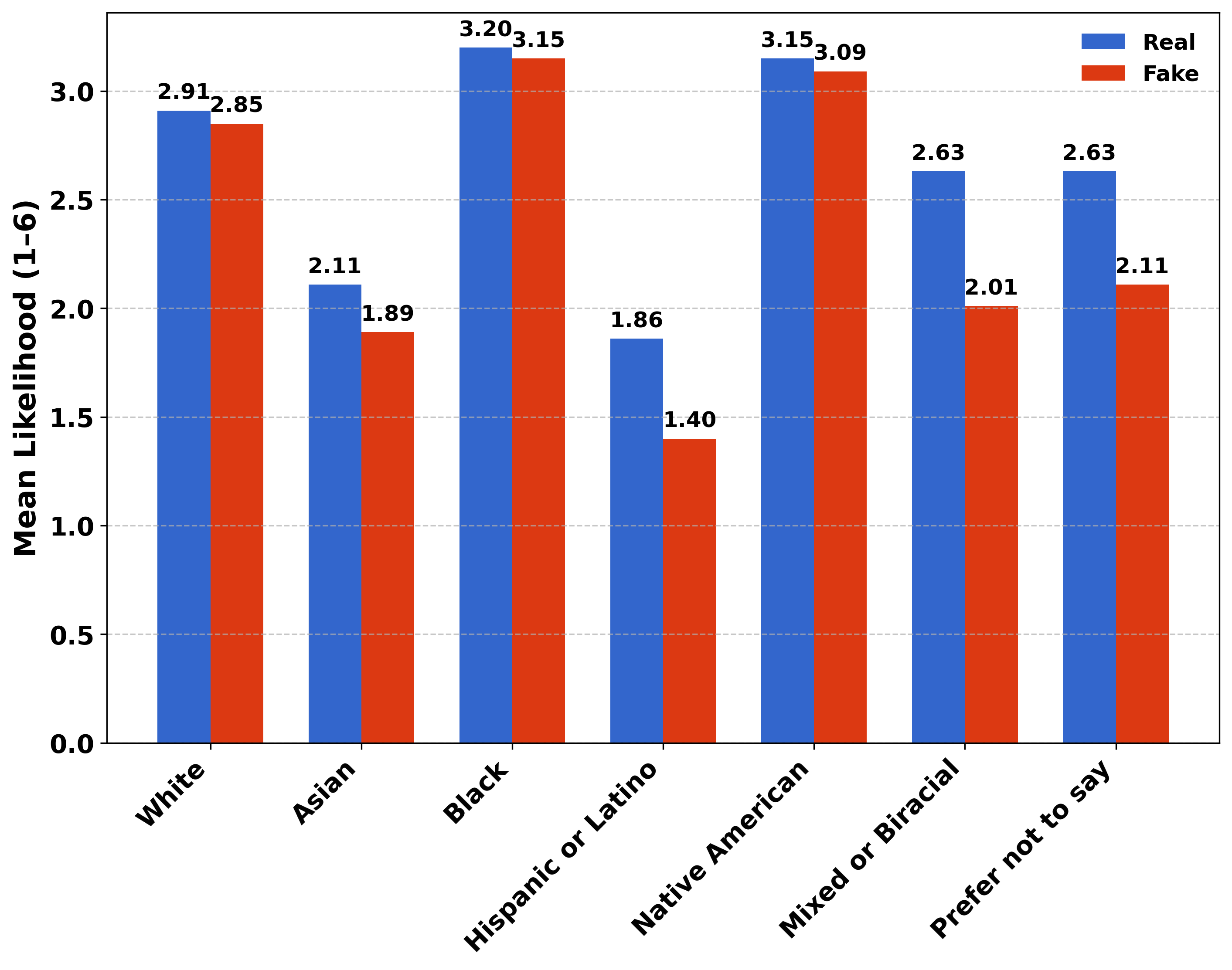}
    \caption{Interaction likelihood (1–6) by ethnicity}
    \label{fig:second}
\end{subfigure}
\hfill
\begin{subfigure}{0.48\textwidth}
    \includegraphics[width=\textwidth]{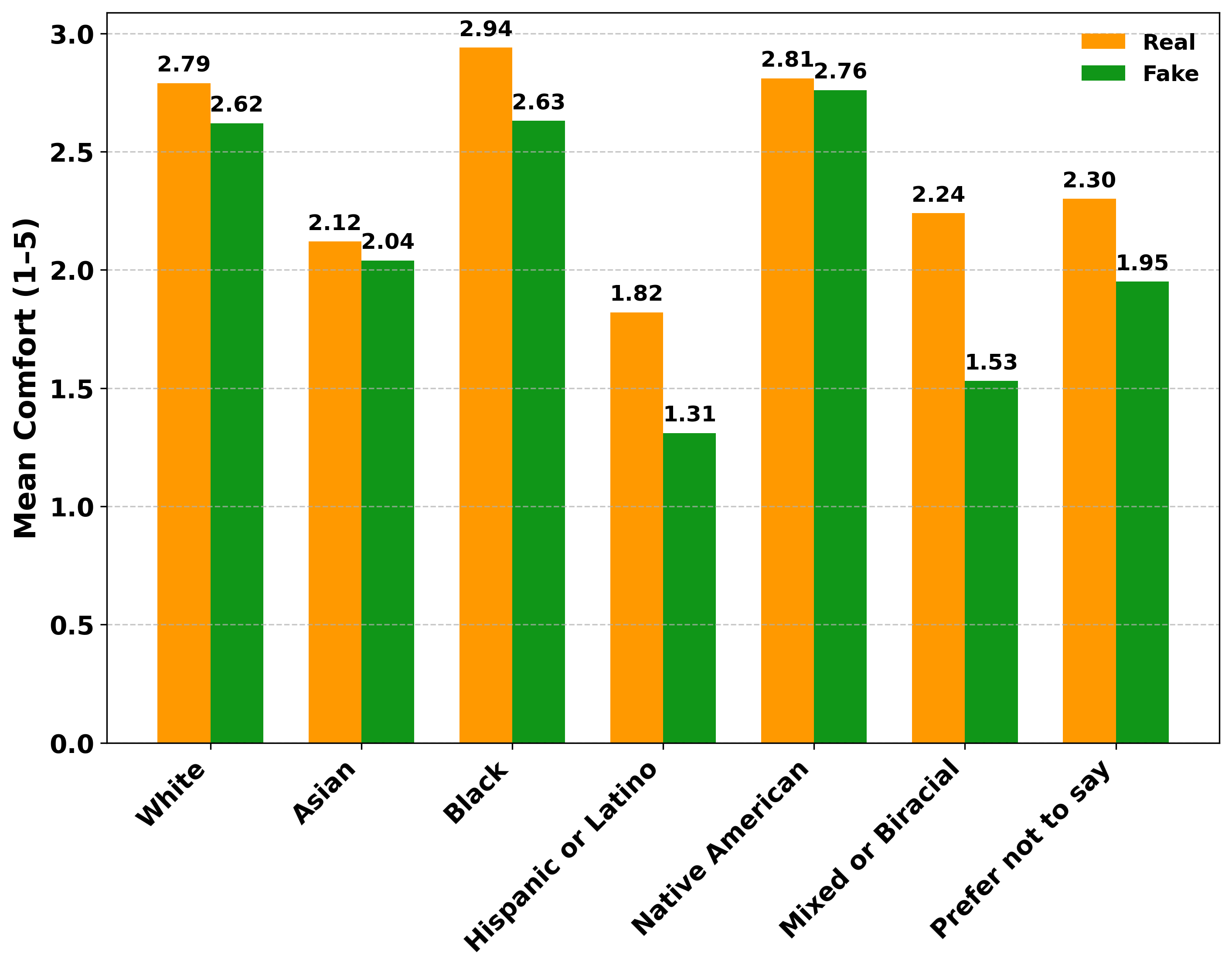}
    \caption{Comfort level (1–5 Likert) by ethnicity}
    \label{fig:third}
\end{subfigure}
        
\caption{Analysis of ethnicity in existing user studies (\cite{timkounderstanding}).}
\label{fig:3-panel-ethnicity}
\end{figure*}

\subsection{Diversity of Participants in Existing Studies}

Here, we discuss various existing user study research, which are dedicated to understand users' perceptions and susceptibility when exposed to smishing. We also examine whether demographic factors impact their susceptibility.

\begin{table*}[t]
\centering
\scriptsize
\caption{Overview of User-Study-Driven Smishing Research}
\label{tab:smishing_study_overview}
\renewcommand{\arraystretch}{1.2}
\setlength{\tabcolsep}{6pt}
\rowcolors{2}{white}{gray!10}
\begin{tabularx}{\textwidth}{p{3.2cm} X c p{2.8cm}}
\toprule
\textbf{Ref.} & \textbf{Key Finding} & \textbf{User Count} & \textbf{Study Type} \\
\midrule
\citep{multi_modal_neupane} &
EEG and eye-tracking show users often miss phishing cues but attend to malware warnings. &
23 &
Experimental user study \\

\citep{clasen2021friend} &
Examines how users distinguish legitimate SMS messages from smishing messages. &
576 &
Survey-based quantitative \\

\citep{blancaflor2021let} &
Evaluates user vulnerability to SMS phishing attacks. &
24 &
Experimental \\

\citep{bach2022targets} &
Analyzes phishing susceptibility across groups from European countries with different educational backgrounds using fuzzy clustering. &
N/A &
Quantitative analysis \\

 \citep{lutfor_23_codaspy_users_smished} &
Studies message and participant factors associated with smishing success. &
256 &
Experimental \\

\citep{edwards2023smishing} & Studying user behaviors and susceptibility to smishing attacks when exposed in controlled settings & 28 & Survey-based Behavioral \\

\citep{yoro2023evidence} & Investigating how demographic factors  and personality traits impact phishing susceptibility to selected university undergraduates in Nigeria & 480 & Quantitative Experimental\\

\citep{faklaris2023preliminary} & Investigating demographic and contextual factors impacting the vulnerability of user participants to smishing attacks & 1,007 & Survey-based Demographic Analysis \\
\bottomrule
\end{tabularx}
\end{table*}

\begin{table*}[t]
\centering
\scriptsize
\caption{Participant Diversity in User-Study-Driven Smishing Research}
\label{tab:smishing_study_diversity}
\renewcommand{\arraystretch}{1.2}
\setlength{\tabcolsep}{5pt}
\rowcolors{2}{white}{gray!10}
\begin{tabularx}{\textwidth}{p{2.8cm} X X X X}
\toprule
\textbf{Ref.} & \textbf{Ethnic/Racial Diversity} & \textbf{Age Diversity} & \textbf{Education Level} & \textbf{Gender Diversity} \\
\midrule
\citep{multi_modal_neupane} &
N/A &
19--22 (44\%); 23--26 (20\%); 27--30 (16\%); 31--34 (8\%); 35+ (12\%) &
Students (72\%); working professionals (16\%); others (12\%) &
Male (64\%); Female (36\%) \\

\citep{clasen2021friend} &
N/A &
18--21 (40.5\%); 22--25 (27.9\%); 26--30 (18\%); 31--40 (9.2\%); 41+ (4.4\%) &
N/A &
Male (50\%); Female (45\%); Others (5\%) \\

 \citep{blancaflor2021let} &
N/A &
N/A &
N/A &
N/A \\

\citep{bach2022targets} &
European countries &
N/A &
 C1 (6 countries) - Low education: 28.33\%, Medium education: 40.00\%, High education: 53.83\%; C2 (18 countries) - Low education: 7.11\%, Medium education: 10.17\%, High education: 16.06\%; C3 (11 countries) - Low education: 20.45\%, Medium education: 30.45\%, High education: 43.45\% &
N/A \\

\citep{lutfor_23_codaspy_users_smished} &
White (48.30\%); African American (17.35\%); Asian American (10.57\%); Mexican American (8.67\%); American Indian (8.30\%); Other Hispanic/Latino (4.90\%); Native Hawaiian (1.13\%); Other (0.75\%) &
18--24 (18.11\%); 25--34 (54.71\%); 35--44 (22.64\%); 45+ (4.5\%) &
$<$High School (1.88\%); High School (9.81\%); Associate (25.28\%); Some College (18.49\%); Bachelor’s (32.45\%); Master’s (9.05\%); Doctoral (3.01\%) &
Male (53.58\%); Female (46.42\%) \\

\citep{edwards2023smishing} &  N/A &  18 - 21
 & 
  High School/GED (53.6\%); Associate (3.6\%); Some College (42.9\%)

&  Male (32.1\%); Female (67.9\%)
 \\

 \citep{yoro2023evidence} &  
 Nigerian
 & 
 Under 21 (21.5\%); 21 - 25 (34.5\%); 26+ (44.0\%)
  & 

 1Y (36.0\%); 2Y (19.8\%); 3Y (19.5\%); 4Y and above (25.7\%)
 & 
     Male (50.9\%);
     Female (49.1\%) 
 \\

 \citep{faklaris2023preliminary} & 

 White (83.4\%); Black/African (10.5\%); Asian (2.38\%); Native American/ Alaska Native (0.9\%); Self-Described (1.5\%); Prefer not to say (1.3\%)
  &  18-24 (23.0\%); 25-34 (6.95\%); 35-54 (31.0\%); 55-64 (17.2\%); 65+ (21.7\%) 
 & 
 No school (53.37\%); In School (No 4Y deg.) (12.82\%); 4Y deg. (23.06\%); Doctorate (10.73\%)

&  Male (62.6\%); Female (36.0\%); Nonbinary (1.0\%); Self-described (0.2\%); Prefer not to say (0.2\%)
\\

\bottomrule
\end{tabularx}
\end{table*}
\subsubsection{Ethnic Diversity and Impacts}

Existing studies show that ethnicity plays an important role in how users perceive and react to smishing messages \citep{lutfor_23_codaspy_users_smished, timkounderstanding}. We have collected this ethnic diversity information from Timko et al. \citep{timkounderstanding} and created Figure \ref{fig:3-panel-ethnicity} to show how users from different ethnic groups interact with smishing messages. In a controlled detection study of 187 U.S. smartphone users, Timko et al. \citep{timkounderstanding} reported that Hispanic or Latino participants were the least likely to interact with both real and fake SMS messages and felt the least comfortable as shown in Figure \ref{fig:third}. 
However, they correctly identified fake messages with the highest accuracy ($93.2\%$). This suggests that Hispanic users tend to be more cautious during message interactions (shown in Figure \ref{fig:first}). 
White and Native American participants showed moderate engagement and accuracy, while Asian participants had low engagement and comfort and struggled to recognize real messages (only $26.6\%$ of real messages), although they were reasonably good at detecting fake messages (with $82.5\%$ of fake messages). Figure \ref{fig:second} shows the probability of interacting with real and fake
messages by ethnicity (on a scale from 1 to 6), which reveals that Black and Native American users are more likely to interact with messages compared to other ethnic groups. 
\begin{figure*}[!t]
\centering
\begin{subfigure}{0.48\textwidth}
    \includegraphics[width=\textwidth]{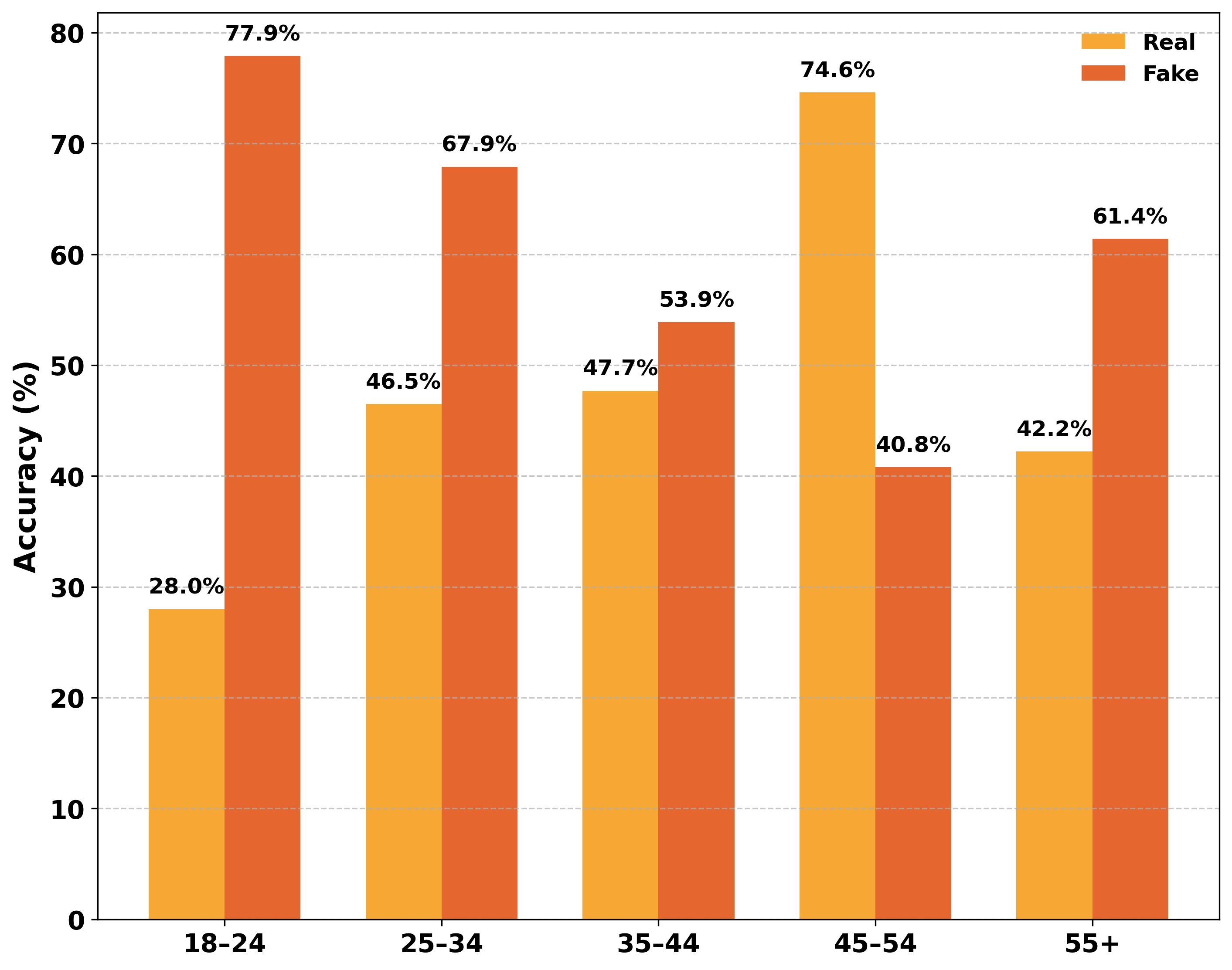}
    \caption{Detection Accuracy across age groups}
    \label{fig:age-accuracy}
\end{subfigure}
\hfill
\begin{subfigure}{0.48\textwidth}
    \includegraphics[width=\textwidth]{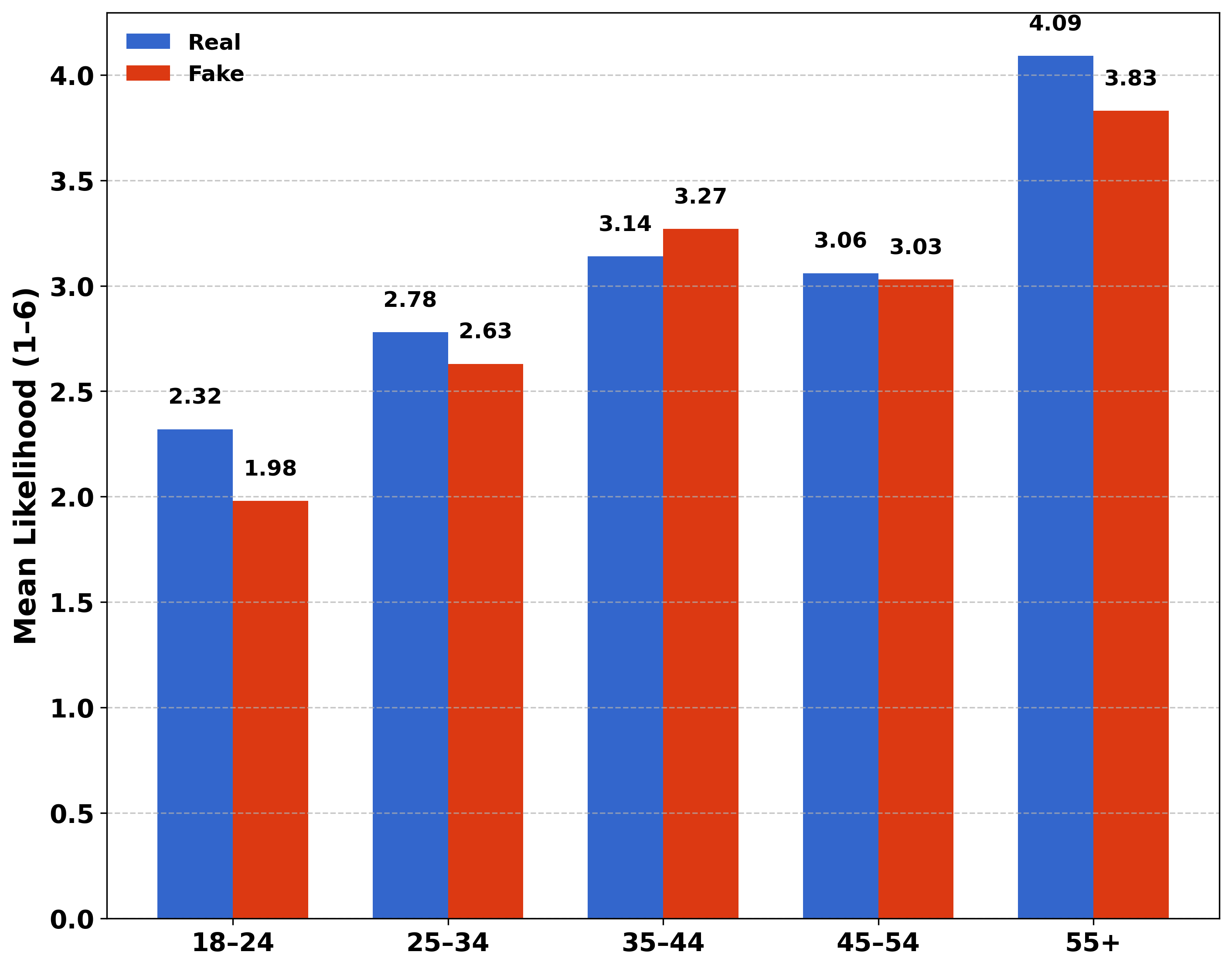}
    \caption{Interaction Likelihood across age groups}
    \label{fig:age-interaction}
\end{subfigure}
\hfill
\begin{subfigure}{0.48\textwidth}
    \includegraphics[width=\textwidth]{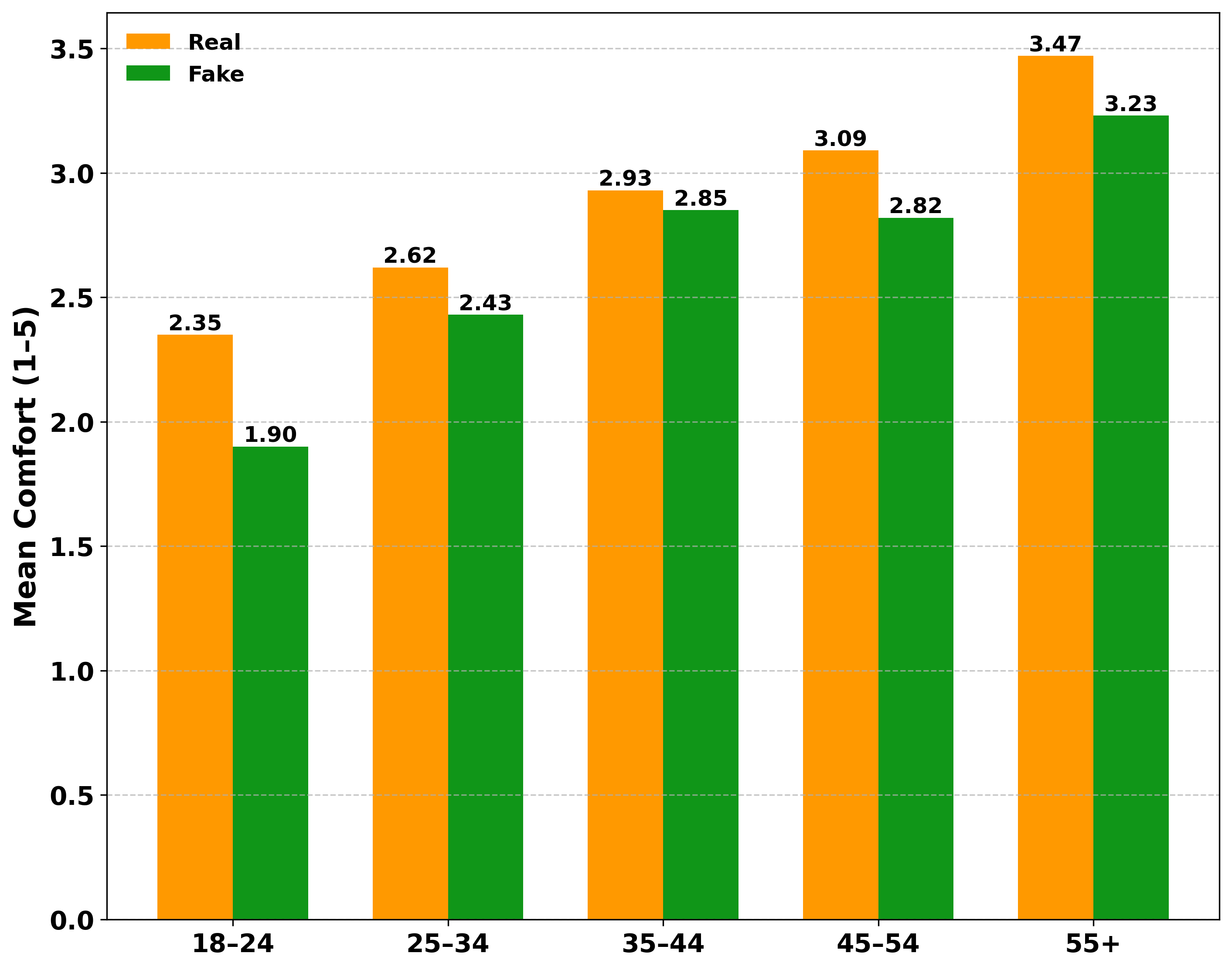}
    \caption{Comfort Level across age groups}
    \label{fig:age-comfort}
\end{subfigure}

\caption{Analysis of age-related patterns in existing user studies \citep{timkounderstanding}.}
\label{fig:3-panel-age-updated}
\end{figure*}


In another user study experiment of smishing messages with 265 human participants, Rahman et al. \citep{lutfor_23_codaspy_users_smished} found that African American ($19.6\%$ fell victim) and Mexican American ($21.7\%$ fall victim) users were much more likely to fall for smishing attacks than White users ($7.8\%$ fall victim), while Asian American users also showed similar results ($7.1\%$ fall victim). Together, these findings reveal a complex picture with some studies contradict in their results. Some groups, like Hispanic/Latino users, showed a 0\% smishing success rate in experimental settings, while others, like African American and Mexican American 
users, were more vulnerable to the smishing attacks conducted in this study's near-realistic environment. We believe that to address these differences, smishing awareness programs should be designed and curated for individualized communities and place emphasis to communities that are systematically more vulnerable than others. 

\ignore {
\begin{figure}[H]
  \centering
  \includegraphics[width=\linewidth]{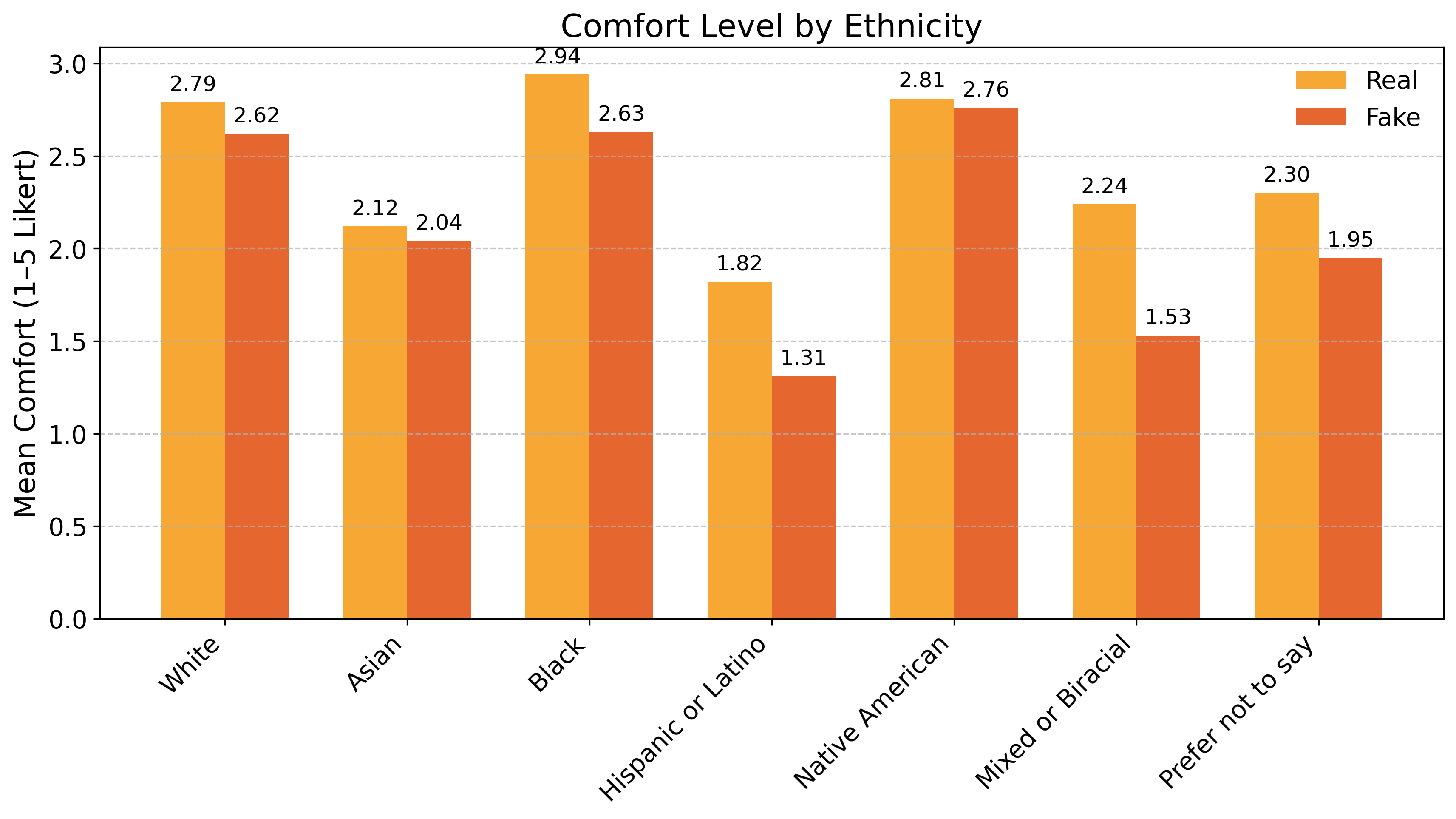}
  \caption{Mean comfort level with real and fake messages across ethnic groups, Data adapted from \citep{timkounderstanding}}
  \label{fig:comfort-ethnicity}
\end{figure}

\begin{figure}[H]
  \centering
  \includegraphics[width=\linewidth]{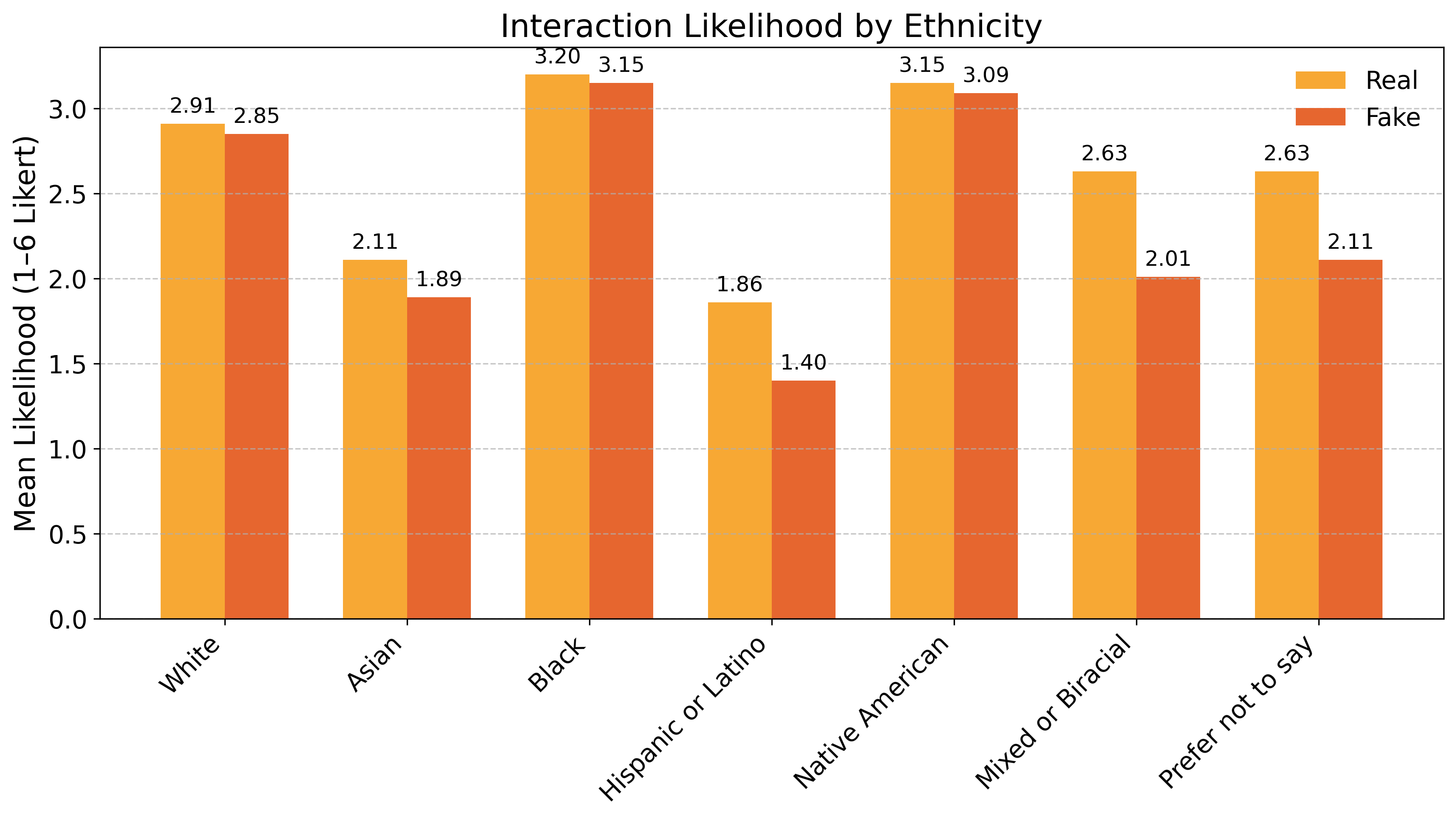}
  \caption{Mean likelihood of interacting with real and fake messages by ethnicity, Data adapted from \citep{timkounderstanding}}
  \label{fig:interaction-ethnicity}
\end{figure}

\begin{figure}[H]
  \centering
  \includegraphics[width=\linewidth]{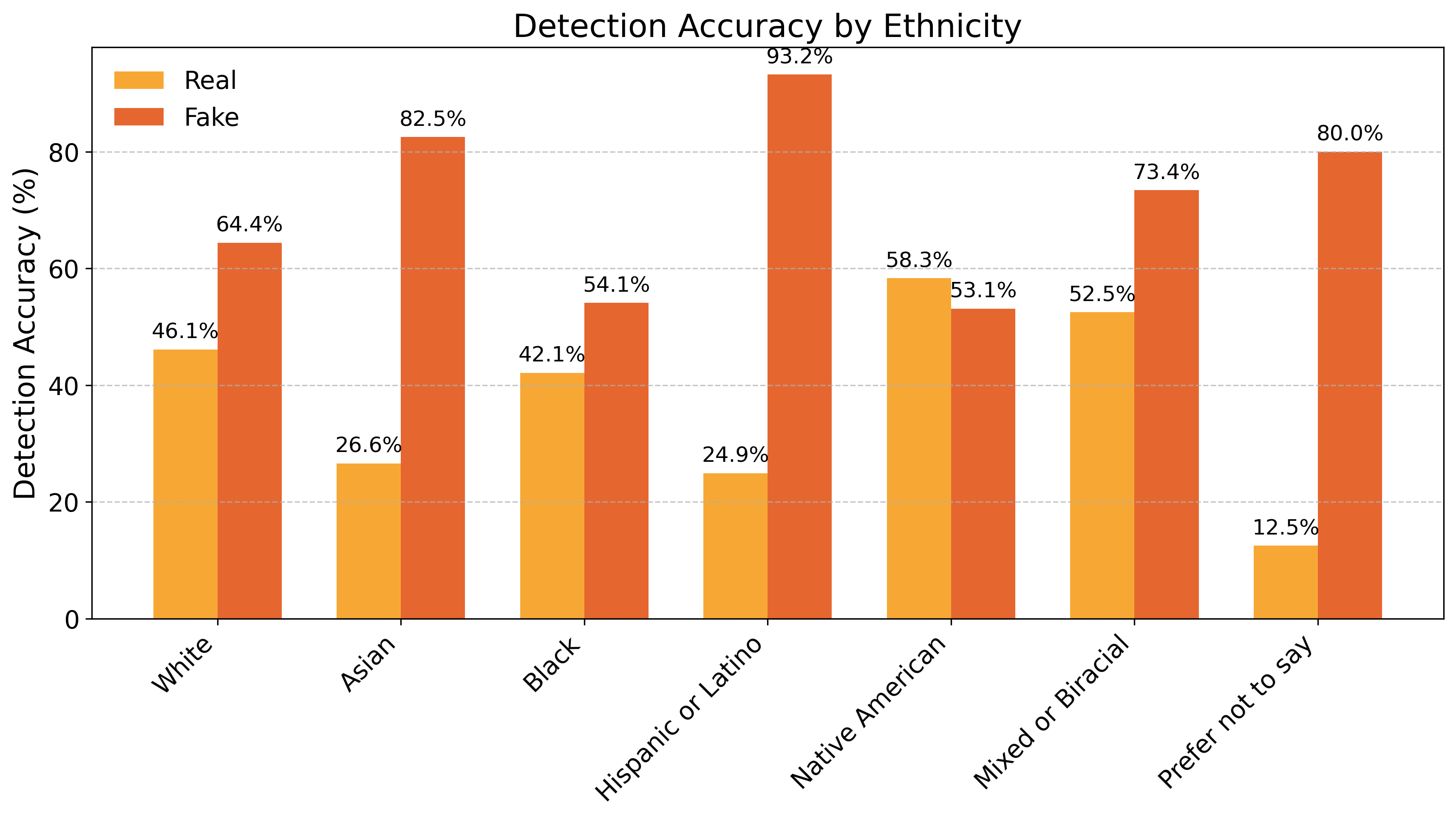}
  \caption{Detection accuracy for real and fake messages across ethnic groups, Data adapted from \citep{timkounderstanding}}
  \label{fig:detection-accuracy-ethnicity}
\end{figure}
}

\subsubsection{Age-related Patterns and Impacts} 
We have also explored how age affects users’ responses to smishing attacks as presented in existing literature. Based on the information gathered from Timko et al. \citep{timkounderstanding}, we have created Figure \ref{fig:3-panel-age-updated} to visualize patterns and impacts of age in smishing. Timko et al. \citep{timkounderstanding} found that people aged between $18$ and $24$ years were less likely to interact and less comfortable with real and fake messages than those aged between $25$ and $34$ years and between $35$ and $44$ years. In contrast, the group between $45$ and $54$ years felt more comfortable with real messages than the younger groups. However, age did not make much difference in how well people could identify real messages from smishing ones. Figure \ref{fig:3-panel-age-updated} 
shows detailed impact of various age groups on smishing attacks.

In another study, Rahman et al. \citep{lutfor_23_codaspy_users_smished} conducted real smishing attacks (using Twilio) with 265 U.S. participants to track `CLICK', `REPLY', and `CALL' responses. They found that users between $18$ to $24$ years had the lowest potential smishing success rate at $4.17\%$, while users over age $45$ were most vulnerable with $25\%$ successful smishing. This rate was in the intermediate level among people in the age group between $25$ and $34$ ($11.72\%$) years and between $35$ to $44$ ($10.00\%$) years.

The above findings indicate that while age does not significantly alter detection accuracy for distinguishing real SMS from smishing in lab settings, it does influence real-world behavioral responses. Younger adults (18–24) report lower comfort with both real and fake messages compared to older age groups and exhibit much lower interaction rates in actual smishing tests (approx. 4.2\% vulnerability). In contrast, adults over 45 express greater comfort with genuine messages, which corresponds with significantly higher vulnerability in practice (approx. 25\%) for smishing messages as well. Finally, although detection skills in controlled tests are similar across age groups, the more cautious behavior of younger users appears protective, while the greater comfort reported by older users correlates with increased real-world susceptibility to smishing. However, we need further studies to identify the detailed causes of these behaviors.

\subsubsection{Gender Impacts} Researchers have shown an interest in considering gender as one of the demographics in their susceptibility to smishing studies. Their findings have changed over time so that there is no significant difference in smishing vulnerability based on gender \citep{faklaris2023preliminary} as both genders showed vulnerability to smishing attacks. Prior findings reveal females had a slightly higher susceptibility (11.38\%) than males (9.86\%) \citep{lutfor_23_codaspy_users_smished}. In terms of categorizing legitimate and smishing messages, male participants had slightly more accurate results (74.8\%) than female participants (71.9\%) \citep{clasen2021friend}. But, in summary, there is no strong indication of gender impacts as per existing literature. 

\subsubsection{Educational Background Related Impacts} Research has also discussed that there is a direct relation between higher education level and susceptibility to fall victim to smishing attacks. In a user study experiment with 265 human participants \citep{lutfor_23_codaspy_users_smished}, the authors reported that people with a doctoral degree had the highest susceptibility at 37.5\% while those with less than a high school degree showed a susceptibility rate of 20\% and those with a high school degree showed 11.54\%. In addition, college students who are currently studying for a four-year degree were found to be highly susceptible \citep{faklaris2023preliminary}. 
It shows a descending distribution where first-year students were the most vulnerable ($82.9\%$) and it is decreasing during the next academic years in which fourth-year or above students had the lowest susceptibility rates ($21.3\%$) \citep{yoro2023assessing}.






\section{Characterization of Smishing Threats and Attack Landscapes}
\label{sec:threat_landscape}


\ignore{
\begin{figure}[!t]
  \centering
  \includegraphics[width=\linewidth]{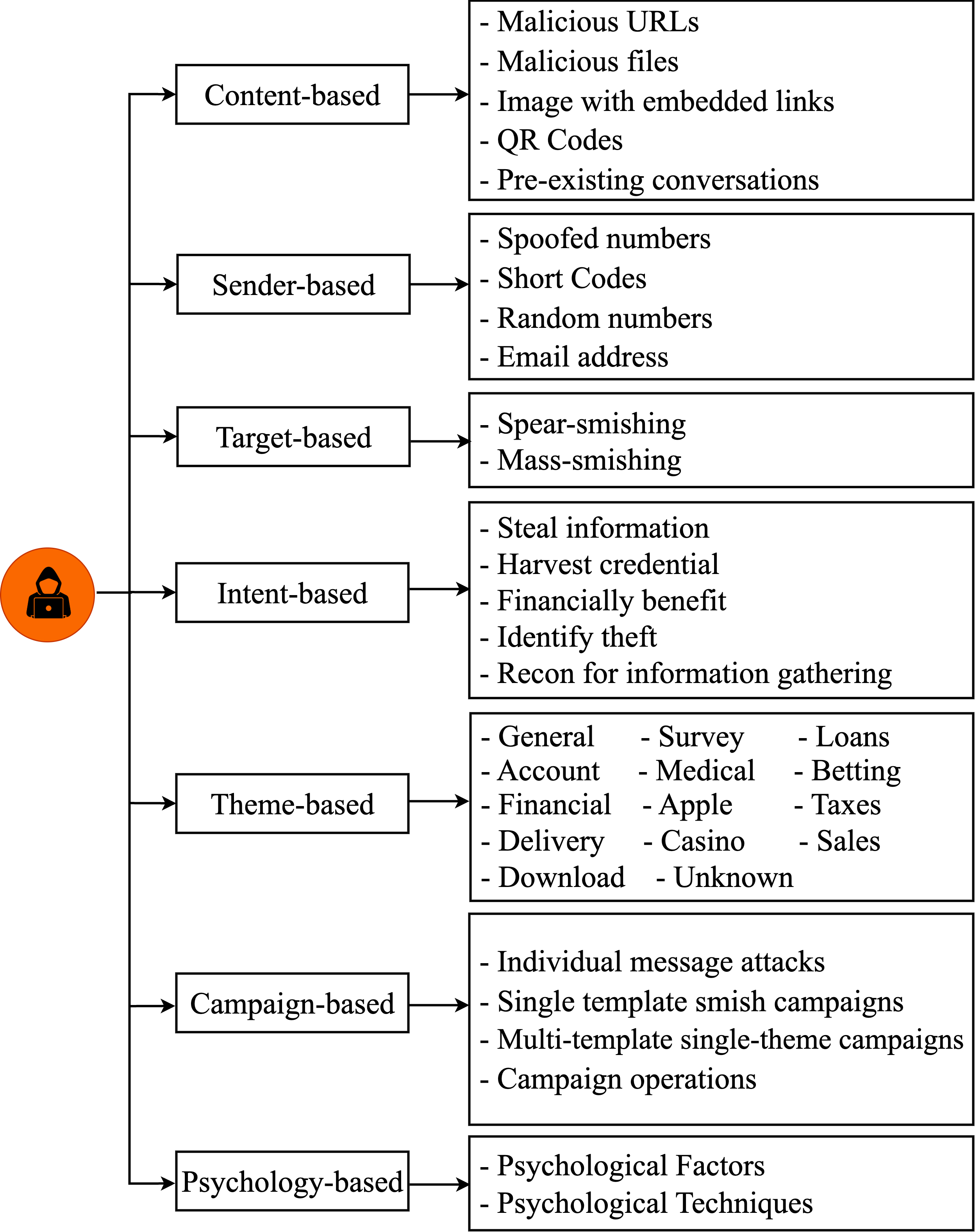}
  \caption{Overview of SMS phishing threat landscapes}
  \label{fig:threat-landscape}
\end{figure}
}
In this research pillar, we systematically categorize the smishing threats and introduce various classification schemes based on contents, senders, target-audience, attacker's intent, attack campaigns, attack themes and PTs. 
Smishing attacks can be classified in several ways based on diverse factors. Figure \ref{fig:threat-landscape} illustrates our defined taxonomy of SMS phishing threats, which includes content-based, sender information-based, target audience-based, intent-based, message delivery technique-based, message topic/theme-based, campaign-size based, and user psychology-based classification schemes. 
Next, we discuss each of these to systematically understand the various types and categories of smishing threats.


 
\subsection{Content-Based Classification}
    \vspace{-0.25mm}In terms of content diversity, we see smishing messages attached with \textit{malicious URLs} \citep{jain2022content}, \textit{malicious files} (linked to an *.apk file) \citep{smishing_detector_security_model_mishra_2020}, \textit{attached image (MMS)}\citep{proofpointGrowingThreat}, or \textit{QR codes}\citep{ftcScammersHide}. 
    Here, malicious URLs can direct users to a fake website or download malicious software (malware or ransomware) onto the mobile device, while malicious attachments included with an SMS may disguise themselves as a legitimate attachment that can infect the device. Moreover, malicious website links embedded within QR codes can make users fall victim to phishing attacks. Furthermore,  
     messages with images and/or graphics can trick people into revealing confidential information or falling for other scams \citep{proofpointGrowingThreat}.
    

\subsection{Sender Information-Based Classification}
    \vspace{-0.25mm}
    We can also classify smishing threats in terms of sender information such as- \textit{spoofed numbers}\citep{tsunoda2024demonstrating} that are used imitating legitimate organizations, \textit{short codes} \citep{haizam2024analysing} used in SMS marketing, \textit{random numbers} \citep{gupta2016exploiting} often broadcast to non-targeted audience, and \textit{email addresses} created algorithmically for mimicking a legitimate entity.    

\noindent\begin{center}
  \includegraphics[width=\columnwidth]{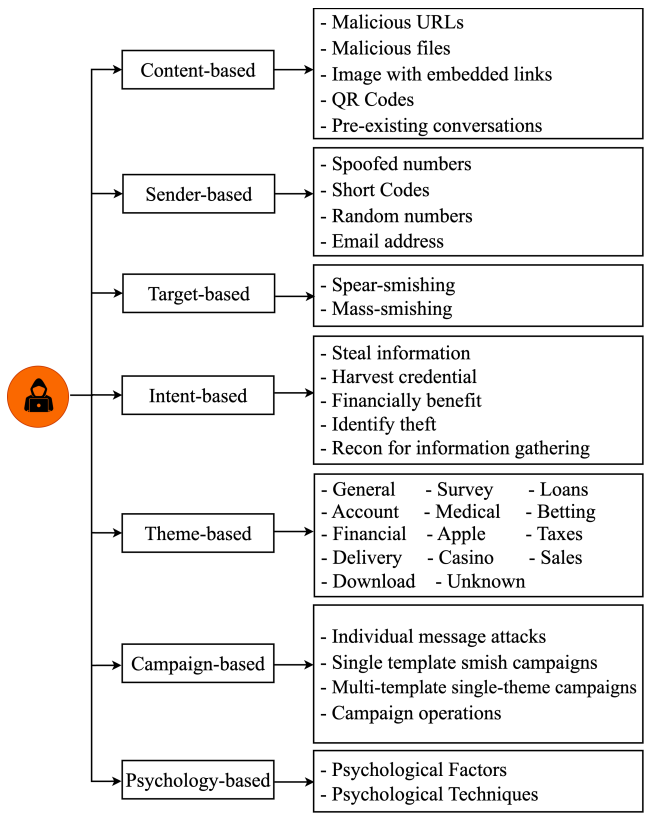}
  \captionof{figure}{Overview of SMS phishing threat landscapes}
  \label{fig:threat-landscape}
\end{center}
   
    \subsection{Target Audience-Based Classification}
    When considering the target audience as the classification factor, we classify smishing attacks as \textit{targeted spear-smishing} \citep{gupta2016exploiting} where specific individuals are targeted based on previous reconnaissance activities and \textit{mass-smishing} where a message is sent to many unknown targets as message broadcast for opportunistic attacks. Usually targeted smishing needs more preparation and sophistication with potentially higher success rates, while the mass smishing is usually less sophisticated to carry on but with lower success rates. 
    
    \subsection{Intent-Based Classification} Based on the intent (or goal) of the attackers, in this paper, we define the following intent-based smishing classes: \textit{stealing and harvesting information}, \textit{recon for information gathering}, \textit{benefiting financially}, and \textit{conducting identity theft}. There can be certain intents which is achieved through subsequent successful smishing attacks with another intent. To illustrate, attackers may first send a smishing message to act as a delivery notification to perform reconnaissance to collect active phone numbers and know which recipients engage with the provided link in the message. When the attackers identify victims, they may follow up with a second message that is designed to harvest sensitive login credentials (e.g., bank or email account details). These stolen credentials can then be used in the next step to facilitate direct financial fraud or conduct identity theft. 
    

    \subsection{Campaign-Based Classification}
    Liu et al.'s work \citep{liu2021detecting} on SMS spearphishing also provides valuable insights into smishing campaigns. Their work has identified a total of $11,475$ unique campaigns within their dataset. Their campaign-level analysis revealed that certain types of spearphishing scams display oligopolistic characteristics, meaning 
   a small number of campaigns are responsible for the majority of the messages in a particular scam category, effectively dominating that market. A prime example of this is seen in the \textit{financial scam} category, where over $50\%$ of the messages were attributed to just the top six campaigns. Similarly, for both the \textit{fortune-telling scam} and \textit{insurance scam}, the largest single campaigns were found to control more than $50\%$ of the messages for those respective types of scams. Additionally, they uncovered several sophisticated attacker strategies employed in these spearphishing campaigns, such as sending test messages to controlled devices to check for detection before a large-scale attack. 

    Recent studies have tried to characterize smishing campaigns using message content and operational infrastructure. For example, authors in {\em SmishViz} \citep{pritom_25_codaspy_smishviz} introduced a graph-based visualization approach that enables defenders to monitor and characterize ongoing smishing campaigns and campaign operations by clustering similar messages and connecting them with common web infrastructures used in the message contents as connected tripartite graphs. 
    Moreover, the {\em SmishViz} system can reveal how large-scale smishing activities can be mapped across topical themes, message templates, and web domains to uncover broader coordinated attack operations. Thus, by considering the structure and reachability of the smishing campaigns introduced in Sanjari et al. \citep{pritom_25_codaspy_smishviz}, we can refer to the following campaign-based smishing classification: (i) individual message attacks, (ii) single template smishing campaigns, (iii) multi-template single theme campaigns, and (iv) campaign operations, which are described in details below. 
   
    
    \subsubsection{Individual Attacks} These are one-time smishing attempts that are unique and do not connect to any known themes. These attacks may be crafted to send to a single user or a very limited number of users and sent from a personal phone number using manual SMS sends. 
    These attacks may be conducted using a few rogue SIM (Subscriber Identity Module) cards. 

    \subsubsection{Single Template Campaigns} 
    In this case, attackers aim to reach many users using a single message template and send the same message to multiple receiver phone numbers. These campaigns often involve impersonation of trusted brands or services \citep{pritom_25_codaspy_smishviz}, and due to their template-based nature, they are easier to detect using pattern recognition techniques such as the Ratcliff pattern recognition algorithm \citep{ratcliff1988pattern}.
    
    \subsubsection{Multi-Template Single-Theme Campaigns} 
    In more complex campaigns, attackers would create multiple templates of similar themes (e.g., IRS tax refund, USPS delivery) to send to diverse target population. Usually these campaigns would have multiple templates for attacking various user groups aiming to maximize the number of victims. However, they have known themes and listed target brands, which would be constant throughout the campaigns. Attack campaigns often use automated tools to send out a massive number of text messages simultaneously. Analyzing attacks at campaign-level can reveal patterns in the language, urgency levels, and impersonated sources used by scammers.
    
    \subsubsection{Campaign Operations} 
    Some attacks may connect multiple single template or multi-template campaigns of different themes using common shared infrastructures, which would be financially more beneficial for the attackers. Sanjari \textit{et al.} \citep{pritom_25_codaspy_smishviz} has referred to these threats as {\em campaign operations} that often involve groups from multiple themes connected through common web entities (e.g., domain names or shortened URLs). This campaign operations often indicate that a common owner is running the coordinated campaign operations that host multiple smishing campaigns to reach mass user. If these attack operations are not dismantled, the owners may create new smishing campaigns with new themes to make them more deceptive to their victims.

   

 \subsection{Theme-Based Classification}
    The topic or themes within a smishing message can vary to connect users and make them trust the SMS for further interactions. The more relevant the theme is, the higher the chance of a successful attack. From existing literature, we find that Nahapetyan et al. \citep{nahapetyanposter_2023} presents the following themes of smishing campaigns: \textit{generic}, \textit{account-theme}, \textit{financial-theme}, \textit{delivery-theme}, \textit{casino-theme}, \textit{surveys-theme}, \textit{medical info-theme}, \textit{Apple-brand-theme}, \textit{download-theme}, \textit{sales-theme}, \textit{loans-theme}, \textit{betting-theme}, \textit{taxes-theme}, and \textit{unknown}. The unknown class is a possibility as new themes may evolve over time that are unique in nature and may not be classified into existing established themes. However, themes can evolve with new events around the globe and be leveraged by attackers. For example, COVID-19 has emerged as a popular theme to carry out cyberattacks \citep{Mir2020_COVID_themed_threats_ISI, Mir2020_data_COVID_websites_ISI,covid-themed-scams-asiaCCS2023} in early 2020 or any global or regional war incidents (e.g., Russia-Ukraine war) \citep{mia2025_warthemed_websites_TPS} can be leveraged as themes to carry out smishing attack campaigns. In another study, Liu et al. \citep{liu2021detecting} 
    highlighted the prevalence of specific themes on SMS spearphishing, including financial scams (40.86\%), lawsuit scams (27.11\%), and even fortune-telling scams (14.43\%), which were found to be particularly active. This indicates that attackers engaging in spearphishing may focus on themes that can be personalized with the victim's information for greater believability. Moreover, their analysis also uncovered that attacker activities can vary across these categories, with financial scams tending to be more active during weekdays and working hours, while fortune-telling scams mainly occur at night. 
    Furthermore, themes can also be seasonal that usually correspond to specific events 
    or timings of the year, while urgency-driven themes such as accounts, subscription or missing a delivery are used as tactics to compromise victims.  
    

\subsection{User Psychology-Based Classification} 

Alike other SE attacks, smishing attacks 
also exploit user psychology to succeed \citep{siddiqi2022study}. To get a better understanding of user psychology for smishing, it is essential to understand two core concepts in the context of smishing attacks: 1) Psychological Factors (PFs) and 2) Psychological Techniques (PTs) \citep{XuPIEEE2024}. 

\subsubsection{Psychological Factors (PFs)}
In the context of smishing attacks, PFs work as human characteristics and vulnerabilities that attackers can exploit. PFs can be identified into 5 major categories of factors, including-- [i] social psychological factors, [ii] personality and individual differences factors, [iii] cognitive factors, [iv] emotional factors, and [v] workplace factors. We want to discuss each of these PF categories and how they are relevant in the context of 
smishing with examples. 
Our examples are based on the definitions provided in \citep{XuPIEEE2024} that shows a clear evidence of how each PF categories can be used in generic social engineering (SE) attacks. 



\noindent\textbf{[i] Social psychological factors:} 
Here we contextualize nine (9) social psychological factors for smishing attacks and oftentimes there are multiple factors that are leveraged together to victimize users. Starting with \textit{Authority}, which refers to the tendency to obey experts or authoritative figures. Next, \textit{Respect} refers to the importance given based on the perceived
worth. Next one is \textit{Disobedience} that refers to the strong refusal to follow authority or rules. Now, see the following example message--
\textit{"CITI: Unauthorized activity was detected, we have moved forward in locking your account. Please verify at http://citibsec[dot]com or at a local branch."}

This message leverages both \textit{authority} and \textit{respect} factors by impersonating CITI Bank, and using the obedience that the user feels to financial institutions, which hold high perceived worth over their assets. Since the message aims to warn the user by threatening a locked account; it also exploits the user's \textit{disobedience} factor by giving a shortcut (malicious link) instead of the official way (i.e., call customer services, waiting on hold, answer security questions) to fix it. 

Next, \textit{Liking} reflects the positive reaction to known individuals, and \textit{Reciprocation} means the feeling of being obliged to return a favor while \textit{Social-proof} is the tendency to replicate others’ behaviors, and \textit{Consistency} refers to the commitment to something (e.g., a task or an individual). To contextualize these factors, we see the following message-- \textit{``Free Msg: Your bill is paid for March. Thanks, here's a little gift for you: mentioningdears[dot]com/nzWjdtrM93"}, which targets \textit{liking} and \textit{reciprocity} by performing an initial and valuable favor (i.e., ``bill is paid'') and then offers a second `gift' to create a double sense of goodwill and obligation to click on the link. This example uses \textit{consistency} factor by pointing out the user's status as a frequent customer (i.e., ``Your bill is paid for March'') and finally by framing the malicious link as a standard and normalized interaction, which is expected from a reliable service provider, triggers the user's \textit{social-proof} factor.

Next, we have \textit{Scarcity} that refers to the value of items that are limited in availability and \textit{Perceptual contrast} refers to a mistaken judgment when comparing two items sequentially. Attackers may target these factors together as evident by the following message-
\textit{``Bank Of America Hurry up! Offers Have arrived! (\$\$90 value). storage[dot]googleapis[dot]com/jzoxubc/1fhg3lf BANK OF AMERICA"}, that exploits \textit{scarcity} factor by using an urgent command ``Hurry up!" and implying a limited-time deal (i.e., ``Offers Have arrived!") and also offers a high value (i.e.., ``\$90 value") with the minimal effort action of clicking the link to secure the reward to exploit the user's \textit{perceptual contrast} factor.


\noindent\textbf{[ii] Personality 
and individual differences factors:} 
Within this PF class, there are several other PFs. First, \textit{Openness} describes a person’s creativity and ability to generate new ideas, while \textit{Curiosity} is the degree of an individual’s desire for knowledge. Next, \textit{Extraversion} is the degree of sociability, assertiveness, and emotional expressiveness in an individual, and \textit{Agreeableness} is a trait associated with trust, altruism, and benevolence. These above 4 PFs rely on the victim's trust in the sender's identity, which is the tendency of individuals to place faith in others. To illustrate, the following message \textit{``Costco: Daniel, the code 42003 printed on your receipt from 10 came in 2nd in our Airpods draw: f2gpy[dot]info/RzNKEws Zve''}, initially relies on the user's trust in a well-known brand and then exploits the user's \textit{curiosity} and \textit{openness} with an incomplete but intriguing narrative (e.g., draw result) and promises a reward. The personalized details, including the name and receipt code, makes the message appear legitimate, helping attackers succeed by exploiting users' high \textit{agreeableness}. A slight change to this message, such as adding a geo-location to the draw, can turn it into a social invite and would also target \textit{extraversion}. 


Next, \textit{Conscientiousness} refers to a person’s self-discipline and focus on achieving goals while \textit{Negligence} is a failure to adequately perform a task, leading to security vulnerabilities. Moreover, \textit{Disorganization} refers to the tendency to act spontaneously and tolerate disorder. These PFs are successful targets because they appeal to a user's lack of effort, or \textit{Laziness} factor, in completing boring tasks. For example, in the following message \textit{``Hi, you still owe UPS \$4.10 USD in customs fees for your previous package. Reply 1 to receive a secure link to pay your invoice''}, attacker tries to trigger \textit{conscientiousness} by presenting a necessary and formal duty (i.e., missing payment) that must be paid. 
By implying that the user failed to handle the task, and creating pressure to fix the mistake quickly, it exploits the user's \textit{negligence} or \textit{disorganization}. Finally, the simple instruction to `Reply 1' for a `secure link' triggers \textit{laziness} by promising an easy shortcut to an administrative task.


Additionally, several other personality factors such as \textit{Neuroticism}, \textit{Impulsivity}, \textit{Self-Control}, \textit{Impatience}, and \textit{Vulnerability} can be targeted together within a specific smishing message. Here, \textit{Neuroticism} refers to a person's tendency toward mood swings and emotional instability. \textit{Impulsivity} is the tendency to act without careful consideration; \textit{Self-Control} refers to the capacity to manage decision-making despite intense emotions. Next, \textit{Impatience} is defined as a state of frustration experienced during delays, and \textit{Vulnerability}, is defined as the need for special care or protection. To illustrate, the following message- \textit{``WEL01[dot]US/R/REST05 WELLS            FARGO(CS): Profile locked because of unusual activities, kindly restore.''}, creates an immediate distress by targeting \textit{neuroticism} and exploiting the user's \textit{vulnerability} to financial harm. It also bypasses the user's \textit{self-control} and triggers \textit{impulsivity} by using authoritative language and requesting a quick action (i.e., ``kindly restore'').

In another scenario, PFs such as \textit{Vigilance}, \textit{Freewheeling}, \textit{Individual Indifference}, and \textit{Submissiveness} can be targeted together where \textit{Vigilance} is defined as the degree to which an individual remains conscious of potential dangers or irregularities. \textit{Freewheeling} is the degree to which an individual ignores rules or social conventions. \textit{Individual Indifference} refers to the level of indifference shown toward a task, especially regarding security. Lastly, \textit{Submissiveness} is the degree of a person’s willingness to comply with others' wishes. For example, in the following message- \textit{``WELLS FARGO: Update>: A decision has been made, your account is suspended. Visit http://VERIFYWELLSFARGO[dot]GA to regain access now! This action takes few seconds.''}, attackers use a command (e.g., ``A decision has been made") that tests the user's \textit{submissiveness} and states the action (e.g., ``takes few seconds''), which exploits \textit{individual indifference} by taking advantage of a user's lack of interest toward the task and guiding them the easiest option. 
It suggests that the user can quickly get the task off their mind. This small effort reduces the need for \textit{Vigilance} (taking less care with malicious URLs). The entire process is attractive for \textit{freewheeling} individuals by offering a quick way to account recovery outside of official bank procedures. This method allows them to disregard standard protocol. 

\noindent\textbf{[iii] Cognitive PFs:} \textit{Cognitive Miser} is under the cognitive PF class as refers to the reliance on a person's mental shortcuts when making decisions. A good example message for this PF is \textit{``It's Nicole again Office: 18665841391 It appears you may be eligible for compensation on your Injury. Can you give me a quick call please? - STOP to STOP"}, which exploits the \textit{cognitive miser} tendency by presenting a complex problem (compensation for an injury) with a simple and deceptive solution: ``Can you give me a quick call please?". The victim's brain tries to find the path with the least cognitive resistance. It relies on a quick and easy mental shortcut (such as a simple phone call) that resolves the entire complex legal or financial problem.

\noindent\textbf{[iv] Emotional PFs:} A group of emotional PFs such as \textit{Greed} and \textit{Fear} often targeted together by the attackers where \textit{Greed} refers to the person's strong craving for wealth or power, 
and \textit{Fear} is the expectation that harm or danger may occur. 
For example, the following message- \textit{``[I.R.S] You will receive a tax refund of \$573.00, but cannot process it due to incomplete personal information. Please confirm your information in the link. https://irs[dot]gov[dot]safe-ordering[dot]com In order to expedite the processing of your refund and avoid potential delays or penalties, please provide accurate personal information immediately to ensure precise payment. Deadline: November 30, 2023. Thank you for your cooperation! Sincerely, [IRS]"}, exploits \textit{greed} by presenting an unexpected financial gain (\$573). It also amplifies the \textit{fear} factor by introducing the sense of losing the money or paying penalties if deadline is past, and made the user to click on the link to save their money. 


Next, \textit{Sympathy} is the affective understanding of other's emotional state without sharing the same emotional experience. For example, the following message- \textit{``I know it’s get late in the day but I could really use your advice on this St. Lucie Co. project. Can you review this before go to bed"} creates a sympathy for the sender. Similarly, \textit{Empathy} is the emotional state in which one identifies with other’s feelings based on personal prior experiences. For example, the message- \textit{``Medicare: You have been in close proximity with somebody who is positive for omicron. Please immediately get a testkit now through: https://medicare[dot]id-53[dot]com"}, points to a universal fear of health issues. Since almost all recipients personally experienced the anxiety of exposure and quarantine, this message can effectively trigger empathy by allowing for immediate emotional   identification as well. 


Moreover, \textit{Loneliness} and \textit{Hopelessness}, these 2 emotional PFs can be targeted together where \textit{Loneliness} refers to a personal sense of the gap between desired and
actual social connections while \textit{Hopelessness} refers to a perceived absence of hope or possibility for improvement. 
For example, the message- \textit{``Hi Julianne long time no see, I'm Aleen, how are you doing''}, initiates a conversation that tries to target the person's \textit{loneliness} by offering a friendly and unexpected social reconnection that the victim may be looking for. This is also effective against users who are vulnerable to \textit{hopelessness}, ‌including people who feel disconnected or that their situation is unlikely to improve. It makes them receptive to any new social link that promises a temporary emotional lift or distraction.

\noindent\textbf{[v] Workplace PFs:} Under this class, \textit{Workload} PF refers to the total number of tasks or amount of work an individual is responsible for completing. Another workplace PF \textit{Stress} refers to the physical, emotional, or psychological pressure experienced by an individual while \textit{Busyness} refers to a condition of extreme workload that reduces
attention to detail. Another workplace PF \textit{Hurry} is the degree to which any user may hurry to complete a task that reduces adherence to secure practices. These above workplace PFs can be targeted by attacker altogether. To illustrate, the message- \textit{``[CITI Alert 05.02.2022] Debit card associated with Acct*8139 is locked. Visit bit[dot]ly/89BNsg81289 at earliest convenience. Stop=end|Help=help"}, exploits \textit{workload} factor by creating an immediate administrative task (i.e., resolving a locked card) that must be added to the user's schedule, and exploits \textit{stress} factor by implying a threat of financial disruption. It also exploits \textit{busyness} and \textit{hurry} factors by forcing the user to act immediately (i.e., ``at earliest convenience").

Next, other workplace factors \textit{Affective Commitment} and \textit{Habituation} can be targeted together where \textit{Affective Commitment} refers to a person’s emotional bond with an organization, while \textit{Habituation} is the tendency to lose responsiveness to a task or
recurring warnings over time. For example, the message- \textit{``Unusual Activity : We've suspend your Amazon account due to unusual activity. Login attempt blocked from : Mozilla Firefox 8.7 [168.254.81.174 - Turkey] Our system has suspend your Amazon account for security reasons. To unlock your Amazon account, please verify with link below: http://www[dot]xprswebsite[dot]com/free/userf111 /redirf111/?=5826865 You need to take action within 2 days before account will be suspended. Regards, Amazon Teams"}, uses technical details (IP address, browser version) to make the routine alert seem legitimate to exploit \textit{habituation}. It also leverages \textit{affective commitment} by threatening the user to lose access to their highly valued Amazon account 
and compels the user to act quickly to protect their connection to that resource using the malicious link.


\subsubsection{Psychological Techniques (PTs):}
\begin{figure*}[t]
    \centering
\includegraphics[width=1\textwidth]{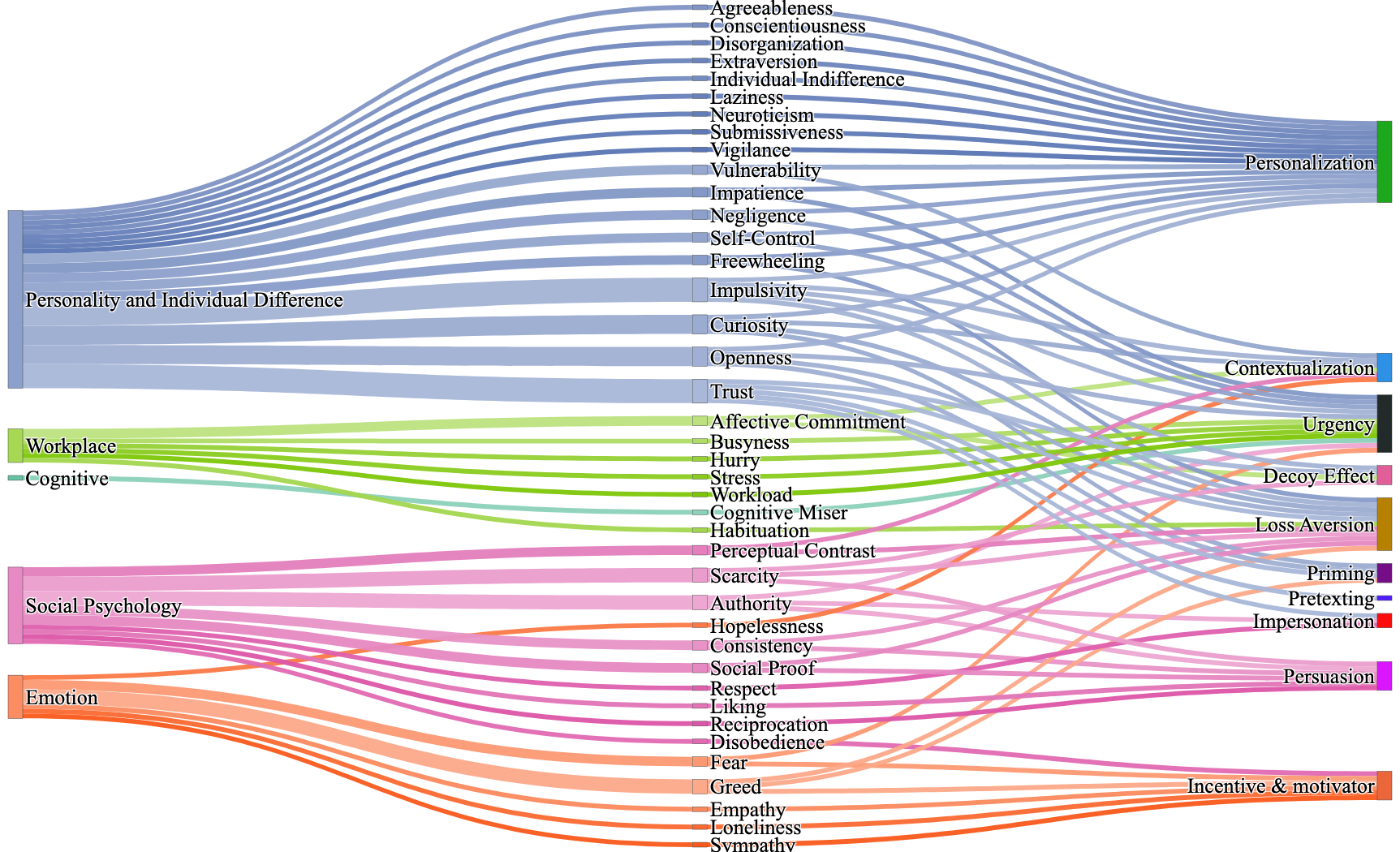}
    \caption{Mapping of PF classes, PFs and Corresponding PTs}
    \label{fig:PF-PT}
\end{figure*}

Psychological techniques (PTs) are techniques adopted by attackers to formulate an attack (i.e., smishing message). Longtchi et al. \citep{XuPIEEE2024} reviewed a widespread range of PTs for generic SE attacks. 
Among these, we have contextualized 10 PTs that can be applicable to any smishing attack scenarios. We have also mapped all these PTs into corresponding PFs as shown in Figure \ref{fig:PF-PT}. 

\noindent\textbf{[i] Personalization:} refers to the customization of messages with personal information to build trust. It exploits the following PFs: \textit{agreeableness}, \textit{conscientiousness}, \textit{disorganization}, \textit{extraversion}, \textit{individual indifference}, \textit{laziness}, \textit{neuroticism}, \textit{submissiveness}, \textit{vigilance}, \textit{vulnerability}, \textit{impatience}, \textit{negligence}, \textit{self-control}, \textit{freewheeling}, \textit{impulsivity}, \textit{curiosity}, \textit{openness}, and \textit{trust}. 

\noindent\textbf{[ii] Impersonation:} meaning impersonating another identity, such as a legitimate organization, to increase the probability of the victim's compliance. It exploits the following PFs: \textit{authority}, \textit{respect}, and \textit{trust}. 

\noindent\textbf{[iii] Incentive \& motivator:} refers to offering external rewards or internal satisfaction to encourage a desired behavior.
It exploits following PFs: \textit{greed}, \textit{fear}, \textit{sympathy}, \textit{empathy}, \textit{loneliness}, and \textit{disobedience}.

\noindent\textbf{[iv] Urgency:} refers to creating a situation that needs the user's immediate action, with the probability of reducing the chance of recognizing an attack. It exploits following PFs: \textit{impatience}, \textit{negligence}, \textit{self-control}, \textit{impulsivity}, \textit{openness}, \textit{busyness}, \textit{hurry}, \textit{stress}, \textit{workload}, \textit{cognitive miser}, \textit{authority}, and \textit{fear}.

\noindent\textbf{[v] Decoy effect:} refers to making users believe they are receiving a good deal. It exploits the following PFs: \textit{impulsivity}, \textit{trust}, \textit{affective commitment}, \textit{scarcity}, and \textit{freewheeling}.

\noindent\textbf{[vi] Loss aversion:} means providing a free good or service to build the victim’s trust and attachment and later charging an exorbitant amount once the victim is attached to it. It can also involve influencing a victim's decisions by gradually providing them with seemingly legitimate information over time before launching the attack. This PT exploits the following PFs: \textit{freewheeling}, \textit{curiosity}, \textit{openness}, \textit{trust}, \textit{habituation}, \textit{perceptual contrast}, \textit{scarcity}, \textit{consistency}, \textit{social proof}, and \textit{greed}.

\noindent\textbf{[vii] Priming:} refers to gradually influencing an individual's subsequent decisions through subtle manipulation. This PT
exploits the following PFs: \textit{curiosity}, \textit{openness}, \textit{trust}, and \textit{greed}. 

\noindent\textbf{[viii] Persuasion:} is a technique that encourages a particular behavior by exploiting various psychological factors. It exploits the following PFs: \textit{liking}, \textit{reciprocation}, \textit{social proof}, \textit{consistency}, and \textit{authority}.

\noindent\textbf{[ix] Contextualization:} posing as a member of the victim's group to create a sense of shared affiliation. It exploits the following PFs: \textit{hopelessness}, \textit{perceptual contrast}, \textit{curiosity}, \textit{impulsivity}, \textit{vulnerability}, and \textit{affective commitment}. 

\noindent\textbf{[x] Pretexting:} refers to creating a potentially possible scenario to increase the victim's engagement with the attacker. This PT exploits the \textit{trust} PF. 

In summary, attackers use a wide range of PTs to manipulate and exploit various user PFs while carrying out smishing attacks. As shown in the Figure \ref{fig:PF-PT}, these $10$ PTs are employed to exploit a total of $40$ different PFs that leads attackers' success.

\subsection{Summary of Smishing Threat Landscape}
The smishing threat landscape presented in Figure~\ref{fig:threat-landscape} is intrinsically dependent on the analytical scope and the specific lens through which an individual study is pursued. A defining characteristic of this landscape is the \textit{many-to-many} relationship that exists between various classification schemes, where a single attack instance can simultaneously occupy multiple categories. For instance, a ``Failed Package Delivery'' attack may be primarily classified under a {\em theme-based} scheme; however, it concurrently intersects with {\em user psychology-based} (by inducing urgency), {\em content-based} (via malicious URL embedding), and {\em intent-based}(targeting credential theft) schemes.  

Consequently, the categorization of a smishing campaign is not mutually exclusive but rather multidimensional. A study focusing on ``Executive Impersonation'' may prioritize the target audience (spear smishing), yet it must also be viewed through the lens of sender information (identity spoofing) and message delivery techniques (personalized social engineering). This overlapping taxonomy ensures that the threat landscape remains comprehensive, allowing researchers to pivot their analysis between different principal categories while acknowledging the interconnected nature of modern smishing schemes.

\section{Evaluating Smishing Defense Landscape}
\label{sec:defense_landscape}
\subsection{Detection-based Defense Measures}
In this section, we synthesize articles on various smishing detection mechanisms using rule-based, ML-based, DL-based, NLP-based, Federated Learning-based, and hybrid approaches. 

\begin{table*}[!t]
\centering
\scriptsize

\caption{Summary Analysis of Existing Smishing Detection Using Various ML and DL Methods}

\label{tab:method_sum22}
\renewcommand{\arraystretch}{1.15}
\setlength{\tabcolsep}{3.5pt}

\rowcolors{2}{white}{gray!10}

\begin{tabular}{|p{2.7cm}|p{4.4cm}|cc|ccccc|cccc|}

\toprule

\multirow{2}{*}{\textbf{Author and Year}} &
\multirow{2}{*}{\textbf{Method}} & 
\multicolumn{2}{c|}{\textbf{Content Analysis}} &
\multicolumn{5}{c|}{\textbf{URL Analysis}} & 
\multicolumn{4}{c|}{\textbf{NLP Techniques}} \\
\cmidrule(lr){3-4}
\cmidrule(lr){5-9}
\cmidrule(lr){10-13}

&
& \textbf{ML} & \textbf{DL} 
& \textbf{APK} & \textbf{SAL} & \textbf{SHRT} & \textbf{BLK} & \textbf{TAG}
& \textbf{TN} & \textbf{TKN} & \textbf{WE} & \textbf{TF-IDF} \\

\midrule

 \citep{goel2018smishing} & Naïve Bayes Classifier & ✓ & . & ✓ & ✓ & . & ✓ & . & ✓ & ✓ & . & . \\

\citep{mishra2019content} & content-based analysis + URL behavior inspection & ✓ & . & ✓ & ✓ & . & . & ✓ & ✓ & ✓ & . & ✓ \\

 \citep{ghourabi2020hybrid} & CNN-LSTM & . & ✓ & . & . & . & . & . & ✓ & ✓ & ✓ & ✓ \\

\citep{sonowal2020detecting} & AdaBoost, Random Forest, Decision Tree, and SVM & ✓ & . & . & . & . & . & . & . & ✓ & . & . \\

 \citep{smishing_detector_security_model_mishra_2020} & Naïve Bayes Classifier & ✓ & . & ✓ & ✓ & ✓ & ✓ & ✓ & ✓ & ✓ & . & ✓
\\

 \citep{mldetect_smishing_frauds_boukari} & Naïve Bayes and Random Forest classifiers & ✓ & . & . & . & . & . & . & . & ✓ & . & ✓ \\

 \citep{smdetector_ghourabi} & Deep Learning (DL) + BERT & . & ✓ & . & . & . & ✓ & . & ✓ & ✓ & ✓ & . \\

 \citep{njuguna2021model} & Naïve Bayes Classifier & ✓ & . & . & . & . & . & . & ✓ & . & . & .
\\

 \citep{ulfath2022detecting} & SVM, Random Forest, AdaBoost, XGBoost, CART & ✓ & . & . & . & . & . & . & ✓ & ✓ & . & ✓
\\

 \citep{spam_filtering_cnn_lstm_hossain} & CNN-LSTM & . & ✓ & . & . & . & . & . & ✓ & ✓ & ✓ & ✓
\\

 \citep{jain2022content}  & XGB, GBDT, RF, BgC, KNN, ETC, DT, LR, AdaBoost, BNB, MNB, SVC, GNB & ✓ & . & . & . & . & . & . &  ✓ & ✓ & . & ✓
\\

\citep{implement_smishing_detector_mishra_2022}  & ANN & . & ✓ & ✓ & ✓ & ✓ & ✓ & ✓ & ✓ & ✓ & . & .
\\

 \citep{smsprotect_akande} & Rule-based + RIPPER + PART C4.5 & ✓ & . & . & ✓ & . & ✓ & . & ✓ & ✓ & . & . \\


 \citep{karhani_phishing_2023} & Domain Checker (Decision Tree) + NLP Checker(SVC with TF-IDF) & ✓ & . & . & . & ✓ & ✓ & ✓ & ✓ & ✓ & . & ✓ \\
\citep{dsmish_mishra_2023} & Domain Checking Phase + SMS Classification Phase & . & ✓ & ✓ & ✓ & ✓ & . & ✓ & ✓ & ✓ & . & . \\

 \citep{shanto2023federated} & FL + LSTM, multi-lingual support & . & \simplecircledmark & . & . & . & . & . & . & . & ✓ & . \\
 \citep{commerial_anti_smishing_timko_2023} & Comparative study & . & . & . & . & ✓ & ✓ & . & . & . & . & . \\
\citep{on_phishing_tactics_nahapetyan2023sms} & Large-scale empirical study & . & . & . & . & ✓ & ✓ & . & ✓ & . & . & . \\
 \citep{sidhpura2023_fedspam:} & FL + DistilBERT & . & \simplecircledmark & . & . & . & . & . & ✓ & ✓ & ✓ & . \\
\citep{kanaoka2023beyond} & Image-based URL extraction + List-based URL matching & . & . & . & . & . & ✓ & . & . & . & . & . \\
 \citep{seo2024device} & Char-CNN & . & ✓ & ✓ & . & . & . & . & ✓ & ✓ & ✓ & . \\

 \citep{anh2024federated} & FL with Non-IID setup + PhoBERT & . & \simplecircledmark & . & . & . & . & . & . & ✓ & . & . \\

\citep{hapase2024telecommunication} & CNN + PCC-PCA & . & ✓ & . & . & . & . & . & ✓ & ✓ & . & ✓ \\

\citep{rose2024next_federated} & Iterative FL setup + ConvLSTM & . & \simplecircledmark & . & . & . & . & . & . & . & ✓ & . \\

\citep{mehmood2024enhancing} & CNN-LSTM & . & ✓ & ✓ & . & . & ✓ & . & ✓ & ✓ & ✓ & ✓ \\

\citep{shinde2024sms} & unsupervised learning(K-means, NMF, GMM, PCA) + deep semi-supervised learning (RNN-Flatten, LSTM, Bi-LSTM) & . & ✓ & . & . & . & . & . & ✓ & ✓ & ✓ & ✓ \\

\citep{llm_shim2024persuasion} & Persuasion-based prompt learning + transformer-based classification & . & ✓ & . & . & . & . & . & . & . & . & . \\

 \citep{uddin2024explainabledetector} & Fine-tuned RoBERTa + LIME & . & ✓ & . & . & . & . & . & ✓ & ✓ & ✓ & . \\
\bottomrule
\end{tabular}%

\vspace{0.8em}
\parbox{\textwidth}{\scriptsize
\textbf{Keys:} 
ML: Machine Learning, DL: Deep Learning, FL\simplecircledmark: Federated Learning, APK: Application Package Analysis(containing download link), SAL: Self-Answering Link, SHRT: URL Shortening Detection, BLK: Blacklist, TAG: HTML Form Tagging, TN: Text Normalization, TKN: Tokenization, WE: Word Embeddings, TF-IDF: Term Frequency-Inverse Document Frequency.
\vspace{0.5em}

XGB: Extreme Gradient Boosting, GBDT: Gradient Boosted Decision Tree, RF: Random Forest, BgC: Bagging Classifier, KNN: K-Nearest Neighbors, ETC: Extra Trees Classifier, DT: Decision Tree, LR: Logistic Regression, AdaBoost: Adaptive Boosting, BNB: Bernoulli Naïve Bayes, MNB: Multinomial Naïve Bayes, SVC: Support Vector Classifier, GNB: Gaussian Naïve Bayes, ANN: Artificial Neural Network, PCC: Pearson Correlation Coefficient PCA: Principal Component Analysis, LSTM: Long Short-Term Memory, NMF: Non-Negative Matrix Factorization, GMM: Gaussian Mixture Models.
}
\end{table*}


\subsubsection{Rule-based Detection} 
Rule-based smishing detection represents a traditional approach that relies on a predefined set of rules to identify suspicious messages \citep{smsprotect_akande}. These rules target specific characteristics commonly found in smishing attempts. For example, rules might flag messages containing specific keywords associated with phishing scams (e.g., `urgent' and `account locked'), suspicious URLs with specific patterns, or based on sender information that appears spoofed or illegitimate. While this method offers a simple and efficient way to catch smishing attempts, its effectiveness can be challenged by attackers adopting evasive tactics and newer attacks. 

In literature, Akande et al. \citep{smsprotect_akande} combine URL blacklist checks, scam number matching, spam keyword thresholds, symbol patterns, and message style cues (emoticons, abbreviations, self-answer prompts) using a \textit{RIPPER/C4.5} rule-learning pipeline to make their detection based on both URL and keywords, with additional structural and sender-based rules. RIPPER and C4.5 are both popular classification algorithms that approach rule-based classification in different ways. RIPPER (Repeated Incremental Pruning to Produce Error Reduction) is an algorithm that directly learns a set of rules. It works by first generating a rule set and then iteratively pruning and optimizing it to minimize the error rate. This process involves growing rules until they become too complex, pruning them back to a simpler state, and then entering an optimization phase where it replaces or modifies rules to improve performance. On the other hand, C4.5 is a decision tree algorithm that can be adapted to produce rules. It first constructs a decision tree and then converts the tree into a set of ``if-then'' rules by creating a rule for each path from the root to a leaf. The rules are then simplified and pruned to reduce complexity and improve accuracy and coverage, which is the proportion of correctly classified instances. 

\begin{table*}[!t]
\centering
\scriptsize
\caption{Comparative Analysis of Smishing Detection Literature (Categorized by Rule-based, ML, DL, and LM approaches)}
\label{tab:review-1}

 \begin{tabular}{|p{11.5em}|p{18.5em}|p{15.5em}|p{14.5em}|}
\toprule
\textbf{Author and Year} & \textbf{Key Idea} & \textbf{Strengths} & \textbf{Weaknesses} \\
\midrule

\rowcolor{gray!45} \multicolumn{4}{|l|}{\textbf{Rule-based and Heuristic Approaches}} \\
 \citep{goel2018smishing} & multi-step logics, Normalizes SMS, applies NB classifier & Multi-feature coverage (URLs, APKs, sender, text cues) & No performance metrics  \\
 \citep{mishra2019content} & Combines text and URL for filtering rules with OneVsRestClassifier & Hybrid and comprehensive detection approach & Relies on keywords, Imbalanced dataset, No performance metrics \\
 \citep{njuguna2021model} & User notification and content filtering with NB classifier & Accuracy of 94\%, real-time user notification  & Reliance on keyword filtering, lacks comparison metrics \\
\citep{smsprotect_akande} & Rule-based detection using PART (C4.5) model, content filtering and user alerts & High accuracy (98.42\%), real-time performance & Relies on pre-defined rules, limited adaptability to novel smishing tactics \\
\citep{commerial_anti_smishing_timko_2023} & Evaluation of commercial anti-smishing tools, comprehensive benchmark of anti-smishing tools & Characterizing smishing attack through a qualitative analysis & Infeasible full testing, time-limited experiments (3 days) \\
 \citep{on_phishing_tactics_nahapetyan2023sms} & Content and infrastructure-based clustering & Large public dataset, evaluation through blacklists & Potential bias in the Data source due to the use of public gateways  \\

 \citep{kanaoka2023beyond} & Cross-device analysis for smishing defense using augmented reality (AR) glasses  & Accuracy of 85.19\%, high System Usability Scale score (74.4) & Simple URL blacklist matching \\

\midrule
\rowcolor{gray!45} \multicolumn{4}{|l|}{\textbf{Machine Learning (ML) Approaches}} \\
 \citep{smishing_detector_security_model_mishra_2020} & 4-part detector for text, URL, code, apk with NB classifier & Covers diverse attacks, 96.3\% accuracy & No malware analysis of downloaded APK \\
\citep{sonowal2020detecting} & Ranks features for ensemble models (AdaBoost) & High accuracy (98.4\%), feature focus & Lacks feature/data details \\
 \citep{mldetect_smishing_frauds_boukari} & Detects via term importance and complaints using RF & 98.15\% accuracy, adapts to vishing/phishing & Low recall, many false negatives \\
 \citep{ulfath2022detecting} & Advanced feature selection via ANOVA test, SVM model for detection & 98.39\% accuracy & Weak adaptability to language evolution \\
 \citep{jain2022content} & Text + URL spam detection using rare words with voting (RF, KNN, ETC) & 99\% accuracy, clean dataset & Ignores URL-less SMSes \\

 \citep{karhani_phishing_2023} & Hybrid domain and NLP feature detection with RF & 99.4\% accuracy, F1-score > 99\%  tested on publicly avaiable data & No independent real-time system, higher processing time complexity \\

\midrule
\rowcolor{gray!45} \multicolumn{4}{|l|}{\textbf{Deep Learning (DL) Approaches}} \\
 \citep{ghourabi2020hybrid} & Bilingual spam detection with CNN-LSTM & Handles Arabic/English, 98.4\% accuracy & Limited feature explanation \\

 \citep{spam_filtering_cnn_lstm_hossain} & CNN-LSTM for spam detection with embeddings & 98.4\% accuracy, 98\% F1-score & Low recall (89\%) for minority spam class \\

 \citep{implement_smishing_detector_mishra_2022} & Feature-focused ANN Backpropagation for smishing & 97.4\% accuracy, domain + message dual checking & Potential bias for Paytm themed URLs \\


\citep{dsmish_mishra_2023} & URL-based detection using authenticity with MLP model & 97.93\% accuracy, consistent results & Limited features, narrow comparison \\

 \citep{hapase2024telecommunication} & Time-efficient feature extraction with Pearson correlation coefficient and Principal Component Analysis (PCA) and CNN model & Competitive accuracy rate (99.8\%) with robust model setup & Intricacy to re-generate features for new sms \\

 \citep{seo2024device} & Model resilient to text evasion & On-device, 99\% accuracy on 1.2M messages & EVA tool generalization unverified \\
 2024\citep{shinde2024sms} & Optical Character Recognition + hybrid model for detecting spam from screenshots & 94.13\% Accuracy, user-adaptive & Limited scalability, language constraints \\

 \citep{rose2024next_federated} & Federated ConvLSTM model & 99.19\% Accuracy, Scalable and adaptable for other phishing scenarios & Low interpretability, domain-limited \\

\midrule
\rowcolor{gray!45} \multicolumn{4}{|l|}{\textbf{Language Model (LM) Approaches}} \\

\citep{smdetector_ghourabi} & App with multilingual smishing detection BERT-FC model & 99.63\% accuracy, cross-language support & Lack of mobile implementation details causes privacy concerns \\

 \citep{sidhpura2023_fedspam:} & Protecting users' privacy using an on-device detection method with DistilBERT model & Privacy-preserving method and high accuracy (98\%) & No user-study validation \\

 \citep{uddin2024explainabledetector} & Explainable SMS spam detection RoBERTa model & 99.84\% detection rate, balanced data & lack of robust explainable approach like SHAP or gradient explainers \\
 \citep{anh2024federated} & Privacy-preserving SMS spam detection with PhoBERT model & 99.38\% accuracy, privacy-preserving via FL, dual language support & Subjective data labeling for non-spam entries, synthetic evaluation data\\

 \citep{mehmood2024enhancing} & Use  CNN-LSTM model for detection & 99.74\% Accuracy, strong evaluation & Overfitting, low data diversity \\

 \citep{llm_shim2024persuasion} & Few-shot RoBERTa with augmentation & Cost-effective learning, 97.8\% accuracy with RoBERTa model in Tenfold augmented data & Needs human checks, attacker misuse risk \\
\citep{wangcan} & Agentic AI , LLMs detection , user-friendly explanation & Uses external context (URLs, WHOIS, HTML, screenshots, brand info) , multi-modal LLM, 98.8 \% accuracy & Potential high computational cost of LLMs, external tool dependency \\

\bottomrule
\end{tabular}

\end{table*}

\subsubsection{Machine Learning (ML) based Detection Approaches} 

Beyond rule-based methods, ML-based approaches offer a more sophisticated and diverse smishing detection. Table \ref{tab:method_sum22} provides a summary of methods and techniques employed in recent smishing detection literature. In this approach, researchers utilize established algorithms such as Neural Networks (NN) \citep{ghourabi2020hybrid, smdetector_ghourabi,spam_filtering_cnn_lstm_hossain,implement_smishing_detector_mishra_2022, shanto2023federated, sidhpura2023_fedspam:, seo2024device, anh2024federated, hapase2024telecommunication, rose2024next_federated, mehmood2024enhancing}, Support Vector Machines (SVMs) \citep{sonowal2020detecting, ulfath2022detecting, jain2022content, karhani_phishing_2023}, Naive Bayes \citep{goel2018smishing,mldetect_smishing_frauds_boukari,njuguna2021model,smishing_detector_security_model_mishra_2020,jain2022content}, Decision Trees \citep{sonowal2020detecting, ulfath2022detecting, jain2022content, smsprotect_akande, karhani_phishing_2023}, and Random Forests \citep{sonowal2020detecting, mldetect_smishing_frauds_boukari, ulfath2022detecting, jain2022content} 
algorithms that are trained on features extracted from message contents, such as keywords, part-of-speech patterns, or sentiment analysis. Some of the important and relevant features from existing literature are highlighted below. We also provide a comparative overview of key works on smishing detection, and highlight their strengths and limitations in Table \ref{tab:review-1}. Clearly, the winners are the language models in terms of accuracy and detection rate as depicted in this table.

\noindent \textbf{Indicators of Detection.} A combination of URL-based and/or text-based features serving as indicators of smishing attacks.

\noindent \textbf{(i) Content Features:}
Oftentimes, the actual content of SMS messages can be used as features. ML algorithms can learn from large datasets of labeled messages (smishing vs. legitimate) where texts are tokenized to root words by applying methods like stemming or lemmatization, or grouped as n-grams. Then, the supervised learning approach can be used to learn from the ground-truth dataset and identify patterns of smishing messages. The ML models need the tokenized text to transfer into vectors of numbers by applying methods like CountVectorizer, bag of words (BoW), Term Frequency-Inverse Document Frequency (TF-IDF), word to vector (Word2Vec), global vectors for word representation (GloVe), or custom tokenized word vectors which represent each token with relevance importance in the vector space of the given corpus. However, the researchers also used heuristic-based extraction from text messages that involve features such as message length, the count of parts of speeches (legitimate words), number of misspelled words, count of special characters, spaces, punctuations, alphabets, and digits, binary flag check of whether the message contains emails, URL or phone numbers, readability check, saved in a tabular format \citep{smsprotect_akande,sonowal2020detecting}. 

\noindent \textbf{(ii) URL Features:}
While analyzing message content is crucial, smishing attacks often rely on malicious URLs to steal user information or deploy malware. In some real-world cases, the only message content is the URL itself. Here, we highlight some of the existing features used in literature that are extracted from the URLs attached to SMS messages. 
Usually, there are two types of features extracted from URLs:
\begin{itemize}
    \item \textbf{\textit{URL lexical information:}} Lexical information is collected directly from the URL strings and is easily available. For example, features such as unusual characters, excessive length, presence of subdomains, or specific keyword patterns within the URL can indicate potential red flags. Other features include the number of letters and digits, a binary flag check of whether the URL is shortened or not, and whether the domain name matches the nouns extracted from the original message. 
    
    \item \textbf{\textit{URL host and web content information:}}  Host and content-based information from the URLs are not directly available and need further collection of data by querying each URL. Thus, host information is more challenging to collect and may have time-limited value, as some domains may not have any meaningful contributions once it is taken down. 
    Techniques involve checking whether the domain name actually represents the corresponding brand name (if any) provided with the message, whether the domain name extracted from the URL belongs within the top 5 URLs resulting from Google search engine  \citep{dsmish_mishra_2023} the website's reputation, age, registration details, domain ranking, redirection status, HTTPS check, TLD reputation, web page contents, external links, internal links, and webpage size \citep{karhani_phishing_2023, mia2024can}. 
    Additionally, analyzing the content itself for phishing elements or malware presence can further strengthen the detection process \citep{smdetector_ghourabi, timko2024smishing}, and this can be done by checking the repository of blacklists such as VirusTotal \citep{virus-total} or PhishTank \citep{phishtank}.
\end{itemize}

However, the use of traditional content and URL features is not effective when facing modern and evolving attack models. For example, in conversational scams, such as the \textit{'Hi Mum and Dad'} scam  \citep{agarwal2025hey}, which has recently been prevalent in the UK region, the initial contact text intentionally bypasses classic cues. This message often contains no malicious URL and uses psychological techniques instead to convince the victim to continue the conversation on a different platform (such as WhatsApp or Telegram) using a new mobile number provided in the message. The fraud request, including the user's account details, is only shared in the next messages. This multi-stage nature reveals that the set of features used to analyze the presence of URLs or keywords in the initial message might fail to identify the threat at its entry step.

\subsubsection{Deep Learning-based Detection Approaches}
Deep Learning techniques have also shown promising results \citep{smdetector_ghourabi}\citep{spam_filtering_cnn_lstm_hossain}\citep{implement_smishing_detector_mishra_2022} in content and URL-based analysis and detection of smishing attempts. Additionally, Mishra and Soni \citep{dsmish_mishra_2023} used a leet words check, which indicates if any letters are replaced with proximity numbers (like the letter `o' is replaced with digit zero `0') and a binary check if the corresponding message contains any smishing-related keywords from a pre-determined keywords list with a back propagation algorithm. Deep neural networks, with their ability to learn complex relationships from data, can be particularly effective in identifying nuanced language patterns and uncovering hidden features within message content that might be missed by classical machine learning approaches. Among these, convolutional neural networks (CNNs) have been explored for extracting features from SMS text\citep{ghourabi2020hybrid}, using spatial hierarchies in the data to detect malicious patterns. Long Short-Term Memory (LSTM) networks have been widely used due to their capability to model sequential dependencies in SMS phishing messages, capturing context beyond simple word-based analysis\citep{spam_filtering_cnn_lstm_hossain}. Additionally, bidirectional autoencoders have been investigated for their potential in learning compressed representations of smishing messages, allowing for anomaly detection by reconstructing normal SMS patterns and identifying deviations indicative of phishing attempts\citep{mehmood2024enhancing}. However, while these models improve classification performance, their hidden layers add complexity, reducing the transparency and explainability of classifier decisions, which remains a challenge for real-world adoption. In addition, deep learning models tend to obtain higher accuracies than traditional algorithms as they automatically learn relevant features, resulting in fewer inaccuracies where real messages are misclassified as malicious. 

\subsubsection{Natural Language Processing-based Detection Approaches}
Different strategies that would integrate NLP technologies along with ML algorithms to investigate smishing and fraudulent content are also observed in the current literature. Methods combining \textit{TF-IDF} and \textit{LDA} (Latent Dirichlet Allocation) have been shown to surpass the performance of weight average \textit{Word2Vec} for both model accuracy and F1 scores, indicating that the information provided by text analysis is often very effective for smish detection \citep{samad2023smishguard}. 
Combining NLP with ML provides substantial ground for addressing the burgeoning challenge of smishing attacks \citep{ulfath2022detecting}. Liu et al. \citep{liu2021detecting} proposed a novel NLP-based detection algorithm with entity recognition using NER (Name Entity Recognition), 
regular expressions for PII, and syntactic parsing that achieved a high precision of 96.16\% in identifying spearphishing messages from a large SMS corpus. In another study, Ghourarbi et al. \citep{smdetector_ghourabi} proposed to use a pre-trained BERT (Bidirectional Encoder Representations from Transformers) model for smishing detection. Moreover, Shim et al. \citep{llm_shim2024persuasion} incorporated the RoBERTa model (an improved variant of the original BERT model) for the detection. They used an augmented dataset created with prompt engineering based on persuasion theory and using the GPT-3.5 model. More recent studies include the work of Lee and Han \citep{expl_lee2024korsmishing} who proposed a framework based on a Korean-centric LLM. Their contribution is an end-to-end approach to classify and explain smishing messages, achieving a 15\% improvement in accuracy over the GPT-4 model.

Additionally, Uddin et al. \citep{uddin2024explainabledetector} used the RoBERTa model with a data augmentation technique called \textit{back translation}, where they reduce the issue of class imbalance while maintaining contextual similarity. The \textit{back translation} method involves translating a native message into one or more non-native language-based messages and then re-translating those back into the original language message by using any ML-based translator. To expand LLM-based approaches, Wang et al. \citep{wangcan} developed SmishX, a prototype that uses LLMs to not only detect SMS phishing but also to generate evidence-based explanations. Addressing the challenge of SMS lacking context for security reasoning, SmishX gathers external contexts such as domain and brand information, URL redirection, and web screenshots to augment the LLMs' chain-of-thought reasoning. This approach achieved an overall accuracy of 98.8\% in real-world SMS datasets and significantly improved users' phishing detection efficacy across age groups through its explanations.

Furthermore, some other defenses are designed to defend against specific PTs. One example is a defense measure that counters attacks using the \textit{Impersonation} PT, called the Online Social Network (OSN) Profile Cloning Protection method. It is reported to have a precision of 97.23\% in detecting OSN-based profile cloning attacks. This defense is effective for social media platforms and relies on users' extensive profiles and friend list data \citep{XuPIEEE2024}. This nature of the short messages 
makes it unsuitable for the smishing context, which involves short and less contextualized text messages. To address this mismatch, we can use an alternative approach like the SmishViz system \citep{pritom_25_codaspy_smishviz} that focuses on finding similarities in message attributes rather than social media profiles. The SmishViz system employs NLP techniques to analyze textual patterns and themes for grouping messages based on shared web infrastructures, which results in the identification of impersonating messages. This methodology helps to mitigate the \textit{Trust} factor by making it harder for attackers to exploit a victim’s trust in a smishing message \citep{XuPIEEE2024}. 

\subsubsection{Federated Learning-Based Detection Approaches}
Researchers are also deploying federated learning (FL) to address the limitations associated with centralized data collection. This approach enhances the privacy and security of user data by eliminating the need to transmit raw data to a central server, thereby facilitating the local model training on client devices. First, one centralized ML model is created, and then it is trained on a global dataset and sent to the other connected client machines to be trained with the private data only incoming to that particular client. After a certain period, the model weights are sent to the central machine from each client, and then the global ML model gets updated with the new aggregated weights from all local models, which are redistributed to every client machine. This is a robust and adaptively iterative process, thus maintaining a contemporary context of the data in each iteration. Sidhpura et al. \citep{sidhpura2023_fedspam:} leveraged a pre-scheduled once-every-15-days training schedule on recent SMS data to allow local models to predict and store SMS labels in a local database system on every client device. The central model in this federated learning system is a DistilBERT model, a lightweight and efficient version of BERT. Following 15-day periods of local training on decentralized data, the updated weights from these local DistilBERT models are aggregated on a global server. 

A pre-trained language model for Vietnamese PhoBERT is used by Anh et al. \citep{anh2024federated} to present a novel FL framework for Vietnamese SMS spam detection that achieves similar or even higher performance than centralized techniques 
that allows collaborative training without disclosing sensitive data. While the non-independent and identically distributed (non-IID) setting in this study, a condition where data points are not drawn from the same underlying probability distribution and are not statistically independent from one another, ensures that each client has a considerably different distribution of SMS messages, reflecting the heterogeneity of the real-world data. The IID counterpart setting indicates that each client has a similar and balanced distribution of spam and non-spam messages in both Vietnamese and English. Next, Shanto et al. \citep{shanto2023federated} developed an LSTM-based global model that is not specific to any particular language. \ignore{In another study, Vats et al. \citep{vats2024federated} developed a custom neural network (NN) global classifier with one input embedding layer, several convolutional layers, and one softmax layer. Although the paper did not discuss the specific dataset used for the model training, the authors reported that the model’s accuracy exceeds 90\% in the first epoch and rises to 98.3\% by the third epoch as client updates are merged through the {\em FedAvg} aggregation process.} Moreover, Remmide et al. \citep{remmide2024privacy_federated} applied the {\em FedAvg} aggregation method with a BiLSTM global model, but it also suffered from low accuracy in the federated setup. On the other hand, Rose et al. \citep{rose2024next_federated} have used {\em ConvLSTM} global model with {\em FedAvg} aggregation and achieved a closer and higher accuracy than the initial non-federated setup after nine federated iterations. 
    
\subsubsection{Hybrid Detection Approaches}
Hybrid methods for smishing detection fully exploit the strengths offered by multiple ML, DL algorithms, and rule-based or a combination of these, instead of a single method, thereby 
reducing the possibility of false positives. A hybrid algorithm comprising CNN and LSTM 
gives an accuracy score of 99.74\%, which efficiently identifies significant features from text messages \citep{mehmood2024enhancing}. Another hybrid method combines Bidirectional Gated Recurrent Units (Bi-GRUs) and CNN 
for enhancing smishing classification, using text embeddings and Word2Vec trained on preprocessed text \citep{mahmud2024enhancing}.


\subsubsection{Explainability and Transparency of Detection Approaches}

Even though ML and DL are widely adopted in smishing detection, a critical aspect—explainability—is often overlooked when interpreting model decisions and enhancing transparency. The black-box nature of DL models, particularly CNN and LSTM architectures, makes it challenging to understand why a certain SMS is classified as smishing. A lack of explainability hinders trust, regulatory compliance, and model adoption in real-world security applications. In recent years, research has begun to explore explainability in smishing detection models. Techniques like SHAP (SHapley Additive exPlanations)\citep{NIPS2017_SHAP} and LIME (Local Interpretable Model-agnostic Explanations)\citep{ribeiro2016should} have been applied in cybersecurity \citep{mia2024can} and text classification \citep{uddin2024explainabledetector} to identify key contributing features in model predictions. These methods allow researchers to visualize which words or URL patterns influence the classification decision. However, few studies have specifically applied these explanation techniques to smishing detection. For instance, Bhagyashree D. Shendkar et al.\citep{shendkar2024enhancing} investigate and evaluate the use of SHAP and LIME to improve interpretability and transparency in phishing attack detection. They note that these models help analysts to understand how the detection system works.

Additionally, researchers have explored attention mechanisms within transformer models like BERT to enhance the interpretability of smishing detection \citep{uddin2024explainabledetector}, but these models are often computationally expensive. Hybrid approaches that combine ML models with interpretable rule-based techniques to balance accuracy and transparency \citep{mohite2024interpretable} can be applied for explainability and transparency of smishing detection. While the aforementioned approaches enhance model transparency, they often fall short of providing explanations that are readily interpretable to end-users. This is because these techniques rely on technical outputs, such as feature relevance scores or attention weights, which are typically comprehensible to domain experts but lack the intuitive clarity required for a non-expert user. 

Lee and Han \citep{expl_lee2024korsmishing} used fine-tune KULLM (Korean University Large Language Model) to generate structured explanations, while Uddin et al. \citep{uddin2024explainabledetector} used LIME \citep{ribeiro2016should} for smishing detection. Comparing these two explainable methods, the prompt-based KULLM explanation method is more precise and generates plausible explanations that provide specific reasoning with high fidelity. Wang et al. \citep{wangcan} significantly contributed to the explainability of smishing detection by developing a system that generates evidence-based explanations for lay users. They achieved this by augmenting the chain-of-thought reasoning of Large Language Models (LLMs) with external contexts, such as domain and brand information, URL redirection, and web screenshots. This comprehensive reasoning process is then summarized into short, semi-structured explanation messages that include the detection decision, key reasons, and actionable advice. This approach has been shown to improve users' phishing detection efficacy and received high usability ratings. Notably, their method also effectively suppresses LLM hallucinations in the generated explanations, ensuring factual consistency.

\subsection{Anti-Smishing Tools and Mobile-based Defense Measures}

\begin{table*}[!h]
\centering
\caption{Critical Analysis of Existing Anti-Smishing Tools and Defense Measures}
\label{tab:dataset_result_sum}
\rowcolors{2}{white}{gray!10}
\scriptsize
\begin{tabular}{|m{7.1em} |m{10.2em} |m{10.55em} |m{5.55em} |m{10.6em}|m{10.5em}|}
\hline
        Defense Strategy  &     Reference &                                      Countermeasure Approach &                     Perf. metrics  &  Strengths & Weaknesses \\
\midrule

     &  \citep{shahriar2015mobile}  & SMS/URL filtering with blacklists and trusted number whitelist  & - & Detects various mobile phishing types, low false positives, and computationally lightweight & Limited detection methods, reliance on user awareness, and operative system or device dependency \\

      & \citep{njuguna2021model}    & Sender authentication + con-
tent filtering + user awareness &  - & Focuses on user awareness,
combined filtering & Naive Bayes reliance lacks
evaluation details  \\

Mobile-Based  &
      \citep{kohilan2023machine} &  Neural Networks (CNN,
RNN, Simple NN) +
Translator API + Floating
Bubble UI &  99.68\% ACC   &  multilingual
detection without language-
specific datasets, real-time
monitoring, user-friendly
interface for social media
platforms & Potential struggle with inten-
tional misspellings/grammar
errors in fraudulent messages,
challenges with hybrid languages (e.g., Singlish, Tanglish) \\

      &  \citep{smsprotect_akande}    & Mobile app for SMS spam,
user-controlled filtering &  98.42\% ACC & Handled
missing data well & Limited features, struggles
with missing data  \\

      &  \citep{wang2024verisms}    & Call-to-verify system with dynamic Message ID and static
Secret Words &  SUS: 79.1/100 & Inclusive (works on feature phones without apps), resilient to omission-based spoofing attacks & High user effort (requires dialing), Scalability issues for multiple services \\

      &  \citep{goel2024machine}    & SMS slang normalization +
Naive Bayes classifier &  96.2\% ACC, 97.14\% TPR, 96.12\% TNR & Effectively handles informal
SMS language. & English-only slang dictionary,
content-only (no URL/app behavior yet)  \\

  \midrule

     &  \citep{siadati2017mind}  & Improving the message
verification mechanisms with
warnings  & 8\% Success Rate (Attack) & Effective multi-factor authentication security (e.g., 2FA)
and scalable tools & only focus on SMS 2FA with
limited use-case scenarios \\

     User Awareness   &    \citep{blancaflor2021let} &  Phishing awareness campaign using simulation &  Smishing campaign success rate: 4.17\% &  Comprehensive platform
comparison, user behavior
analysis & Small sample size, static secu-
rity, single campaign \\

      & \citep{clasen2021friend}    & Phishing detection from
URLs using ML techniques &  73.4\% OA (87.6\% phishing vs. 59.2\% genuine) &  User-centric focus, cue-based behavioral analysis & Demographic bias, age imbalance, artificial survey environment  \\

 \midrule

      Law Enforcement &   \citep{zielinski2024evolving_law} & ACAEC law: SMS abuse reporting system + senderID/CallerID spoofing controls + operator blocking

of fraudulent texts/domains &  - & Direct legal basis for prosecution, compels industry action, provides public reporting
mechanism & Implementation and compliance burden; relies on user reporting only  \\

 \midrule

    Infrastructure-Based   &    \citep{on_phishing_tactics_nahapetyan2023sms} &  Analysis of SMS phishing
campaigns, infrastructure, and illicit markets (derived from public gateway data) to inform investigations and disruption efforts against cyber-criminals &  Smishing dataset with 68K records   &  Offers detailed insights into phishing, infrastructure, and
supply chains to aid targeted disruption and actor identification. & Data source bias (e.g., Western focus, no MMS), reliance on limited external detection tools, and reveals current regulatory gaps leading to abuse shift to unmonitored channels. \\
 \midrule
    Cross-Device   &    \citep{kanaoka2023beyond} &  Cross-device approach using
AR and URL analysis &  SUS: 74.4/100   &  Reduced device vulnerability,
user engagement & Relies on AR tech, camera
limitations \\
 \midrule
    Social Media Analysis   &    \citep{twitter-clues-ccs2022-spam} &  SpamHunter pipeline (image detection, tweet classification, text recognition) & 95\% Prec, 87\% TPR     &  Up‑to‑date, multi‑lingual corpus, surfaces spam URLs days–weeks before VirusTotal & Relies on voluntary Twitter
reports/Twitter API, sampling
bias, no direct SMS blocking \\
 \midrule
    Detection and Filtering   &  \citep{commerial_anti_smishing_timko_2023} &  Crowd-sourced smishing data collection &  55 zero-day samples, 0\%–61.6\% Detection Rate   &  Used real smishing data, novel dataset approaches & Small sample size (20 tested) and selection bias (user-reported samples only)\\ 
\hline
\end{tabular}

\vspace{0.8em}
\parbox{\textwidth}{\small

\textbf{Keys:} 
ACC: Accuracy, SUS: System Usability Scale, TPR: True Positive Rate, TNR: True Negative Rate, OA: Overall Accuracy, Prec: Precision, 
}
\end{table*}

\subsubsection{Third-party Anti-Smishing Tools} 
In literature, Timko et al. \citep{commerial_anti_smishing_timko_2023}  provide a comprehensive benchmark of anti-smishing tools and a qualitative analysis of smishing attacks. The role of third-party anti-smishing solutions is supplementary protection. These apps can be used alongside the filtering mechanism of mobile carriers. Carriers have the capability to use built-in message filtering or refuse to deliver messages via their networks. These apps increase security by providing additional layers of detection and blocking phishing messages that bypass the carrier-level filters \citep{commerial_anti_smishing_timko_2023}. 
These apps mostly work as filters and classifiers to categorize messages into spam or phishing. For zero-day smishing attacks, content-based filtering, including heuristics and ML approaches, is also popular for phishing detection and classification. A unique feature of content-based filtering over list-based filtering is its ability to detect zero-day smishing attacks. The impacts of these third-party anti-smishing apps are different and vary. Some apps, like Robokiller  \citep{appleRobokillerSpam} and Textkiller \citep{appleTextkillerSpam}, showed higher smishing hit rates. The study tested various anti-smishing apps, including Key Messages, Anti Nuisance, Call Control, and Calls Blacklist for Android, and SpamHound \citep{appleSpamHoundSpam}, Robokiller, Malwarebytes \citep{appleMalwarebytesMobile}, and NomoRobo \citep{appleNomoroboRobocall} for iOS. Among these tools, Textkiller had a significant smishing hit rate, and Robokiller demonstrated the highest smishing hit rate among the tested apps while blocking a high percentage of benign messages. NomoRobo also showed a good performance in managing to block a large portion of smishing messages while minimizing false positives. There are some other tools called Bulk Messaging Services, designed for legitimate mass communication, but can be exploited by attackers. These tools attempt to filter out malicious content. Among them, SimpleTexting \citep{appleSimpleTexting} and Text-Em-All \citep{appleTextEmAll} achieved higher smishing hit rates than other services, for example, Twilio \citep{twilioCommunicationsAPIs}, SlickText \citep{googleSlickTextApps}, and TextSpot \citep{textspotTextSpotMass}. This shows a better impact in identifying and blocking phishing messages \citep{commerial_anti_smishing_timko_2023}. 
Another proposed solution for mitigating smishing attacks on mobile platforms (MMSAMP) \citep{njuguna2021model} introduces an Anti-Smishing Application (ASAPP) to authenticate sender IDs, preprocess data, filter smishing content, and classify messages as safe or unsafe. This model, which is developed using Python, MySQL, and Naïve Bayes, uniquely informs users of potentially harmful content, a feature not found in existing solutions. This model combines ASAPP with user awareness, anti-malware programs, and information security policy enforcement for a comprehensive defense. The MMSAMP achieved 94\% efficiency in detecting, filtering, and classifying smishing content.


\subsubsection{Mobile-based Solutions and User Alerts}
Mobile-app-based solutions and user alerts are crucial defense mechanisms against smishing attacks. These solutions address the unique vulnerabilities of mobile devices. As highlighted by Shahriar et al. \citep{shahriar2015mobile}, the small screen size makes it difficult for users to identify legitimate websites from spoofed ones. To prevent this, mobile security solutions often employ content-based filtering, which examines the context and suspected URLs within SMS messages to identify phishing attempts. Additionally, user awareness and training are important, as users are frequently less aware of security options on mobile platforms compared to desktop environments. Best practices, such as downloading only official applications, utilizing safer browsers with built-in security features, and bookmarking frequently accessed legitimate sites, can significantly reduce the risk of falling victim to smishing. Furthermore, \citep{shahriar2015mobile} suggested that app stores should implement more strict controls over application uploads to prevent malicious apps as well as mobile security applications, similar to desktop antivirus programs for detecting and eliminating malicious activity.

Researchers have explored the use of security features on mobile devices, such as two-factor authentication 
in order to prevent smishing attacks. 
This method adds a layer of protection beyond standard credentials, and during this process, a one-time verification code will be sent to the user through SMS. While this method aims to enhance user safety in case of credential compromise, Hossein Siadati et al. \citep{siadati2017mind} 
demonstrates that SMS-based two-factor authentication can be vulnerable to social engineering techniques, specifically Verification Code Forwarding Attacks (VCFA). These attacks can force users to forward authentication codes, with an experimentally obtained 50\% success rate against Google's SMS-based authentication. 
Moreover, the authors discussed two more strategies 
to decrease smishing attacks  \citep{siadati2017mind}: (1) \textit{malicious request detection:} which involves profiling users using ML with location data from SMS providers to limit attackers, (2) \textit{sender verification:} enhance user interface usability to highlight sender information to SMS recipients. 
Furthermore, \citep{siadati2017mind} also highlights the use of effective warning components as countermeasures, such as: (1) \textit{abuse-proof:} meaning the content of the warning should be carefully mentioned to prevent attackers from abusing it to their advantage for social engineering attacks, (2) \textit{worry-free:} meaning the warnings should include the risk alerts by telling users that no immediate action is required if they have not requested the code, reducing unnecessary anxiety, (3) \textit{actionable and practical:} it should provide clear instructions on how to avoid the hazard, and (4) \textit{concise and clear:} it should be brief, easily understandable, and avoid technical jargon. It is interesting to highlight that the study mentions risk alerts stating ``Please ignore this message if you did not request a code'' before the verification code significantly reduced the attack success rate to just 8\% \citep{siadati2017mind}.

Another novel approach, introduced by Kanaoka et al. \citep{kanaoka2023beyond}, is a cross-device solution for smishing detection and prevention that extends beyond traditional mobile device limitations. This method involves using a separate device, such as augmented reality (AR) glasses, to capture images of displays (smartphones, laptops, televisions, or walls) and analyze embedded URLs for malicious content. The system evaluates URL maliciousness and provides risk indications to the user on the AR glasses, and encourages them not to access high-risk links. A prototype system using AR glasses demonstrated a significant improvement in correct decision-making, with users achieving an 85.19\% correct rate with the prototype compared to 46.3\% without the glasses. This indicates the potential of image analysis for smishing detection. However, the requirement of an additional device (e.g., AR glasses) is not a generic or practical solution. 

In the realm of mobile-based solutions, Akande et al. \citep{smsprotect_akande} developed SMSPROTECT, a mobile application that automatically detects smishing. It intercepts incoming SMS messages and sends them to an ML-based model for analysis 
This model uses \textit{RIPPER} and \textit{C4.5} classifiers to formulate rules that analyze the message content and classify it as spam or legitimate. The user is then notified of the classification and can decide whether to keep or discard the message. This system provides a practical way to prevent smishing.

Other research also highlights that ML models, particularly in combination with NLP techniques, can be trained to identify patterns and language in SMS messages \citep{siddiqi2022study}. These trained models can be used as the core back-end of mobile applications or on-device solutions to identify language patterns 
in SMS messages to detect abnormalities and flag potential smishing attempts, and finally generate user alerts. Despite all these advantages, using ML as the core of mobile-based defense still requires updated training data to deal with the multi-dimensional nature of smishing attacks, which exploit human vulnerabilities and psychological decision-making \citep{siddiqi2022study}.


Another approach proposed in the literature is to use a two-factor scheme consisting of a dynamic 5-digit Message ID that is generated for each message and two static Secret Words that are specifically generated for each user to verify messages. This approach is presented in \textit{VeriSMS} \citep{wang2024verisms}, which is a solution that healthcare providers can use to contact their patients. In this system, patients receive a card during their in-person visit that contains a list of their Secret Words and the clinic’s verification phone number to maintain trust. The VeriSMS providers encourage patients to call an agent to verify a message whenever they are unsure. During the call, the user types the Message ID and hears the Secret Words read back. If they match the card, the message is genuine. This method is designed to be inclusive and accessible without requiring extra hardware. However, VeriSMS has several shortcomings. 
The use of static Secret Words represents an intentional trade-off between usability and security. Their user study showed that most users would not call to verify every message and some would only call when they felt suspicious, and others would not call at all. As a result, using dynamic Secret Words was not considered practical and was therefore avoided. This shows that the system strongly depends on users' interaction, and because the user action is not guaranteed, the highest level of security is not consistently applied. Furthermore, participants described the process of switching screens to enter the Message ID as `tedious'. Mistakes like typing the ID incorrectly sometimes caused legitimate messages to be flagged as fraudulent. In addition, if an attacker were to obtain a user's static \textit{Secret Words} in other ways, the system's effectiveness would be significantly reduced, especially for users who do not use voice calls to verify the messages \citep{wang2024verisms}.


 There are other proposed methods that incorporate mobile applications for smishing detection by analyzing incoming messages for suspicious message contents, links, or sender information \citep{goel2024machine, shinde2024sms, chichwadia2024detecting, chithambaramani2024sms}. These apps can leverage ML algorithms to flag messages based on known smishing indicators and real-time threat intelligence. 
However, the effectiveness of mobile app-based solutions depends on user adoption and the ability to keep the respective detection models updated against evolving smishing techniques. The concern with user privacy in the process of analyzing incoming messages still remains a key challenge. 

\subsection{User-Training, Awareness, and Regulation-based Defense Measures}
\begin{figure*}
    \centering
    \includegraphics[width=1\textwidth]{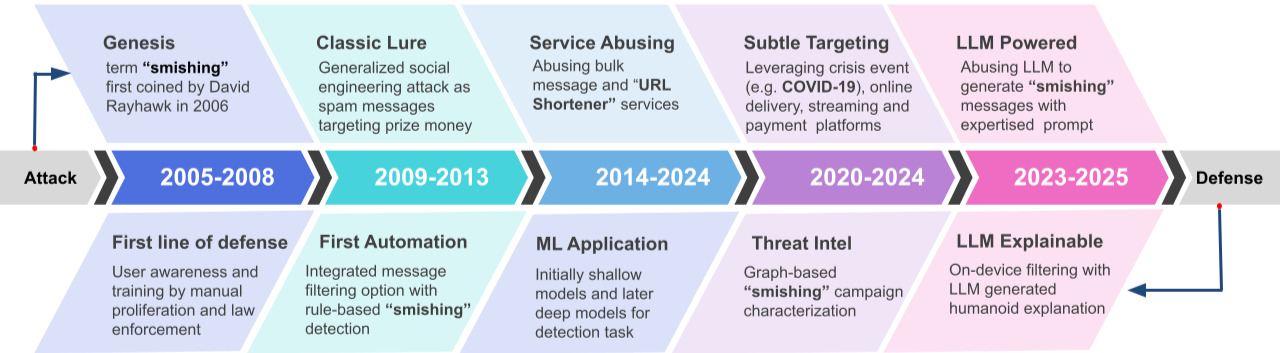}
    \caption{Evolution in smishing attack (upper boxes) and defense (lower boxes) strategies over the years}
    \label{fig:smsih_evolution}
\end{figure*}

\subsubsection{User Training and Awareness Practices }
    User awareness emphasizes the importance of educating users to identify various red flags while interacting with SMS messages \citep{katsarakes2024users}. Clasen et al. \citep{clasen2021friend} used an online questionnaire to examine which features help people spot smishing, which shows that user awareness training plays a vital role in empowering individuals to detect and avoid smishing. Training programs can educate users about common smishing tactics using suspicious message elements such as shortened URLs, inconsistent message content, and urgency style \citep{clasen2021friend}. 
    This study also suggests that interactive simulations like the questionnaire used in this study and real-world scenarios can enhance user engagement and retention of key defensive strategies \citep{clasen2021friend}. 
    The study by Blancaflor et al. \citep{blancaflor2021let} further highlights the importance of user awareness through a phishing campaign that included smishing. Although their smishing campaign had a low success rate (4.17\%), the overall findings from their post-experience survey revealed that trust is the most significant factor influencing whether targeted users click on malicious links. This suggests that even with existing security measures on devices, the human element of trust and curiosity remains a primary vulnerability, underscoring the critical role of continuous user education in recognizing and resisting social engineering tactics. Finally, we find evidence from literature that user awareness training, combined with technological solutions, can aid the multi-layered defense strategies against smishing attacks.

\subsubsection{Social Media Analysis and Awareness}
People often report smishing messages on social media to warn others, seek help, or notify impersonated services. This behavior creates an opportunity to use platforms like X (previously Twitter) as a real-time and collective user reporting crowd source, which enables early detection and rapid response against ongoing smishing campaigns. For instance, researchers have proposed a solution that allows users to verify smishing attempts from social media platforms such as Facebook and WhatsApp using ML-based English translated content analysis \citep{kohilan2023machine}. 
Furthermore, SpamHunter enables researchers to collect spam, scam, and smishing messages derived from X (previously Twitter) posts where users shared screenshots of these messages \citep{twitter-clues-ccs2022-spam}. The authors highlighted the strength of this approach, evidenced by the fact that more than 15\% of the dataset's reported URLs were subsequently flagged by threat platforms (e.g., VirusTotal).  

\subsubsection{Policy Enforcement and Regulatory Efforts}

As smishing continues to rise, policymakers need to enforce coordinated measures such as effective policies and online monitoring efforts. 
Law enforcement agencies are increasingly tasked with tracking cybercriminals across jurisdictions, while regulatory bodies strive to define and mitigate these threats within coherent current legal structures
\citep{zielinski2024evolving_law, sugunaraj2022cyber_laws}.

Within the United States, several federal and non-profit organizations offer distinct services that support prevention, reporting, and recovery in response to cyber frauds such as smishing. The \textit{Federal Trade Commission} (FTC) allows consumers to report incidents of fraud and guides for victims to recover funds or halt unauthorized charges. To reduce the risk of follow-up fraud, the FTC clarifies that it typically does not contact individuals after a report is filed. The \textit{Internet Crime Complaint Center} (IC3), operated by \textit{Federal Bureau of Investigation} (FBI), enables individuals to report Internet-related crimes, including smishing, and collects key data such as the victim's contact and transaction information during the submission. While IC3 compiles these details for law enforcement purposes, the FTC's role focuses on informing affected consumers about possible remedies \citep{sugunaraj2022cyber_laws}. Other organizations also play crucial roles. The \textit{Department of Justice's Office for Victims of Crime} (OVC) provides financial aid for victim assistance programs, which may include services to assist older adults in reporting financial crimes and connecting them with appropriate support services \citep{ojpElderFraud}. The \textit{AARP Fraud Watch Network} provides scam alerts and technological tools to help users aged 50 years or older to protect themselves, along with information on dozens of fraud types. 
The \textit{National Center on Elder Abuse} (NCEA) offers educational materials about elder abuse, including financial abuse, and outlines when \textit{Adult Protective Services} (APS) may intervene to investigate reported incidents if they fall under state-level elder abuse laws \citep{sugunaraj2022cyber_laws}.

\begin{table*}[!t]
\centering
\caption{Comparison of Common Features Across Regulatory Frameworks in the US and EU Regions}
\label{tab:regulatory-comparison}
\scriptsize 
\rowcolors{2}{white}{gray!10}

\resizebox{1\linewidth}{!}{
\begin{tabular}{|m{11.75em}|m{12.5em}|m{12.5em}|m{12.5em}|m{15em}|}
\hline
\toprule
\textbf{Feature} & \textbf{US (TCPA \& State Laws) \citep{gdprlocalEssentialGuide}} & \textbf{California (CCPA) \citep{yotpoNeedKnow,gdprlocalEssentialGuide,caCCPAEnforcement}} & \textbf{Poland (ACAEC) \citep{zielinski2024evolving_law}} & \textbf{EU (GDPR) \citep{idtexpressDataPrivacy}}
 \\

\midrule
\textbf{Scope} & Governs telemarketing \& SMS as calls & Privacy \& data selling restrictions for CA residents & Focused on smishing  \& telecom abuse & Comprehensive data privacy regulation\\
\hline
\textbf{Consent} & Prior express consent needed & Consent for data use; opt-out for data selling & Not main focus, targets fraud & Explicit opt-in required \citep{idtexpressDataPrivacy} \\
\hline
\textbf{Penalties} & State \& federal fines (up to \$20K/msg) & Up to \$7,500 per violation, lawsuits allowed & Criminal (3 months--5 years) & Up to EUR 20M or 4\% revenue\\
\hline
\rule{0pt}{3ex}\textbf{Reporting} & FCC + state rules & Complaint \& lawsuit mechanisms & Toll-free reporting (8080) & General breach notification \\
\hline
\textbf{Data Handling} & Disclosure of identity \& opt-out options & Must disclose data practices \& third-party sharing & Allows telecom data sharing to combat fraud & Strict transparency \& privacy policies\\
\hline
\textbf{Enforcement} & Federal, state laws & CA Attorney General & National body (CSIRT NASK) & National supervisory authorities\\
\hline
\rule{0pt}{3ex}\textbf{Cross-Border Coordination} &  \rule{0pt}{3ex}Complex US compliance & \rule{0pt}{3ex}Limited to California only & \rule{0pt}{3ex}Minimal, national focus & \rule{0pt}{3ex}EU-wide harmonization\\
[2ex]
\hline
\textbf{Consumer Empowerment} & Do-Not-Call registry, opt-out & Rights to know, delete, opt-out & Reporting channel & Broad data rights (access, erase)  \\
\hline
\end{tabular}
}
\end{table*}

\begin{table*}[!h]
\centering
\scriptsize
\caption{Legal and Policy Categories Related to Smishing}
\label{tab:legal_policy_efforts}
\rowcolors{2}{white}{gray!10}
\begin{tabular}{|m{3.2cm}|m{4.2cm}|m{5.2cm}|m{3.2cm}|}
\toprule

\textbf{Category of Law/Policy} & 
\textbf{Primary Governing Body \& Relevant Acts} & 
\textbf{How It Applies to Smishing} & 
\textbf{Typical Penalties} \\ 

\midrule
\textbf{Consumer Protection} &
Federal Trade Commission (FTC): Enforces consumer protection laws that prevent fraudulent, deceptive, and unfair business practices\citep{ftcAbout}. The CAN-SPAM Act (2003)\citep{ftcCANSPAMAct} also applies to some commercial texts.
&
FTC investigates and penalizes smishing activities that fall under its broad mission to protect users. They focus on how these messages trick people and are used to commit fraud. They can use the CAN-SPAM Act (2003) to stop and punish these commercial text scams \citep{ftcNoticesPenalty, ftcCANSPAMAct}.
& FTC mentions about civil penalties and fines of up to \$50,120 per violation \citep{ftcNoticesPenalty}. \\ 
\textbf{Unsolicited Communications} &
The Federal Communications Commission (FCC) regulates interstate and international communication\citep{Aboutfcc}. The Telephone Consumer Protection Act (1991) \citep{FCC_TCPA_Rules} restricts the use of auto-dialed calls and text messages \citep{gdprlocalEssentialGuide}. &
The FCC fights smishing by mandating that mobile providers should block illegal robotexts, especially those from invalid or spoofed numbers. Under the Telephone Consumer Protection Act (1991) \citep{FCC_TCPA_Rules}, the FCC requires companies to get consumer consent before sending marketing texts. The agency also uses consumer complaints on spam and reports forwarded to 7726 to inform its enforcement actions and to help providers block illegal messages \citep{fcc_scam_texting_2023, fcc_robocalls_and_texts_2025}. &
FCC mentions that civil penalties and fines could be up to \$10,000 for each violation of illegal spoofing \citep{fcc_robocalls_and_texts_2025}. Moreover, The Telephone Consumer Protection Act
(1991) imposes statutory damages of \$500 to \$1,500 per violation \citep{cote_tcpa_2024}.
\\ 
\textbf{Fraud \& Identity Theft } &
Department of Justice (DOJ) and Federal Bureau of Investigation (FBI) are the primary law enforcement and legal bodies for fraud and identity theft protection \citep{fbi_spoofing_and_phishing, ftc_identity_theft_2004}. Further details can be found at Title 18, U.S. Code, Section 1343 (Wire Fraud Statutes, 2008)
\citep{USC18-1343,doj_criminal_manual} and Title 18, U.S. Code, Section 1028 (Identity Theft and Assumption Deterrence Act, 1998) \citep{USCode18_1028,doj_criminal_manual_1512}.

&
Smishing is often just the initial step in a larger criminal campaign. The Wire Fraud statute (2008)
\citep{USC18-1343} is a powerful policy tool since it criminalizes the use of an electronic medium, such as a text message, to execute a plan to defraud \citep{doj_criminal_manual,leppard_law_wire_fraud}. The Identity Theft
and Assumption Deterrence Act (1998)\citep{USCode18_1028} criminalizes the knowing transfer or use of another person's identifying information to commit any unlawful activity.  
&
In case of wire fraud, fines and/or imprisonment for up to 20 years. Penalties can increase to 30 years for scams targeting financial institutions
\citep{leppard_law_wire_fraud}. In case of identity theft, fines and/or imprisonment for up to 15 years, with higher penalties for aggravated identity theft \citep{doj_identity_theft}.
\\ 
\textbf{State-Level Action} &
State Attorneys General and various state laws. For example, California's Anti-Phishing Act (2005)-- specifically targets phishing and spoofing \citep{california_bpc_22948_2005}. Virginia's Computer Crimes Act (1984) contains provisions that can be applied to smishing \citep{virginia_code_18_2}. 
& 
States create specific legislation to complement federal laws and prosecute crimes that occur in their areas. These laws extend the definitions of computer and identity theft to include smishing and allow state prosecutors to pursue cases independently of federal agencies.
\citep{werksman_jackson_anti_phishing, virginia_code_18_2}
&
Penalties by state and often include major fines and felony charges, which can lead to imprisonment. For example, Virginia's laws can result in felony charges with up to 10 years in prison \citep{virginia_code_18_2}. 
\\ 

\bottomrule
\end{tabular}
\label{tab:smishing-law}
\end{table*}

Recent research \citep{on_phishing_tactics_nahapetyan2023sms} emphasizes the challenges faced by these policy efforts as indicated by the 70\% increase in SMS and voice phishing attacks in 2022. 
Studies have observed that while regulatory measures, such as the Federal Communications Commission (FCC)'s efforts adopted in March 2023 to block robotext messages, are implemented, they do not always address the fundamental issues of weak identity verification in telephony communication or the difficulty in distinguishing legitimate from fraudulent messages. Regulatory efforts by network carriers have focused on making it more difficult to send bulk application-to-person (A2P) traffic. While these measures may have been effective in closing off the A2P route, they inadvertently pushed malicious actors toward exploiting unmonitored person-to-person (P2P) channels. 
A notable spike in phishing message volume in public SMS gateways between the FCC's March 2023 order and its September 2023 implementation deadline indicates that the problem is not self-resolving. Furthermore, an illegal market that openly advertises bulk SMS services on public platforms such as LinkedIn and Telegram facilitates the transit of annoying or illegal traffic \citep{on_phishing_tactics_nahapetyan2023sms}.

From a legal perspective, both the United States and the European Union (EU) have introduced frameworks to mitigate smishing efforts by regulating the collection and utilization of consumer data in SMS communication. Table \ref{tab:regulatory-comparison} provides a
a high-level, global comparison of broad regulatory frameworks in the US and EU, and highlights differences in consent requirements, penalties, and enforcement agencies. The \textit{General Data Protection Regulation} (GDPR) governs all electronic communications in the EU, including SMS marketing. It requires businesses to get clear permission from users, explain how their data will be used, and give them an easy way to stop receiving messages, which helps prevent fraudsters and unapproved marketers from misusing SMS services \citep{gdpr-walker, idtexpressDataPrivacy}. Furthermore, while the \textit{EU NIS 2 Directive} (Directive (EU) 2022/2555) does not explicitly refer to smishing or SMS phishing, it considers providers of electronic communications networks and services as critical infrastructure and requires them to implement robust cybersecurity risk-management measures such as prevention, detection, response, and user-awareness mechanisms against social engineering attacks. It also requires them to report significant incidents, defined as cyber incidents causing substantial service disruption, financial loss, or large-scale user impact, within strict timelines. These rules apply directly to handling and reducing the impact of mass smishing attacks \citep{eu_nis2_2022}. 
Beyond the EU, the United Kingdom addresses smishing primarily through the \textit{Fraud Act 2006}\citep{uk_fraud_act_2006}, particularly Section 2 (Fraud by False Representation), which criminalizes the dishonest use of false or misleading representations made with the intent to obtain financial gain or to cause loss. Unlike the predominantly civil enforcement model observed in the United States \citep{cote_tcpa_2024}, the UK framework establishes explicit criminal liability for deceptive conduct itself, even when financial harm is not immediately realized. Although smishing is not explicitly named in the legislation, it falls directly within the scope of fraud by false representation, as the Act recognizes that fraudulent representations can be made through any electronic communication system. Inspired by this, Poland took a more specific approach by introducing the \textit{Act on Combating Abuses in Electronic Communication} (ACAEC) in 2023, which provides a legal definition of smishing and permits regulatory authorities to block malicious messages and penalize caller ID spoofing \citep{zielinski2024evolving_law}. In the Asia-Pacific region, they adopted a more hybrid approach. Frameworks like the \textit{Personal Data Protection Act (PDPA)} in Singapore\citep{singapore_pdpa_2012} and Thailand\citep{thailand_pdpa_2019} combine privacy consent(EU) with telecommunications restrictions(US). Singapore’s PDPA enforces a strict Do Not Call (DNC) Registry, where sending unsolicited telemarketing messages to registered numbers incurs financial penalties of up to \$10,000 per offense\citep{singapore_pdpa_2012}. Thailand’s PDPA goes further and allows for punitive fines and potential criminal imprisonment for data handlers who facilitate fraud\citep{thailand_pdpa_2019}. In contrast to these unified national frameworks, the United States relies on sector-specific state laws. The \textit{California Consumer Privacy Act(2018)} (CCPA)\citep{CACCPA2018} provides a state-level framework that gives consumers rights to access, delete, and restrict the use of their personal data, particularly in marketing communications \citep{yotpoNeedKnow, gdprlocalEssentialGuide}. However, unlike the GDPR or ACAEC, the CCPA does not directly address smishing as a unique threat vector. This limits its capacity to empower enforcement agencies or telecom providers to take early action against smishing. While the US has comprehensive privacy rights, it does not have the same centralized rules and prevention measures that Europe uses. In conclusion, while both the US and Europe recognize the critical importance of consumer protection against smishing, their legal responses differ significantly in scope and execution. Europe emphasizes proactive regulation and state-provider collaboration, especially with GDPR and Poland’s ACAEC. While the US is robust in privacy protection, its laws and responses are more scattered. As smishing techniques grow more complicated, aligning legal mandates with operational enforcement will be key to ensuring digital trust and safety across borders.

Here, Table \ref{tab:regulatory-comparison}, compares the scope and execution of consumer data protection laws across geographical jurisdictions (US and EU) and shows the fundamental difference between the EU's unified and proactive regulatory approach (like GDPR) and the US approach, which is mostly scattered across different states and sectors. Moreover, Table \ref{tab:legal_policy_efforts} demonstrates the fragmentation of current laws and policies related to smishing among US regulatory agencies (e.g., Federal Trade Commission, Federal Communications Commission, Department of Justice, etc.). It also reveals that none of the US agencies or laws address all aspects of smishing comprehensively, which poses further challenges for coordinating law enforcement. Next, we also highlight a couple of real-world cases where policies reflected into law enforcements within US and UK, respectively.

\noindent \textbf{Real-World Cases with Policy Enforcements}

\noindent \textbf{United States:} Despite the robust legal frameworks, a gap remains between policy implementation and effective prosecution. In the US, the enforcement landscape is bifurcated. For major threats, the Department of Justice pursues criminal indictments, such as the November 2024 charges against five defendants (e.g., United States v. Elbadawy) who used mass smishing campaigns to steal credentials from employees at major U.S. companies \citep{doj_phishing_scheme_2024}. However, for general consumer smishing, the system struggles to recoup losses. The FTC reports show that US consumers are constantly losing significant dollar amount sdue to text message scams \citep{FTC_latest_report_2025}, which suggests that current civil blocking measures are not enough to prevent the tide of smishing.

\noindent \textbf{United Kingdom:} In contrast to the US reliance on fines, the UK actively enforces penalties, with the Information Commissioner’s Office (ICO) issuing over £2.5 million in penalties for nuisance calls and texts since April 2023
prioritizes direct physical intervention. In June 2024, the City of London Police arrested two individuals for operating a homemade mobile antenna (an illegitimate SMS blaster) from a vehicle. This device was used to bypass network filters and transmit thousands of smishing messages directly to people. One of the operators was charged with possession of articles for use in fraud and sentenced to 21 weeks in prison \citep{colp_smishing_arrest_2024}. This case demonstrates a strategy of immediate physical disruption and incarceration rather than delayed financial penalties.

\subsubsection{Summary of Smishing Defense Landscape}
In summary, the evolution of smishing attack and defense strategies is a sophisticated ``arms race'' between cybercriminals and security researchers, transitioning from crude bulk spam to highly personalized, AI-driven deception as shown in figure \ref{fig:smsih_evolution}. In its infancy (2005–2008), smishing was a novelty defined by manual outreach and basic law enforcement intervention \citep{Steele_smish_term}. As mobile adoption surged through 2013, attackers pivoted to ``classic lures,'' such as lottery or prize winning scams, which necessitated the first wave of automated, rule-based filtering. The subsequent decade (2014–2024) saw an escalation where bad actors weaponized bulk-messaging infrastructure and URL shorteners \citep{content_url_analysis_jain, spamhunter_tang2022clues}. Defenders countered by integrating machine learning, moving from shallow statistical models to complex deep learning architectures to identify malicious patterns at scale \citep{anh2024federated, spam_filtering_cnn_lstm_hossain, expl_BERT_uddin2024explainabledetector, mishra2019content, goel2018smishing}.

In recent years, the landscape has shifted toward extreme contextualization and technical stealth. From 2020 to 2024, attackers capitalized on global crises (e.g. Covid-19 \citep{covid-themed-scams-asiaCCS2023}) and the rise of the ``delivery economy'' \citep{timko2024smishing} to craft hyper-relevant messages that bypassed traditional skepticism. During this same window, defenders adopted threat intelligence frameworks, utilizing graph-based analysis to characterize entire smishing campaigns rather than treating messages as isolated incidents \citep{pritom_25_codaspy_smishviz}. Now entering a frontier (2023–2025) which is dominated by Large Language Models (LLMs). While attackers use generative AI to automate flawless, expert-level social engineering \citep{shibli2024abusegpt,gupta_threatgpt}, the defensive community is responding with on-device LLMs that provide humanoid, explainable insights, thus transforming mobile security from a passive filter into an interactive user-centered defense and awareness tool \citep{wangcan, Sanjari2025_spPoster}.

\section{Evaluating Existing Smishing Datasets}
\label{sec:datasets}
\subsection{SMS Phishing Datasets For Research}

We have closely observed a total of 14 SMS datasets and 1 email spam dataset, which are publicly available and used in existing literature as listed in Table \ref{table:sms_dataset_details}.  
Among them, 11 datasets contain data with corresponding labels such as {\em Enron Email Spam} \citep{enron_vangelis2006spam}, {\em British English SMS} \citep{british_sms_dataset2011}, {\em UCI Machine Learning Repository SMS Collection} \citep{UCI_sms_dataset}, {\em Mendeley Smishing}, \citep{salman2024investigating} Super Dataset, \citep{mishra2022sms_mendeley}, {\em Korean Smishing}, {\em RevisedIndian} \citep{githubGitHubShshnk158MultilingualSMSspamdetectionusingRNN}, {\em Bengali SMS Spam} \citep{kaggleBengaliSpam}, {Bangla Barta} \citep{Shahriyar2025Bangalabarta} and MOZ-Smishing \citep{ali2025moz} dataset. The other four datasets--  {\em National University of Singapore (NUS) SMS Corpus} \citep{nus_chen2013creating}, {\em SMS Gateways dataset} \citep{on_phishing_tactics_nahapetyan2023sms}, Spam Hunter \citep{twitter-clues-ccs2022-spam}, {\em Smishtank} \citep{timko2024smishing} have no label information. However, some of these unlabeled datasets, by definition, only contain smish or spam messages. It has also been observed that some researchers have used these two terms very loosely when labeling their datasets. Table \ref{table:sms_dataset_details} provides details on the dataset size, message content length, and the label information for each of the datasets. We provide further details on corresponding datasets in Table \ref{table:sms_dataset_website_details}, where it shows the number of messages that contain URLs, the number of unique fully qualified domain names (FQDNs), the count of unique live websites during our analysis time frame, and the count of unique parked websites during our analysis time frame. The FQDN count corresponds to the original URL found in the message. Thus, if any URL shortener is used, the expanded link is not considered here. We have extracted the URLs from corresponding messages using a custom-built extractor with a rule-based pattern-matching process. 
In some cases, messages contain multiple URLs (mostly in Enron, Super, and Spam Hunter datasets), and we only considered the first URL while extracting the live website information. For future research, we have provided both the refined and the original versions of all available datasets in one unified Smishing Data Hub at \url{https://github.com/MarazMia/SMISH_DT}. Further details for each of the datasets are discussed below:  

\subsubsection{Public Datasets:} 

\noindent \textbf{(i) Enron Email Spam \citep{enron_vangelis2006spam}:} Although this dataset was curated from spam emails, it had been applied to detect SMS phishing as the textual contents share characteristics similar to smishing messages. Being published in 2006, it has 33,715 data instances labeled as either {\tt ham} or {\tt spam}, almost balanced. The average message length of 1446.84 characters is comparatively higher than the other smishing datasets. From Table \ref{table:sms_dataset_website_details}, around 33.08\% of the entire dataset instances have a URL in the message, from which the majority (80.19\%) are within the spam messages. In addition, there are 3,433 unique fully qualified domain names (FQDNs) and 3,456 unique links from which 564 are live and 193 websites are parked. Surprisingly, there are more website links from spam messages than from ham messages.

\noindent \textbf{(ii) English SMS \citep{british_sms_dataset2011}:} This dataset was a collection of two corpora and manually selected ham and spam messages with 875 instances. The average message length is 106.89 for all data instances and 78.89 characters for the ham messages, which is smaller than the average size of the spam ones (136.4). Due to its smaller size, this dataset also contains the lowest number of messages with URLs(only 64 instances), all of which are found within the spam messages. There are no live websites from the ham messages, while only 10 websites are live from 41 unique FQDNs, and 7 of them are parked.

\noindent \textbf{(iii) UCI ML Repo \citep{british_sms_dataset2011}:} The dataset was the result of a combination of three datasets which were the {\em English SMS dataset}(full), the {\em National University of Singapore SMS dataset} \citep{nus_chen2013creating} (manually selected 3,375 SMS) and the currently obsolete {\em SMS Spam Corpus v.0.1 Big dataset}
It is also labeled with binary categories of {\tt ham} and {\tt spam} with $5,574$ total instances ($4,827$ ham and $747$ spam messages). Again, it follows a closer average length of ham messages ($71$ characters) and spam messages ($139$ characters) compared with the English SMS dataset. Moreover, only 1.94\% of all the messages contain URLs, of which 98.14\% are spam. A total of 54 unique domains are found, and only 13 of them are live, while 10 belong to spam messages. Surprisingly, $10$ of the live websites are parked, and here also the majority is from spam messages ($9$ out of the $10$ websites).

\noindent \textbf{(iv) ExAIS SMS Spam \citep{onashoga2015adaptive}:} From the research study conducted at the Federal University of Agriculture, Abeokuta, Nigeria, the dataset consists of SMS messages from 20 voluntary participants with  2,350 {\tt spam} messages and 2,890 {\tt ham} messages. From Table \ref{table:sms_dataset_details}, we can see that the average message length in both classes is quite similar, with 132.42 characters for {\tt ham} class and 142.37 for {\tt spam} class. A total of 193 message instances contain URLs, which is about 3.99\% of the whole dataset. Among these messages, the majority (138 instances) lie within the spam class. In the website analysis part from Table \ref{table:sms_dataset_website_details}, we observe 115 unique links, of which 79 are the live ones, and within that, 59 belong to the spam category. Also, 35 websites are in the parked state, and again, 22 of them are from spam messages.

\noindent \textbf{(v) RevisedIndian \citep{githubGitHubShshnk158MultilingualSMSspamdetectionusingRNN}:} This dataset contains real-world messages in three languages: English, Hindi, and Telugu, totaling 4,567 entries. However, the multilingual nature of the dataset is limited, as only 400 messages are in Hindi or Telugu texts, leaving approximately $91.24\%$ of the message instances in the English language. The dataset is labeled into ham and spam categories, with the majority ($73.48\%$, or 3,356 entries) belonging to the ham class. Notably, the average message length across all instances is approximately 140 characters. A deeper analysis revealed 1,441 messages containing URLs, of which 867 ($60.17\%$) are classified as spam. From the 720 unique websites identified, 450 are found to be live. Of these live websites, 249 belong to the spam class. Furthermore, 180 of the live websites are categorized as parked, with almost half ($96$) of these parked websites being associated with spam messages.

\noindent \textbf{(vi) Mendeley Smishing \citep{mishra2022sms_mendeley}:} This dataset is a partial clone of the UCI dataset with 5,971 total instances of 4,844 being {\em ham}, 489 {\em spam}, and 638 {\em smish} messages. The Mendeley dataset has 4,753 common data instances with the UCI dataset, with 4,199 ham and 554 spam messages. The smish messages were collected from the Pinterest images and later converted into textual form. Being published in 2022, this dataset specifically used the label {\tt smish} for the first time in the literature. As most of the data instances are from the UCI dataset, the average length of the messages follows a similar weight as the UCI dataset, and also the average length of the smish messages is 139.59, which is still closer to the spam messages (133.90). Also, it has more number of messages with URLs (233), from which all of the messages belong to spam and smish categories, with a higher number of unique (125) and live (39) websites, from which no messages of the ham class have any URLs.

\noindent \textbf{(vii) Super Dataset \citep{salman2024investigating}:} Salman \textit{et al.} amassed a substantial dataset of over $60,000$ messages by augmenting from existing repositories (UCI, NUS, Spamhunter, and 3 other multilingual datasets) and incorporating another $4,904$ new messages from diverse origins such as Twitter, Scamwatch. This dataset stands out as one of the largest and most varied of its kind, encompassing SMS from numerous sources. Notably, they contributed $2,920$ labeled spam messages, marking a considerable expansion in the quantity of spam data compared to prior investigations. This comprehensive dataset comprises a total of $67,008$ retrievable SMS messages, of which $40,837$ ($60.9\%$) are categorized as legitimate and $26,181$ ($39.1\%$) are labeled as spam. The average message length over the entire dataset is about $96$ characters, and again, this measurement is higher for the spam counterpart, which is about $144$ characters. We have identified a total of $10,492$ ($15.66\%$) data instances with URL inside the messages from which $10,472$ ($99.81\%$) belong to the spam class. From the $7,759$ unique URLs, we found $1,645$ live websites, and most instances ($1,641$) are from the {\tt{spam}} class. Quite surprisingly, $903$ of the websites are found to be parked and almost all of them ($902$, which is $54.83\%$ of the total live websites) reside within the {\tt{spam}} messages.

\noindent \textbf{(viii) Kor-Smishing \citep{kor_smsphishing_info15050265}:} This is the only public dataset that has Korean messages labeled as $1$ (smish) or $0$ (ham/benign). This is also a highly imbalanced dataset with 42,594 ham and only 615 smish data instances. As the messages are in the Korean language, the average message length of the ham instances is lower (41.20), while this measurement for the smish messages is 296.1. Although it is a comparatively larger collection, only 0.31\% of all messages contain URLs, which is the second lowest count among all the other datasets. Also, only 75 unique websites are found along with 120 unique URLs, from which 66 (63 from the smish messages) are live and 23 are parked websites. This dataset contains only 6 entries with only non-Korean texts, mostly in English. 

\noindent \textbf{(ix) Bengali SMS Spam \citep{kaggleBengaliSpam}:} This dataset, sourced from the Kaggle data repository, primarily consists of 2,602 message instances in Bengali language texts, with only occasional ($0.58\%$, or 15) occurrences of fully English messages. The messages are categorized into ham ($46.08\%$) and spam ($53.92\%$), indicating a relatively balanced distribution between the two classes. The average message length for the entire dataset is 91 characters, with a slight variation between classes: ham messages average $100.39$ characters, while spam messages average $83.97$ characters. Only $6.88\%$ messages (179 messages) contain URLs, of which 89 ($49.72\%$) messages are labeled as spam. The URLs found in ham messages are predominantly government websites or promotional links for a specific carrier provider. Analysis of the website URLs revealed 58 unique websites, with 55 of them found to be live; 38 of these live websites belong to the spam class. Furthermore, 23 websites were identified as parked, with the vast majority (20) being associated with spam messages.

\noindent \textbf{(x) Bangla Barta \citep{Shahriyar2025Bangalabarta}:} This Bengali SMS dataset features three distinct categories: \texttt{normal}, \texttt{promo}, and \texttt{smish}. The normal messages are benign and represent daily life conversation, while the promo category messages denote legitimate promotional news from a specific carrier provider. For simplicity, we have combined the normal and promo categories into a single ham class. The dataset is balanced, totaling 2,772 instances with 924 messages originally in each of the three categories. Only 14 messages ($0.51\%$) are found to be in completely English texts, often mixed with Bengali phrases written in English letters, a practice commonly referred to as \textit{Banglish} (i.e., mixture of English with Bengali). The average message length for the smish class is notably high ($114.75$) compared to the combined ham messages ($55.59$). A total of 410 messages contain URLs, with 272 ($66.34\%$) of these belonging to the smish class. Consistent with other datasets, most of the URLs found in ham messages are primarily from government or specific SIM card providers' promotional websites. We have identified 151 unique websites, 128 of which are live, and 119 of these live sites belong to the smish category. Furthermore, 26 websites are found to be parked, with only 2 of them belonging to the ham class.

\noindent \textbf{(xi) MOZ-Smishing \citep{ali2025moz}:} This dataset contains SMS messages in the Portuguese language, especially focused on the Mozambican context. 
The dataset comprises 2,561 entries. 
Among these messages, 1,816 instances (approximately $71\%$) have used only English language alphabets. 
The dataset features two labels: Legitimate (considered as `ham') and Smishing. The average message length across the corpus is $102.68$ characters, with smishing messages being, on average, slightly shorter than legitimate ones. Surprisingly, only 31 messages ($1.21\%$) contain a URL, yet no smishing message is found to have URLs. 
Furthermore, 768 of the messages ($29.99\%$) contain phone numbers, and 324 of these ($42.18\%$ of the phone number-containing messages) are smishing. This observation indicates that in the specific area where the dataset was curated, contact numbers were considered a more viable medium for attacks than URL propagation. Finally, of the 9 unique links identified, 8 are found to be live, with 4 of these being parked websites.

\noindent \textbf{(xii) NUS Corpus \citep{nus_chen2013creating}:} Although the original dataset \citep{nus_chen2013creating} was published in 2013, the corpus was initially submitted to arXiv in 2011, and the most recent version was published in 2015. The data repository has two modes of languages-- English and Chinese. Thus, two files are provided in XML, SQL, and JSON formats. Moreover, both datasets are unlabeled. The Chinese one has $31,465$ instances and contains only descriptive messages with very few cases of messages with URL links, while the English one has a total of $55,835$ data instances with an average message length of around 52 characters. Only $0.04\%$ of the messages have URLs, which is the lowest among the datasets, and of these URLs, there are only $18$ unique website domains, of which $10$ are live and $3$ are parked.

\noindent \textbf{(xiii) Spam Hunter \citep{twitter-clues-ccs2022-spam}:} 
This dataset was published as a spam message dataset in 2022, which is the result of querying the Twitter API from early January 2018 to late December 2021. The authors collected tweets consisting of {\em spam, phishing, malicious, smish, fraud} and {\em scam} SMSes that have at least one screenshot of the original message. Further, they re-verified the images using an object detector and the spam messages by a simple neural network-based sentiment detector with an overall accuracy of 91\%. These purification steps ended up providing 21,918 SMS spam messages; however, their latest data repository included 25,826 instances in the English corpus in text file format, following the average message length of 147.36 characters (from Table \ref{table:sms_dataset_details}). Also, a large portion of the data (60.14\%) instances include URLs inside the messages from which 8,615 are unique domains with 7,269 live and 4,052 parked websites (see Table \ref{table:sms_dataset_website_details}).

\noindent \textbf{(xiv) SMS Gateways Smishing \citep{on_phishing_tactics_nahapetyan2023sms}:} Being recently released in 2023, this is so far the largest collection of SMS, with 68,029 messages by crawling 11 SMS gateways accessed through a web browser. However, all messages in this dataset have been confirmed as smishing by checking URLs with the VirusTotal, APWG, and Google Safe Browsing results. The original paper \citep{on_phishing_tactics_nahapetyan2023sms} provides two separate files, one with all the phishing messages (68,029 instances) and the other one with the phishing campaigns (35,128 instances) along with the URLs, time stamp and how many times the messages were seen by the gateway crawler (i.e. the campaign where the linguistic contents of the message in different cases remains the same but other identifiers such as sender, URLs, or the one time code got changed). The information for the phishing messages file is given in Table \ref{table:sms_dataset_details}
and \ref{table:sms_dataset_website_details}. The average message length is about 112.45 characters, and almost 63.64\% of messages contain URLs. Surprisingly, only 857 unique FQDNs are found out of 43,297 unique links (100\% of the found URLs), which is very small in terms of the observed URLS. However, even after being confirmed as phishing links, 6,945 websites are still live, and 572 of them are parked. 


\noindent \textbf{(xv) SmishTank \citep{timko2024smishing}:}
As the name of the data set suggests, SmishTank is an active system and the one of the most recent smishing dataset published in 2024, and that's still collecting user-submitted messages in the form of images and text through their website. The Smishtank system currently verifies user input from VirusTotal and extracts other information, such as URLs, sender, and brand name usage from the text or image of the given message automatically. The current downloadable snapshot of the dataset has 1,091 rows provided in {\em CSV} format, but no confirmed label is associated with the message, even though the overall structure and the approach of the system suggest that the messages are highly likely to be smishing messages. The average length of the messages is about 177 characters. Almost 86\% of the messages contain URLs inside, and 740 unique FQDNs are present. Of 740 unique FQDNs from 860 unique links, 193 are still live, and 82 of them are in the parked state.

\subsubsection{Private Datasets:}
As provided in Table \ref{tab:dataset_result_sum}, several studies have used private datasets on smishing in two ways: (a) full private data \citep{mldetect_smishing_frauds_boukari, njuguna2021model, seo2024device} or (b) a mix of both private and public datasets \citep{ghourabi2020hybrid, smdetector_ghourabi, jain2022content, karhani_phishing_2023}. The KISA dataset \citep{seo2024device} is managed by the corporation KISA, which primarily collects smish messages in the Korean language, but only the messages with URLs are considered for smishing detection. Although researchers in \citep{seo2024device} used multiple publicly available normal messages for more generalizability of the normal message entries from three different open-access datasets, their final augmented dataset is not public. Their work achieved a significant accuracy of $99.60\%$ and an F-1 score of $99.44\%$ using the character-level CNN model \citep{seo2024device}. On the other hand, Gourabi et al. \citep{ghourabi2020hybrid, smdetector_ghourabi} worked on both English and Arabic messages with a hybrid mode integrated into the mobile environment. The two highlighted studies appear to have been conducted with similar datasets, achieving $99.63\%$ accuracy and an F-1 score of $95.58\%$ using the BERT model.

\begin{table*}[!t]
    \centering
    \caption{Summary Information of Publicly Available SMS/Smishing Datasets (labels are defined as: \textbf{ham} - legitimate message, \textbf{spam} - unsolicited and unwanted message, \textbf{smish} - phishing message containing malicious URL)}
    \vspace{-0.2em}
    \rowcolors{4}{white}{gray!10} 
    \resizebox{\linewidth}{!}{
    \begin{tabular}{|llll|*{4}{c}|*{4}{c}|}
        \hline
        \multirow{2}{*}{Dataset (year)} &  \multirow{2}{*}{Language} & \multirow{2}{*}{Format} & \multirow{2}{*}{Labeled} & \multicolumn{4}{c|}{\# Instances} & \multicolumn{4}{c|}{Avg. Message Length}\\
        \cline{5-12}
         & & & & full & ham & spam & smish & full & ham & spam & smish\\
        \hline

        Enron Email spam \citep{enron_vangelis2006spam} (2006) & English & .txt & yes & 33,715 & 16,545 & 17,170 & - & 1446.84 & 1651.67 & 1249.47 & - \\
        English SMS  \citep{british_sms_dataset2011} (2011) & English & .doc & yes & 875 & 450 & 425 & -  & 106.83 & 78.89 & 136.4 & - \\
        UCI ML Repo \citep{UCI_sms_dataset}  (2011) & English & .txt & yes & 5,574 & 4,827 & 747 & -  & 80.48 & 71.47 & 138.68 & - \\
        ExAIS SMS Spam\citep{onashoga2015adaptive}  (2015) & English & .csv & yes & 5,240 & 2,890 & 2,350 & -  & 136.88 & 132.42 & 142.374 & - \\
        RevisedIndian\citep{githubGitHubShshnk158MultilingualSMSspamdetectionusingRNN}  (2018) & Multiple & .xls & yes & 4,567 & 3,356 & 1,211 & -  & 140.64 & 140.86 & 140.04 & - \\
        Mendeley Smishing \citep{mishra2022sms_mendeley} (2022) & English & .csv & yes & 5,971 & 4,844 & 489 & 638 & 83.24 & 70.70 & 133.90 & 139.59\\
        Super Dataset \citep{salman2024investigating} (2024) & English & .csv & yes & 67,008 & 40,830 & 26,178 & - & 96.09 & 65.19 & 144.30 & - \\
        Kor-Smishing \citep{kor_smsphishing_info15050265} (2024) & Korean  &  .csv & yes & 43,209 & 42,594 & - & 615  & 44.83 & 41.20 & - & 296.1 \\
        Bengali SMS Spam\citep{kaggleBengaliSpam}  (2024) & Bengali & .csv & yes & 2,602 & 1,199 & 1,403 & -  & 91.00 & 100.39 & 82.97 & - \\
        Bangla Barta\citep{Shahriyar2025Bangalabarta}  (2025) & Bengali & .csv & yes & 2,772 & 1,848 & - & 924  & 75.31 & 55.59 & - & 114.75 \\
        MOZ-Smishing\citep{ali2025moz} (2025) & Portuguese & .csv & yes & 2,561 & 2,009 & - & 552 & 102.68 & 104.57 & - & 95.79 \\

        \hline
        NUS Corpus \citep{nus_chen2013creating} (2011) & English & .json & no & 55,835 & - & -  & - & 52.01 & - & - & -  \\
        Spam Hunter \citep{twitter-clues-ccs2022-spam} (2022) & English & .txt & no & 25,826 & - & 25,826  & - & 147.36 & - & 147.36 & -   \\
        SMS Gateways Smishing \citep{on_phishing_tactics_nahapetyan2023sms} (2023) & English & .csv & no & 68,029 & - & - & 68,029 & 112.45 & - & - & 112.45 \\
        SmishTank \citep{timko2024smishing} (2024) & English & .csv & no & 1,091 & - & - & 1,091 & 176.96 & - & - & 176.96 \\
        \hline
    \end{tabular}
    \label{table:sms_dataset_details}
  }
\end{table*}

\begin{table*}[!t]
    \centering
    \caption{Analysis of Live and Parked Websites Among Datasets (Between Mid-January 2025 and Early-October 2025)}
    \vspace{-0.2em}
    \rowcolors{4}{white}{gray!10}
    \resizebox{\linewidth}{!}{
    \begin{tabular}{|p{3.3cm}p{2.40cm}p{2.5cm}p{2.5cm}p{1.25cm}p{1.25cm}|p{1.0cm}p{1.0cm}p{1.0cm}p{1.0cm}|p{1.0cm}p{1.0cm}p{1.0cm}p{1.0cm}|}
        \hline
        {Dataset}
        & {\#Messages}
        & {\#Spam Messages}
        & {\#Smish Messages}
        & {\#Unique}
        & {\#Unique}
        & \multicolumn{4}{c|}{\#Live Websites}
        & \multicolumn{4}{c|}{\#Parked Websites} \\
        \cline{7-14}
        & {with URL} & {with URL} & {with URL} & {Link} & {FQDN} & full & ham & spam & smish & full & ham & spam & smish \\
        \hline
        Enron Email spam & 11,152 (33.08\%) & 8,943 (54.05\%) & - & 3,456 & 3,433 & 564 & 198 & 366 & - & 193 & 30 & 163 & -\\

        English SMS  & 64 (7.31\%) & 64 (15.01\%) & - & 45 & 41 & 10 & 0 & 10 & - & 7 & 0 & 7 & -\\
        UCI ML Repo  & 108 (1.94\%) & 106 (14.19\%) & - & 59 & 54 & 13 & 1 & 12 & - & 10 & 1 & 9 & -\\
        ExAIS SMS Spam  & 209 (3.99\%) & 138 (5.87\%) & - & 115 & 66 & 79 & 20 & 59 & - & 35 & 13 & 22 & -\\
        RevisedIndian  & 1,441 (31.55\%) & 565 (46.66\%) & - & 720 & 113 & 450 & 201 & 249 & - & 180 & 84 & 96 & -\\
        Mendeley Smishing  & 223 (3.73\%) & 93 (19.01\%) & 130 (15.20\%) & 125 & 109 & 39 & 0 & 20 & 19 & 23 & 0 & 14 & 9\\
        Super Dataset  & 10,492 (15.66\%) & 10,472 (40.00\%) & - & 7,759 & 5,610 & 1,645 & 4 & 1,641 & - & 903 & 1 & 902 & -\\

        Kor-Smishing & 134 (0.31\%) & - & 127 (20.65\%) & 120 & 75 & 66 & 3 & - & 63 & 23 & 2 & - & 21\\
        Bengali SMS Spam  & 179 (6.88\%) & 89 (6.34\%) & - & 88 & 58 & 55 & 17 & 38 & - & 23 & 3 & 20 & -\\
        Bangla Barta  & 410 (14.79\%) & - & 272 (29.44\%) & 151 & 42 & 128 & 9 & - & 119 & 26 & 2 & - & 24\\
        MOZ-Smishing  & 31 (1.21\%) & - & 0 (0.00\%) & 9 & 5 & 8 & 8 & - & - & 4 & 4 & - & -\\
        \hline
        NUS Corpus & 21 (0.04\%) & - & - & 18 & 18 & 10 & - & - & - & 3 & - & - & -\\
        Spam Hunter  & 15,533 (60.14\%) & - & - & 15,531 & 8,615 & 7,269 & - & - & 7,269 & 4,052 & - & - & 4,052\\
        SMS Gateways Smishing  & 43,297 (63.64\%) & - & - & 43,297 & 857 & 6,945 & - & - & 6,945 & 572 & - & - & 572\\
        SmishTank  & 938 (85.98\%) & - & - & 860 & 740 & 193 & - & - & 193 & 82 & - & - & 82\\
        \hline
    \end{tabular}
    \label{table:sms_dataset_website_details}
  }
\end{table*}

\begin{table*}[!h]
\centering
\caption{Evaluation of Existing Detection Models Using Both Public and Private Datasets}
\label{tab:dataset_result_sum}
\rowcolors{2}{white}{gray!10}
\resizebox{\linewidth}{!}{
\begin{tabular}{|c|c|c|cc|cc|}
\hline
        Data Availability  &     Reference &                                      Datasets Used &                     Feature Space  &  Best Algorithm Model & Accuracy &  F1 Score \\
\midrule

     &  \citep{mldetect_smishing_frauds_boukari}& Private &  TF-IDF  &  RF  &      0.9815 &       0.9257 \\

      &    \citep{njuguna2021model} & Private & Word Vectorizer & NB & 0.9375 & 0.9333 \\

      &  \citep{mambina2022classifying} & Private (Swahili) & TF-IDF & RF & 0.9986 & 0.9986 \\

     Private  &     \citep{seo2024device} &  Private (KISA) &  Word2Vec   &  Char-CNN & 0.9960 & 0.9944 \\

      &     \citep{expl_lee2024korsmishing} &  Private (Korean) & Tokenized Text  &  KUBERT & 0.9900 & 0.9900 \\

      &  \citep{anh2024federated} & Private (English + Vietnamese) & Tokenized Text &   PhoBERT (FL)  &    0.9938  &   - \\

      &   \citep{shinde2024sms} & Private & Tokenized Text &   RNN  &    0.9413  &   0.9400 \\

  \midrule

        &    \citep{ghourabi2020hybrid} & UCI ML Repo + Private (Arabic) &  Word Embedding  & 
 CNN-LSTM & 0.9837 & 0.9148 \\

   &  \citep{smdetector_ghourabi}&                                            UCI ML 
  Repo + Private (Arabic) &                                   Tokenized Text      &    BERT-FC &   0.9963 &    0.9558 \\

   Public + Private    &   \citep{jain2022content}& UCI ML 
  Repo + Private (URL Phishing) &        TF‐IDF                           &  Voting (RF, KNN, ETC) &   0.9903 &       - \\

     &   \citep{karhani_phishing_2023}& NUS Corpus + Private (TELUS) & Heuristic + TF‐IDF &   RF + SVC &    0.9940  &    0.9901 \\

     &   \citep{llm_shim2024persuasion}& Mendeley + GPT-3.5-Turbo Augment & Tokenized Text &   RoBERTa  &    -  &    0.9820 \\

 \midrule

      &   \citep{sonowal2020detecting} & UCI ML Repo &  BoW + Heuristic  &  AdaBoost & 0.9867 & 0.9486 \\

      &   \citep{smishing_detector_security_model_mishra_2020}& Mendeley Smishing &       TF-IDF                            &    NB  &   0.9629 &       0.9200 \\

      &   \citep{implement_smishing_detector_mishra_2022}& Mendeley Smishing &  Word2Vec (ANN filter)  &   ANN &    0.9740 &    0.8672 \\

     &    \citep{dsmish_mishra_2023}&                                 Mendeley Smishing &    Heuristic   &     MLP Back Propagation &   0.9793 &       0.8150 \\

    &  \citep{remmide2024privacy_federated} & Mendeley Smishing &    GloVe   &     BiLSTM (FL) &   0.8878 &       0.8700 \\

   Public   &    \citep{rose2024next_federated} & UCI ML Repo &  Word Embedding  &   ConvLSTM (FL) & 0.9919 & - \\

   &    \citep{ulfath2022detecting} & UCI ML Repo &  TF-IDF  &   SVM & 0.9839 & 0.9637 \\

       &    \citep{spam_filtering_cnn_lstm_hossain}&        UCI ML 
  Repo  &     Word Embedding                                 &     CNN-LSTM &    0.9840 &    0.9800 \\

  &  \citep{sidhpura2023_fedspam:}& Enron + UCI ML Repo + English SMS &                             Tokenized Text            & 
  DistilBERT (FL) &     0.9800 &       - \\

    &  \citep{smsprotect_akande} &         UCI ML Repo  &                       Heuristic    &   PART (C4.5) &   0.9842 &       0.9840 \\

    &  \citep{uddin2024explainabledetector} &         UCI ML Repo  &   Tokenized Text                        &   RoBERTa  &   0.9984 &       0.9984 \\

    &  \citep{salman2024investigating} &         Super Dataset  &   Word2Vec                        &   SVM  &   0.9780 &       0.9900 \\
 
\hline
\end{tabular}}
\end{table*}

\subsection{Performance of Various Detection Models Using Datasets}
In the literature of smishing detection, there has been the use of private only, public only, and a mixture of both private and public datasets, as summarized in Table \ref{tab:dataset_result_sum}. The feature space column indicates the used feature vector in each study. Along with popular text vectorization, several studies used custom-extracted features from SMS text, URL lexical, and even from the associated websites' contents based on the domain-specific heuristics defined by the researchers \citep{karhani_phishing_2023, sonowal2020detecting, dsmish_mishra_2023, smsprotect_akande}. With the Korean language-based smishing detection on the KISA private dataset \citep{seo2024device}, researchers were able to achieve a significant accuracy of $0.996$ and an F-1 score of $0.9944$ using the character-level CNN model. Along with the public UCI dataset, Gourabi \emph{et al.} used a privately curated SMS dataset on the Arabic language, thus making their system enable detecting smishing in both language modes. In their first paper \citep{ghourabi2020hybrid}, they used a hybrid CNN-LSTM model and got an accuracy of 0.9837, but with a low F-1 score of 0.9148. However, in their second attempt on the same dataset, they incorporated two other detection methods of checking malicious URLs from VirusTotal and a regular expression-based malicious phone number, email, and other patterns from the text message. With a third BERT-FC-enabled smishing detection technique, they achieved the highest accuracy of 0.9963 with an improved F-1 score of 0.9558. Later, the work of Karhani \emph{et al.} \citep{karhani_phishing_2023} on the NUS corpus and private data from a telecommunication company achieved a significant accuracy of 0.994 and an F-1 score of 0.9901 by using a hybrid model of Random Forest and Support Vector Classifier. The original ExAIS dataset paper \citep{onashoga2015adaptive} provides a detection module with a server-based artificial immune system and a rule-based immune mechanism, and a tokenization process. With this experimental setup, the authors showed that they achieved an overall accuracy of 99\% on this dataset. Among the public datasets, the most frequent one is the UCI ML Repo, which was published in 2011. So far, the best model on this dataset is the RoBERTa as proposed in the study of Uddin et al. \citep{uddin2024explainabledetector}  with a phenomenal accuracy and F-1 score of 0.9984. The overall observation reveals that the researchers have utilized diverse sets of ML and DL models to make reliable smishing detection systems using various datasets. The DL models show better performance, as evidenced by the findings from Table \ref{tab:dataset_result_sum}. Even in the federated setup, DL based NLP models like BERT and its other variants achieve noteworthy detection accuracy in almost every dataset that was used, except the work of Remmide et al. \citep{remmide2024privacy_federated}. In fact, their system achieved the lowest accuracy of $0.8878$ among all the mentioned studies. 
In the original dataset paper \citep{salman2024investigating}, the authors achieved the highest accuracy (97.8\%) and  F-1 score (99\%) on the dataset by using a two-class SVM model. They also applied other deep learning models and obtained the best performance on the RoBERTa model (an accuracy of 97.4\% and F-1 score of 98\%) using the RoBERTa-base feature embedding.

\begin{figure}
    \centering
    \includegraphics[width=0.91\linewidth]{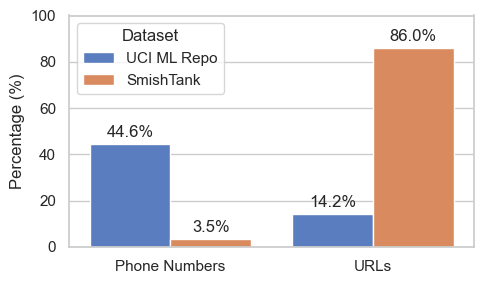}
    \caption{Percentage of phone numbers and URLs inside spam/smish messages in UCI ML Repo (2011) and SmishTank (2024) datasets}
    \label{fig:old_vs_new_dataset_phone_url}
\end{figure}

\subsection{Data Drift Analysis Within Existing English Datasets Based on Time and Geolocation of Data}

Out of the 14 candidate datasets evaluated, the UCI SMS Spam Collection (2011) and the SmishTank (2024) corpus were selected for this data drift analysis. This specific pair was chosen to maximize the temporal and geographical variance in the study. By comparing a legacy UK-based dataset (UCI SMS Spam) with a modern US-based repository (SmishTank), we provide a rigorous test for data drift, ensuring that the identified lexical shifts represent a fundamental evolution in global smishing tactics rather than minor seasonal fluctuations. To measure the drift without relying on model performance, which is not in the scope of our study, we applied the Chi-square ($\chi^2$) test of independence \citep{mchugh2013chi} on the top 500 features. This allows us to isolate words that are statistically unique to each era. The filtering process was implemented, where terms were only retained if their p-value were below 0.05, ensuring a 95\% confidence level that the observed frequency differences between these two datasets. This step isolates the words that provide the strongest evidence of distributional shift between the 2011 and 2024 datasets. As found in our results, the legacy 2011 UCI dataset is characterized by top 20 keywords \{`landline', `draw', `win', `won', `free', `awarded', `box', `tone', `ur', `txt', `urgent', `claim', `mobile', `send', `text', `cash', `prize', `po', `guaranteed', `nokia'\}, reflecting the premium-rate (i.e., high cost) SMS era in the UK. 

Conversely, the 2024 US dataset shows a significant shift toward credential harvesting, dominated by top 20 terms \{`safari', `account', `confirm', `browser', `delivered', `package', `link', `tax', `arrived', `information', `warehouse', `refund', `address', `incomplete', `open', `provide', `usps', `click', `irs', `copy'\}. Geographic and temporal context significantly influences the lexical features of each corpus; the 2024 US dataset reflects a move toward administrative and logistics-based phishing (USPS, IRS), while the 2011 UK UCI dataset relies on traditional social engineering tactics centered on immediate financial gain (cash, draw). 

To ensure computational reproducibility, we define our statistical configuration as follows. Preprocessing involved a deep cleaning pipeline where URLs were stripped using regex patterns, words containing numeric digits were removed, and all non-alphanumeric punctuation was discarded before tokenization. For the $\chi^2$ drift analysis, we utilized the \texttt{CountVectorizer} with a binary frequency configuration to focus on term presence versus absence across messages. The test was conducted using a $2 \times 2$ contingency table for each feature (present/absent vs. 2011/2024 corpora) with Yates' continuity correction applied. 

\subsection{Structural and Linguistic Evolution}
Our structural analysis further reveals a profound shift in message composition between the 2011 and 2024 corpora. The average punctuation density increased from 4.10\% in the legacy UCI dataset to 6.76\% in the modern SmishTank collection ($p < 0.001$). This significant rise is primarily attributed to the transition from text-based social engineering to link-centric phishing, where URL delimiters (e.g., '/', '.', '?') contribute to a higher punctuation-to-character ratio. Conversely, a slight but statistically significant decrease in Shannon Entropy \citep{shanon_vajapeyam2014understanding} was observed ($\mu_{2011\_corpus} = 4.69$ vs. $\mu_{2024\_corpus} = 4.64$; $p < 0.001$). This downward drift suggests a move away from the high-entropy, `noisy' linguistic patterns of legacy SMS spam which is often characterized by non-standard abbreviations and irregular capitalization, toward a more professionalized and low-entropy style that mimics legitimate institutional correspondence. These structural comparisons were verified using an independent two-sample t-test, assuming unequal variances where applicable, with a significance threshold set at $\alpha = 0.05$. This finding is also backed by our manual analysis as shown in figure \ref{fig:old_vs_new_dataset_phone_url} with a radical inversion in call-to-action modalities where legacy 2011 messages relied heavily on telephony-based triggers (44.6\% phone density), and 2024 smishing has almost entirely transitioned to a URL-centric architecture (85.98\%). This shift signifies a technical evolution from premium-rate text fraud to digital credential harvesting via malicious web links. We have also generated word clouds in English texts for each datasets which are provided in the Github portal.

\subsection{Critical Assessment of Dataset Limitations}
Beyond the temporal drift analyzed above, our review identifies distinct structural limitations in the current smishing dataset landscape:
\subsubsection{URL Decay:}
A major challenge for researchers is URL Decay. Phishing domains are short-lived, often surviving less than 24 hours before being blocked \citep{oest2020phishtime}. Furthermore, sinkholing, where security vendors redirect malicious traffic to a safe server, can pollute datasets, as researchers may inadvertently analyze a sinkhole page, thinking it is the attack payload. In addition to a population study involving 50 randomly selected websites in total, restricted to a maximum of five per dataset, we conducted a secondary validation of the currently active domains to evaluate their status. Our findings from manual website queries, the Wayback Machine, and WHOIS look-up reveal that nearly all selected websites are now benign, despite their original inclusion in smishing or spam messages. In rare instances, WHOIS reports indicated that a domain was registered several years after the initial dataset was compiled; this discrepancy likely stems from different versions of the dataset aggregating messages from various time periods, with our current version containing all such entries. Given the highly ephemeral nature of these websites, we recommend that future researchers avoid relying on these datasets for URL or website analysis in their studies.

\subsubsection{Anglocentric Bias and Multilingual Scarcity}
As detailed in Section \ref{sec:dataset-limits}, there is a severe anglocentric bias in public repositories. While detection efforts exist for languages like Arabic \citep{ghourabi2020hybrid}, Vietnamese \citep{anh2024federated} and English-Chinese \citep{nus_chen2013creating}, high-quality public benchmarks remain predominantly English. 
\subsubsection{Benign-to-Malicious Imbalance}
Public datasets rarely reflect real-world traffic distributions. In the real-world, the number of legitimate ones (ham) is more than smishing. However, repositories like SmishTank contain only smishing messages. This prevents the training of robust binary classifiers that can distinguish attacks from benign notifications (e.g., real delivery alerts vs. fake ones).


\ignore{
\begin{table*}[!h]
\caption{Critical Analysis of Smishing Dataset Literature}
\label{tab:reveiw-4}
\resizebox{\textwidth}{!}{
\begin{tabular}{|l|l|l|l|}
\hline
\textbf{Reference} & \textbf{Key Attributes} & \textbf{Strengths} & \textbf{Weaknesses} \\

\hline

\multicolumn{1}{|p{4em}|}{
\citep{UCI_sms_dataset}} &
\multicolumn{1}{p{17em}|}{{\color{red}add details here.}} & 
\multicolumn{1}{p{17em}|}{{\color{red}add details here.}} & 
\multicolumn{1}{p{17em}|}{{\color{red}add details here.}} \\

\hline

\end{tabular}
}
\end{table*}
}

\section{Discussion on Research Gaps and Future Directions}
{\color{teal}






}
\label{sec:future_directions}
With this comprehensive study of recent smishing research, we aim to systematically understand the current research trends, gaps in existing approaches or methods, and direct future researchers to enhance the trustworthiness and security of the SMS communication system.  
We envision the following major areas for future exploration: (A) smishing attacks, (B) user perception and susceptibility studies, (C) smishing defenses, (D) smishing datasets, and (E) Law enforcement, policy, and regulatory studies. We have listed the specific directions and open problems under these major areas below. 


\subsection{Smishing Attacks}
\subsubsection{Abuse of Generative AI for Creating Smishing Threats}
With the ongoing rapid development of LLMs and large foundation models, LLMs can be abused by attackers through jail-breaking and prompt injections to create innovative smishing campaigns that can themselves be a new attack vector in this area \citep{shibli2024abusegpt, gupta_threatgpt, roy2024chatbots}. The ease of generating these diverse attack templates and propagating them through an anonymous bulk medium has made it more concerning to the defender. We envision conducting more in-depth analysis on various generative models to assess the potential abuse cases for smishing threats. 

\subsubsection{Automated Monitoring and Characterization of Smishing Attacks}
\hypertarget{anchor:auto-monitor}{}
Establishing systems and methods for \textit{real-time threat intelligence} generation and sharing among parties (e.g., service providers, security agencies) can be beneficial. Collaborative efforts can enhance the industry's ability to respond promptly to emerging smishing campaigns and adapt countermeasures accordingly. Brands can also benefit from this kind of monitoring and tracking, where they can easily warn their users to safeguard them with awareness instructions. Research efforts like the SmishViz system/tool \citep{pritom_25_codaspy_smishviz} for monitoring and tracking large smishing campaign-operations to dismantle coordinated attacks will be valuable. We envision that continuous and collective security hardening efforts, like large-scale campaign monitoring, are needed for a more secure SMS messaging ecosystem. Additionally, further usability research is required to understand how analysts can effectively use these kinds of tools for defense.


\subsection{User Perception and Susceptibility Studies}
In literature, even though there has been some steady progress in systematically understanding users' susceptibility and behaviors when encountering smishing messages, further continuous studies in this direction would help develop effective defense mechanisms that are relevant to a diverse population. For instance, as mentioned in the 2022 FBI Elder Fraud report \citep{ic3}, victims over the age of 60 experienced a total loss of \$3.1 billion to fraud in 2022, reflecting an 84\% overall increase in losses from 2021, with an average loss of \$35,101 per victim. This highlights a critical vulnerability in older demographics. Specifically, large-scale user studies are needed to understand vulnerabilities in populations like users of niche/emerging communication platforms (e.g., Discord, Telegram, Threads, Twitch, Mastodon, BeReal), individuals in high-risk professions, and those with lower digital literacy. 
Hence, we recommend that researchers explore these open problems and ensure practical personalized defense mechanisms, including evaluating advanced user alert systems through comprehensive user studies among various population groups. For example, a recent work has successfully designed and evaluated visual trust indicators that help users rapidly assess message credibility \citep{zare2025improving}. Furthermore, we recommend that the community should investigate the psychological factors and sophistication of these attacks \citep{montanez2024Quantifying,XuPIEEE2024,Longtchi_2025_Phsychological_tactics} and develop cyber social engineering kill chains 
\citep{montanez2022cyberkillchain} for various types of smishing threats. This investigation should be directly linked to the development of standardized taxonomies and labeling methods that can categorize smishing messages by their psychological techniques, which enable us to conduct more granular and realistic user susceptibility studies.

\subsection{Smishing Defenses}


\noindent\textbf{[i] Advanced Multi-Modal Techniques for Feature Extraction and Detection:}
In the current literature, most of the SMS phishing detection models extract features based on human intelligence or NLP-based features. These limit the existing research to text only smishing problem. However, with the current developments of large image and video models, in the coming days, smishing messages with multimedia attachments will increase. Thus, to effectively defend against those threat vectors, we need to collect feature data from not only texts but also other multimedia items (e.g., images, audio, video). More research efforts to automatically extract these important multi-model features can be an interesting research avenue in the coming days. 

\noindent\textbf{[ii] Explainability and Reasoning for Detection:} Explainable AI is rapidly replacing traditional AI approaches due to its added benefits of transparency and trustworthiness. Like many other detection engineering problems, we envision a series of ongoing research where explainability would be an important aspect for detecting smishing and explaining why certain messages are smishing rather than benign. Additionally, XAI should be integrated even with the usage of LLMs to ensure users can trust the model outcomes and rely on a decision-making process with certain reasoning. Drawing from these concepts, Wang et al. \citep{wangcan} have demonstrated the value of LLM-based explanations in improving users' understanding and efficacy in detecting phishing SMS. Existing studies show that focusing on user-facing design is essential. For example, visual trust indicators successfully enhanced user confidence and decision-making because they clearly communicated security cues generated by the detection system \citep{zare2025improving}. This approach of providing evidence-based explanations has proven effective and highly usable for end-users. Future research should focus on practical ways to improve these explanations. This includes making them more personalized to individual users' needs and understanding, developing interactive tools that allow users to explore the AI's reasoning, and finding better ways to communicate the AI's confidence in its decisions, especially for ambiguous messages. The goal is to ensure users can consistently rely on these explanations without becoming overly dependent on the AI. 

\noindent\textbf{[iii] Adversarial AI Robustness for Detection Models \& Methods:} Future research should also focus on comprehensive \textit{adversarial robustness} testing for smishing detection models using ML, DL, NLP, and LLM-based systems. Developing standardized evaluation metrics and benchmarks for assessing models' resilience against adversarial attacks will contribute to more secure and reliable defense solutions. These directions of research are more applied here in the context of smishing defense, given that the field of adversarial robustness for ML and AI models is also advancing.  


\noindent\textbf{[iv] Usage of LLMs and SLMs for Detection:} LLMs and other large models can be utilized for defensive actions, such as reasoning-based detection of smishing messages within a user device. However, the main concerns with these larger models are that they need extensive computational capacity, which is generally better for cloud computation rather than in-device (i.e., mobile device) computation. Thus, researchers can explore smaller language models (SLMs) that are compressed to effectively run within the mobile device's storage constraints and evaluate their detection and reasoning performances \citep{Sanjari2025_spPoster}. The in-device computation ensures that the user's private message data remains in the end-device and does not go to the cloud at any time for decision-making. This would be a promising research direction given the faster development of smaller language models that are intelligent and showing reasoning capabilities. 

\subsubsection{Systemic and Architectural Defense Strategies}

\noindent\textbf{[i] Cyber Deception for Defense:} SMS phishing messages mostly contain URLs through which users can interact. Oftentimes, these links can lead to mobile malware rather than just a phishing website. Thus, we envision that future research should be conducted on the feasibility of analyzing dynamic mobile malware \citep{qbeitah2018dynamic} through file analysis attached to any spam or suspicious message links. We also envision that mobile cyber deception research \citep{sajid_dodgetron2020,sajid_2021_soda} can be an exciting and challenging direction to explore, where we can find malware types by analyzing the attached contents and send back fake data towards the attacker, as it expects to deceive them.

\noindent\textbf{[ii] User and Data Privacy in Defense:} We also envision a rise in the adoption of privacy-preserving ML/AI approaches, such as federated learning, to address the concerns of privacy while analyzing private incoming messages \citep{vats2024federated, remmide2024privacy_federated}. Moreover, the Homomorphic Encryption mechanism \citep{sun2018private} can enable collaborative model training without compromising user privacy, and thus, researchers can explore the feasibility of such approaches to solve the privacy concerns. 

\noindent\textbf{[iii] Blockchain for Defense:}
Exploring the integration of secure and attack-resilient proof-of-stake \textit{blockchain} technology \citep{bappy2024conchain_attackresilient,bappy2024securing_pos_blockchain} for message authentication and integrity verification is a promising research avenue. A decentralized and tamper-evident ledger can provide an additional layer of trust in SMS communications. Moreover, blockchain can be leveraged while sharing common cyber threats across multiple parties to enable trust and ensure tamper-resistant feeds \citep{2022blockchain_csm_mir_xu}. 

\subsection{Smishing Datasets}
\label{sec:dataset-limits}
Expanding research to include more \textit{diverse public datasets} while conducting cross-cultural and cross-lingual analyses can enhance the understanding of regional variations in smishing tactics. This approach contributes to the development of context-aware detection models with improved accuracy. Currently, the majority of public datasets are in the English language, with limited representation from other languages, despite some detection efforts in Arabic \citep{ghourabi2020hybrid}, Bengali \citep{Shahriyar2025Bangalabarta,kaggleBengaliSpam}, Swahili \citep{mambina2022classifying}, Portuguese \citep{ali2025moz}, Hindi \citep{githubGitHubShshnk158MultilingualSMSspamdetectionusingRNN}, and Chinese \citep{nus_chen2013creating}. 
Thus, a critical research gap is the extensive lack of high-quality and labeled smishing datasets for non-English languages, particularly for widely spoken but underserved languages (e.g., Mandarin, Spanish, French, Persian, etc.). We believe smishing datasets from more languages might enrich the current detection research (alike multi-lingual SPAM detection research \citep{ramanujam2022review_spam_mutlilingual}) to assess the effectiveness of existing defense in these regions. Specifically, while existing literature (including this paper) discusses psychological factors and phishing techniques, current available public datasets lack a standardized method for classifying and labeling messages based on these PFs/PTs. Thus, future research should focus on developing systematic classification and automated data labeling methods for messages based on these real-world attack characteristics. Also, in the current literature, all the public dataset only contains text as their major content of smishing, which may not be sufficient as attacks can happen through multi-media items (e.g., images, videos, audios) embedded within the SMS. Specifically, with the deepfake video generation capabilities, in the future, we may experience more fake video lures by attackers attached to the messages to convince users. Thus, we feel public datasets containing multimedia (image, video) attachments can also be a very important contribution in this field. These types of datasets can also excel in research in the direction of deepfake video or image detection within smishing messages. 



\subsection{Law Enforcement and Regulatory Efforts}
Currently, there is a lack of uniform and comprehensive research dedicated to the legal, regulatory, and law enforcement aspects of smishing. While automated monitoring and threat intelligence sharing are valuable for technical defense, as mentioned in Section \hyperlink{anchor:auto-monitor}{VII-A-2}, the community requires a deeper understanding and enforces more rigorous regulatory efforts on this type of fraud and scams. We find a significant lack of academic research that focuses on the penalties and specific laws related to smishing. We suggest that future work in this particular area should follow several key thematic recommendations. First, one primary theme is the need for a systematic regulatory landscape analysis with a focus on existing federal and state laws related to mobile communication fraud and their specific applicability in smishing attacks, such as impersonation, malware distribution, and financial theft. Future work can explore clearly identifying organizational structures and hierarchies that are mandated to fight and penalize these kinds of threats. Second, researchers can collect, review, and analyze previous court cases and legal precedents related to smishing with comprehensive case studies to provide concrete evidence of current legal effectiveness, jurisdictional challenges, and the enforcing penalties. Finally, we recommend exploring international and comparative law research focused on smishing regulation around the globe. This would compare the different regulations and enforcement against smishing between regions (like Europe and Asia) by looking at diverse legal frameworks (such as GDPR and other country-wise cybersecurity laws). This comparison would identify best practices to develop a stronger and more coordinated global response to find a uniform and coordinated action against smishing threats.

\section{Conclusion}
\label{sec:conclusion}
This paper comprehensively investigates the recent advancements in user susceptibility to smishing, attack landscape, defense landscape, and available datasets. Specifically, this work evaluates the existing literature to systematically characterize the current attack types and evaluate available defense countermeasures, including policy and regulatory efforts, to identify gaps in existing research and directions towards a more secure SMS communication ecosystem. 
We also conduct data analysis on all the available datasets, report summary statistics, and provide a cleaned dataset in a unified GitHub repository to aid future research. Our investigation suggests some promising future research directions to protect users from smishing threats and also leverage novel advanced techniques and tools for effective smishing defense.



\appendix



\bibliographystyle{cas-model2-names}

\bibliography{cas-refs}

\vspace{1cm}
\begin{wrapfigure}{l}{0.175\textwidth}
    \centering
    \includegraphics[width=0.15\textwidth]{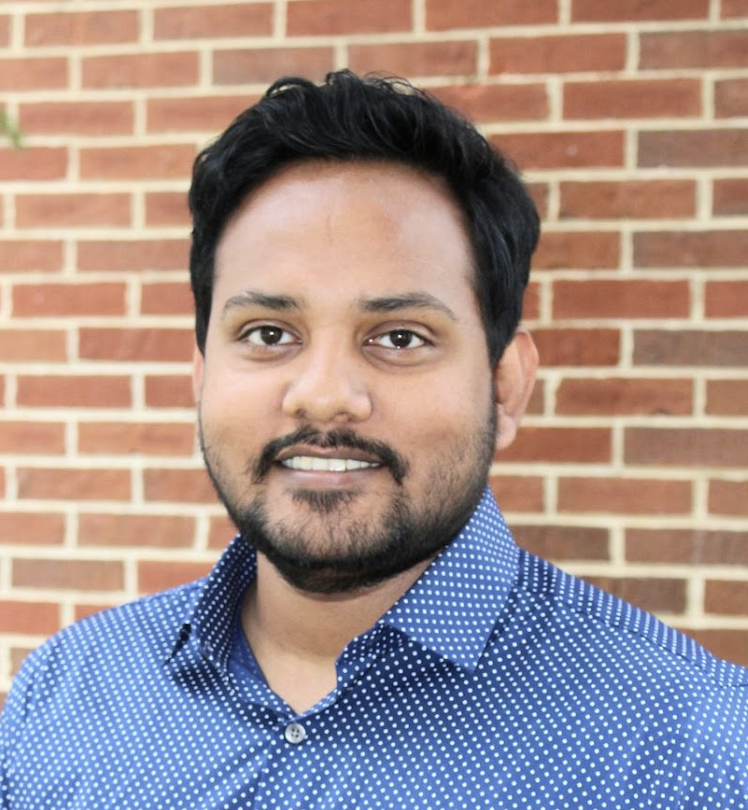}
\end{wrapfigure}

{\footnotesize \noindent \textbf{Mir Mehedi A. Pritom} is an Assistant Professor at Tennessee Tech University in Cookeville, Tennessee, where he is leading the Resilient AI-enabled Cybersecurity and Trustworthiness (ReACT) lab. He earned his B.Sc. in Computer Science and Engineering from the University of Dhaka in 2014, followed by an M.Sc. in Information Technology, specializing in information security, from the University of North Carolina at Charlotte (UNC Charlotte) in 2018. In 2022, he completed his Ph.D. in Computer Science at the University of Texas at San Antonio (UTSA). Dr. Pritom has published a number of high-impact peer-reviewed papers and posters in leading conferences, journals, and workshops such as IEEE S\&P (Oakland), IEEE CNS, IEEE GlobeCom, IEEE Blockchain, ACM CODASPY, IEEE ISI, IEEE ICCCN, and the Journal of Parallel and Distributed Computing. He has also authored a US patent. Furthermore, he has been serving as a TPC member of various prestigious security venues including ESORICS and ACM CODASPY. He is also an organizing co-chair for the ACM Secure and Trustworthy Cyber-Physical Systems (SaT-CPS) workshop that is co-located with the ACM CODASPY. \par}

\vspace{1cm}
\begin{wrapfigure}{l}{0.175\textwidth}
    \centering
    \includegraphics[width=0.15\textwidth]{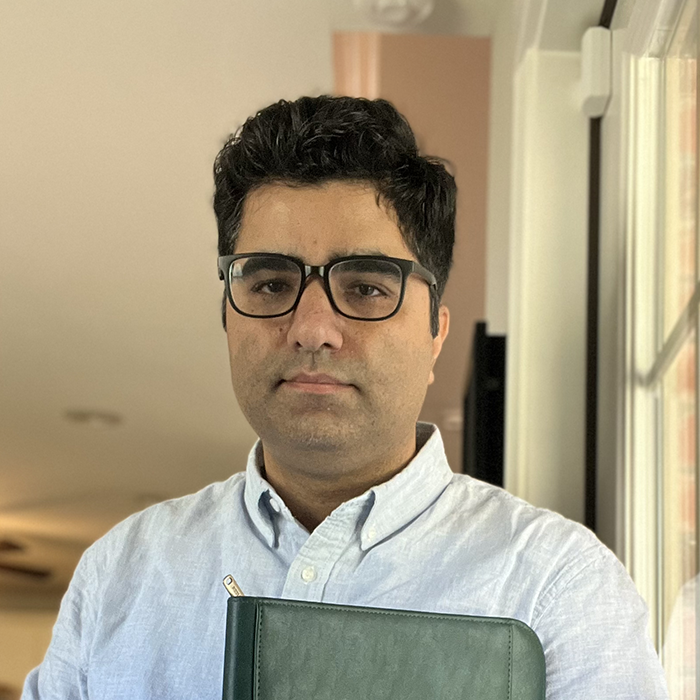}
\end{wrapfigure}

{\footnotesize \noindent \textbf{Seyed Mohammad Sanjari} is currently pursuing a Ph.D. in Computer Science at Tennessee Technological University, where he is a research assistant within the ReACT lab led by Dr. Pritom. He completed a master’s degree in Information Technology at Arkansas Tech University in 2024, a Master's of Software Engineering at the University of Guilan, Iran, in 2016, and a Bachelor’s degree in Computer Engineering at Payame Noor University, Iran, in 2012. His research primarily focuses on advanced AI approaches for tackling complex social engineering cyber threats. \par}

\vspace{1cm}
\begin{wrapfigure}{l}{0.175\textwidth}
    \centering
    \includegraphics[width=0.175\textwidth, height=0.15\textwidth]{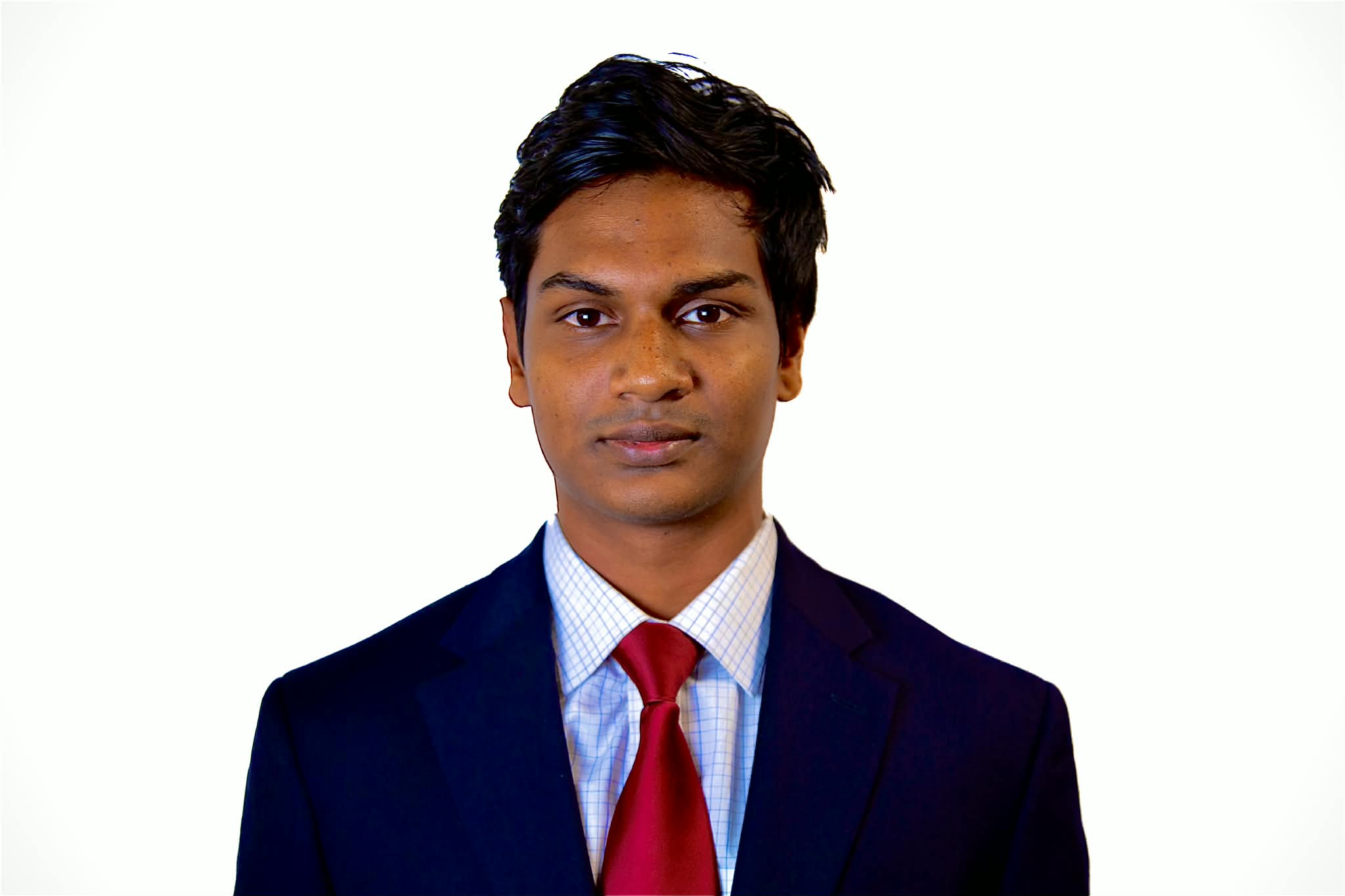}
\end{wrapfigure}

{\footnotesize \noindent \textbf{Maraz Mia} is a Ph.D. student at Tennessee Tech University, where he works as a research assistant in the ReACT lab led by Dr. Pritom. He completed his  B.Sc. in Software Engineering from Shahjalal University of Science and Technology (SUST), Bangladesh, in 2023. His research interests comprise the application of Machine Learning (ML), Neural Networks (NN), Natural Language Processing (NLP), Artificial Intelligence (AI), and explainable AI (XAI) to mitigate various cyber attacks and threats and enhance trust in cyber decision-making. \par}

\vspace{1cm}
\begin{wrapfigure}{l}{0.175\textwidth}
    \centering
    \includegraphics[width=0.15\textwidth]{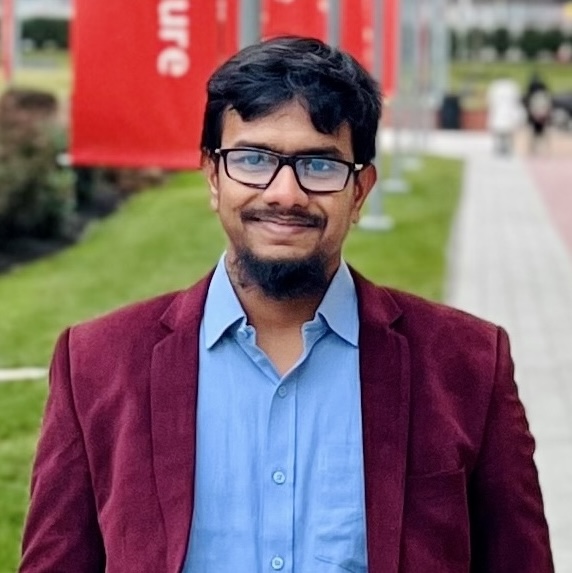}
\end{wrapfigure}

{\footnotesize \noindent \textbf{Ashfak Md Shibli} is currently working as a software engineer at Athlete Den to democratize intelligent athletic activity detection and analysis. He has completed his MS in Computer Science from Tennessee Tech University under the supervision of Dr. Pritom in Summer of 2024. Previously, he graduated from Chittagong University of Engineering and Technology (CUET), Bangladesh with B.Sc. in Computer Science and Engineering in 2017. He has published his research articles and posters in IEEE S\&P, IEEE ISDFS, IEEE ICOCO, and the Springer Emerging Technologies in Data Mining and Information Security (IEMIS) conference. 
He also worked as a lead software engineer (Mobile DevOps) at Samsung R\&D Institute Bangladesh from 2018 to 2022. \par}

\vspace{1cm}
\begin{wrapfigure}{l}{0.175\textwidth}
    \centering
    \includegraphics[width=0.15\textwidth]{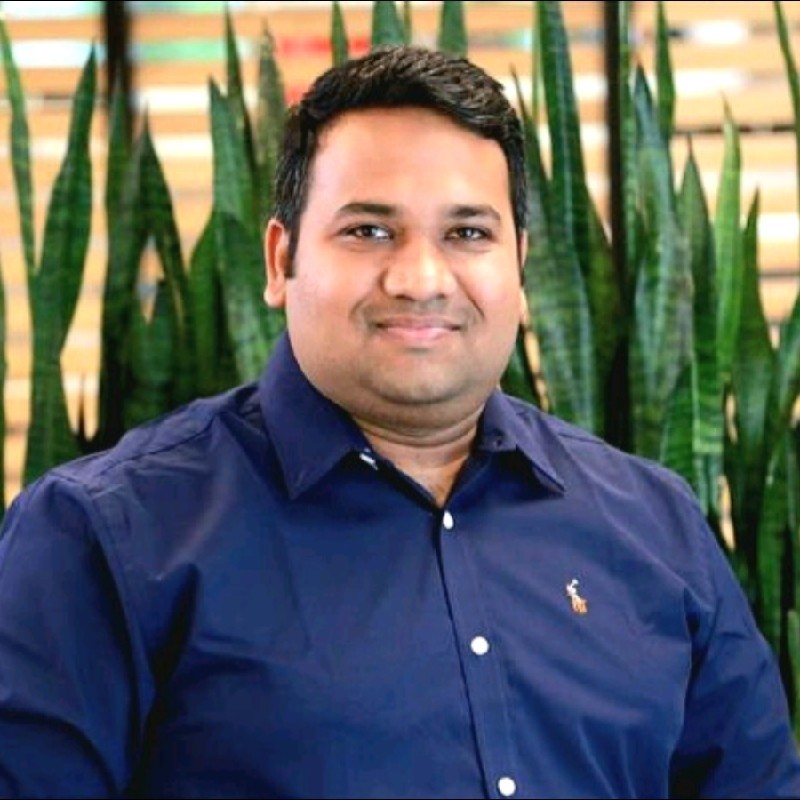}
\end{wrapfigure}

{\footnotesize \noindent \textbf{S M Mostaq Hossain} is a PhD student at Tennessee Tech University. He is currently working as a Research Assistant in the Cybersecurity and Blockchain Research Group with a keen focus on cybersecurity. He earned his MS from Tennessee Technological University and his B.Sc. in Computer Science from the Chittagong University of Engineering and Technology (CUET), Bangladesh. His research encompasses a variety of advanced techniques, including Blockchain Smart Contract, Machine Learning, Explainable AI, and Natural Language Processing, all geared toward strengthening digital security and enhancing transparency in cyber decision-making. Prior to his academic pursuits, he held a position at Samsung R\&D Institute Bangladesh Ltd., where he contributed from 2016 to 2022 as an application developer for the cross-platform environment. \par}

\vspace{1cm}
\begin{wrapfigure}{l}{0.175\textwidth}
    \centering
    \includegraphics[width=0.15\textwidth]{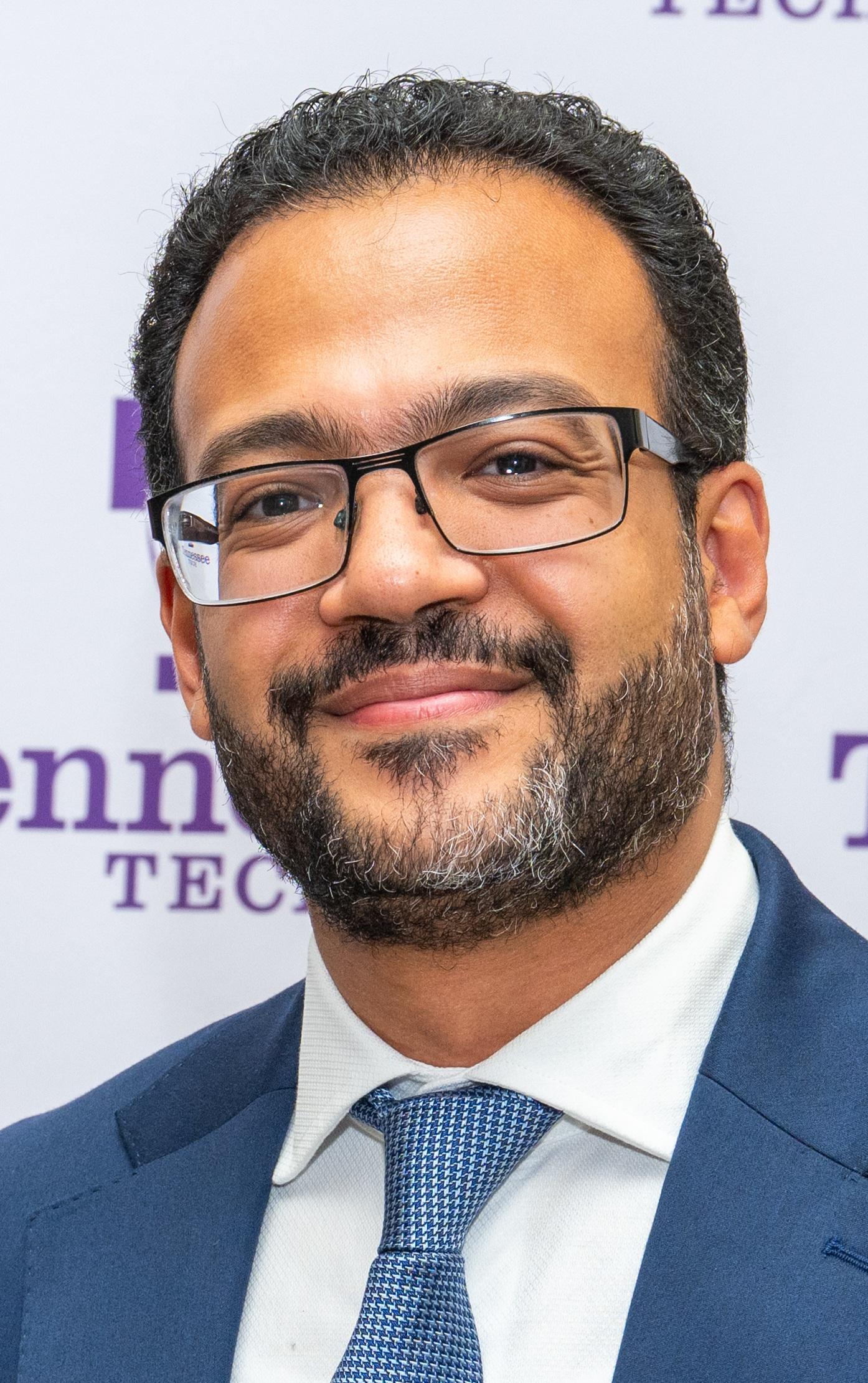}
\end{wrapfigure}

{\footnotesize \noindent \textbf{Muhammad Ismail} (S’10-M’13-SM’17) received the B.Sc. (Hons.) and M.Sc. degrees in electrical engineering (electronics and communications) from Ain Shams University, Cairo, Egypt, in 2007 and 2009, respectively, and the Ph.D. degree in electrical and computer engineering from the University of Waterloo, Waterloo, ON, Canada, in 2013. He is the Director of the Cybersecurity Education, Research, and Outreach Center (CEROC) and an Associate Professor with the Department of Computer Science, Tennessee Technological University, Cookeville, TN, USA. He was a co-recipient of the Best Paper Awards in the IEEE ICC 2014, the IEEE GLOBECOM 2014, the SGRE 2015 and 2024, the Green 2016, and the IEEE IS 2020, and the Best Conference Paper Award from the IEEE Communications Society Technical Committee on Green Communications and Networking for his publication in IEEE ICC 2019. He was the Track Chair in the IEEE Globecom 2024, the Track Co-Chair in the IEEE SmartGridComm 2023, the Workshop Co-Chair of the IEEE Greencom 2018, the Track Co-Chair of the IEEE VTC 2017 and 2016, the Publicity and Publication Co-Chair of the CROWNCOM 2015, and the Web-Chair of the IEEE INFOCOM 2014. He was an Associate Editor of the IET Communications, PHYCOM, the IEEE Transactions on Green Communications and Networking, the IEEE Internet of Things Journal, and the IEEE Transactions on Vehicular Technology. He was an Editorial Assistant of the IEEE Transactions on Vehicular Technology, from 2011 to 2013. He has been a technical reviewer of several IEEE conferences and journals. \par}

\vspace{1cm}
\begin{wrapfigure}{l}{0.175\textwidth}
    \centering
    \includegraphics[width=0.15\textwidth]{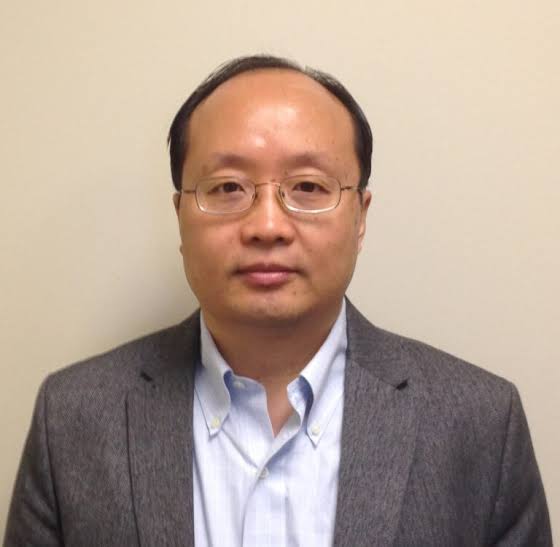}
\end{wrapfigure}

{\footnotesize \noindent \textbf{Shouhuai Xu} (Senior Member, IEEE) 
received the Ph.D. degree in computer science from Fudan University, Shanghai, China, in 2000. He is the Gallogly Chair Professor with the Department of Computer Science, University of Colorado Colorado Springs. He pioneered the Cybersecurity Dynamics approach as foundation for the emerging science of cybersecurity, with three pillars: first-principle cybersecurity modeling and analysis (the $x$-axis); cybersecurity data analytics (the $y$-axis); and cybersecurity metrics (the $z$-axis). He co-initiated the International Conference on Science of Cyber Security (SciSec) and is serving as its Steering Committee chair. He was an associate editor of IEEE Transactions on Dependable and Secure Computing (IEEE TDSC), IEEE Transactions on Information Forensics and Security (IEEE TIFS), and IEEE Transactions on Network Science and Engineering (IEEE TNSE). He is a Distinguished Member of ACM. More information about his research can be found at \url{https://xu-lab.org/} \par}

\end{document}